\documentclass[prl,amsmath,amssymb,notitlepage,twocolumn,nofootinbib,superscriptaddress,longbibliography]{revtex4-2}
\usepackage{amsfonts}
\usepackage{textcomp}
\usepackage{times}
\usepackage{graphicx}
\usepackage{float}
\usepackage{latexsym,amsmath,amssymb,bm,euscript}
\usepackage{color}
\usepackage{subfigure}
\usepackage{epstopdf}
\usepackage[colorlinks=true,linkcolor=darkblue,citecolor=darkblue,urlcolor=darkblue]{hyperref}
\usepackage{hyperref}
\usepackage{soul}
\usepackage[normalem]{ulem}
\usepackage{mathrsfs}
\usepackage{amsmath}
\usepackage{xspace}
\usepackage{natbib}
\usepackage{ulem}
\usepackage{etoolbox}
\usepackage{tikz}

\usepackage{xcolor}
\definecolor{darkblue}{rgb}{0.8,0.,0.}

\newcommand{\opr}[1]{\operatorname{#1}}
\newcommand{\ket}[1]{|#1\rangle}

\newcommand{\nc}{\newcommand}
\nc{\zjx}[1]{{\color{orange}{\textbf{[JX: #1]}}}}
\usepackage[normalem]{ulem}
\usepackage{comment}
\newcommand{\avg}[1]{\langle#1\rangle}
\newcommand{\up}{\uparrow}
\newcommand{\dn}{\downarrow}

\begin{document}

%%%
\title{Competing and Intertwined Orders in Boson-Doped Mott Antiferromagnets}

%%%
\author{Xin Lu}
\thanks{These authors contributed equally to this work.}
\affiliation{School of Physics, Beihang University, Beijing 100191, China}

\author{Jia-Xin Zhang}
\thanks{These authors contributed equally to this work.}
\affiliation{French American Center for Theoretical Science, CNRS, KITP, Santa Barbara, California 93106-4030, USA}
\affiliation{Kavli Institute for Theoretical Physics, University of California, Santa Barbara, California 93106-4030, USA}

\author{Lukas Homeier}
\thanks{These authors contributed equally to this work.}
\affiliation{JILA, Department of Physics, University of Colorado, Boulder, CO, 80309, USA}
\affiliation{Center for Theory of Quantum Matter, University of Colorado, Boulder, CO, 80309, USA}

\author{Shou-Shu Gong}
\email{shoushu.gong@gbu.edu.cn}
\affiliation{School of Physical Sciences, Great Bay University, Dongguan 523000, China, and \\
Great Bay Institute for Advanced Study, Dongguan 523000, China}

\author{D. N. Sheng}  
\email{donna.sheng1@csun.edu}
\affiliation{Department of Physics and Astronomy, California State University Northridge, Northridge, California 91330, USA}

\author{Zheng-Yu Weng}
\email{weng@mail.tsinghua.edu.cn}
\affiliation{Institute for Advanced Study, Tsinghua University, Beijing 100084, China}

\date{\today}% It is always \today, today,

\begin{abstract}
Inspired by the recent experimental advances in cold atom quantum simulators, we explore the experimentally implemented bosonic $t$-$t'$-$J$ model on the square lattice using large-scale density matrix renormalization group simulations. 
By tuning the doping level $\delta$ and hopping ratio $t'/t$, we uncover six distinct quantum phases, several of which go far beyond the conventional paradigm of phase-coherent superfluidity (SF) expected for bosonic systems. In particular, in the presence of antiferromagnetic (AFM) order, doped holes are tightly bound into pairs, giving rise to a pair density wave (PDW) phase at low doping and small $|t'/t|$, which is suppressed on the $t'<0$ side, resulting in a disordered PDW state that lacks coherence of either individual bosons or pairs. 
Upon further doping, bosons can regain phase coherence and form a SF* state, characterized by condensation at emergent incommensurate momenta concurrent with an incommensurate magnetic order.
On the $t'>0$ side, the sign-induced kinetic frustration inherently disfavors local AFM correlations, leading to a phase separation in which doped holes cluster into ferromagnetic (FM) domains spatially separated by undoped AFM regions. Upon further doping, this inhomogeneous state evolves into a uniform SF + $xy$-FM phase. 
Finally, we propose a concrete experimental scheme to realize both signs of $t'/t$ in Rydberg tweezer arrays, with an explicit mapping between model parameters and experimentally accessible regimes. 
Our results reveal competing and intertwined orders in doped antiferromagnets, which are relevant to central issues in high-$T_c$ superconductivity, reflecting the frustrated interplay between doped holes and spin background.
\end{abstract}
 
\maketitle

{\it Introduction.---} Developing a comprehensive understanding of doped antiferromagnetic (AFM) Mott insulators stands as one of the central challenges in modern condensed matter physics, with direct relevance to unconventional superconductors~\cite{Lee2006,Keimer2015,Proust_ARCMP_2019}. A widely used theoretical framework for capturing essential physics is the paradigmatic fermionic $t$–$J$ model~\cite{Zhang1988,Masao_RPP_2008}.
During the past three decades of intensive studies, significant progress has been achieved in the square-lattice fermionic $t$–$J$ model through unbiased numerical simulations. Although some conclusions remain controversial, there is a broad consensus that the next-nearest-neighbor hopping $t'$ plays a pivotal role for superconductivity (SC)~\cite{Jiang_PRL_2021,Gong_PRL_2021,White_PNAS_2021,Jiang_PRB_2023,lu2024emergent,tJ_Feng_2023}.

In parallel with numerical simulations, ultracold atom quantum simulators based on optical lattices and Rydberg tweezer arrays have emerged as powerful platforms to explore doped Mott antiferromagnets~\cite{Koepsell_nature_2019,Hartke_PRL_2020,Bohrdt_AP2021,Gall_nature_2021,Pan_nature_2024,Prichard_nature_2024,Annabelle_2024,bakr2025,Muqing_nature_2025,Qiao_2025}.
Recently, the bosonic $t$-$J$ model with AFM interactions has been proposed and experimentally realized in a Rydberg tweezer platform with three highly excited atomic states~\cite{Lukas_PRL_2024,Qiao_2025}, whose dipole-dipole and van-der-Waals interactions naturally realize a $t$–$t^\prime$–$J$ model with hard-core bosonic hole dopants and AFM spin interactions~\cite{Lukas_PRL_2024}. Alternative implementations have also been explored, including Bose-Hubbard systems with spin-dependent interactions~\cite{Wolfgang_PRX_2021}, staggered fields~\cite{Sun_np_2021} or negative temperature states~\cite{Annabelle_2024,Duan_PRL_2003,Timothy_2024}. Related bosonic physics has also been proposed in solid-state systems through exciton doping in van der Waals heterostructures~\cite{YHZhang2022, Yang2024}. Intriguingly, despite the fundamental difference in statistics, the bosonic $t$–$J$ model still exhibits remarkable similarities to its fermionic counterpart, including the emergence of SC~\cite{Zhang_2024} and stripe order~\cite{Timothy_2024}. However, current experimental implementations are limited to $t'>0$, which hinders a comprehensive exploration of the global quantum phase diagram, including possible phase transitions as well as competing and intertwined orders that may emerge in different regimes. 

In this Letter, we investigate the bosonic $t$–$t'$–$J$ model on the square lattice using large-scale density matrix renormalization group (DMRG) simulations. By tuning the doping level $\delta$ and the hopping ratio $t'/t$, we map out the global phase diagram on a four-leg cylinder, supplemented by eight-leg results. We uncover a rich landscape of unconventional quantum phases that depart significantly from the standard picture of simple superfluidity (SF) at commensurate momenta. These include, among others, a pair density wave (PDW) phase with spatially modulated pairing, a disordered PDW (dPDW) phase lacking both single-boson and pair coherence, and an exotic SF* phase characterized by condensation at emergent incommensurate momenta.
Finally, we propose an experimental scheme using Rydberg tweezer arrays to realize both $t'>0$ and $t'<0$ regimes in doped Mott antiferromagnets. Our work thus provides a solid theoretical foundation for future Rydberg tweezer experiments.

{\it Model and method.---}The bosonic $t$-$t'$-$J$ model is defined as
\begin{eqnarray}
\begin{aligned}\label{H-tJ}
        \mathcal{H} =& -t\sum_{\left\langle i,j\right\rangle ,\sigma } {\hat{\mathcal{B}} }_{i,\sigma }^{\dagger } {\hat{\mathcal{B}} }_{j,\sigma }- t^{\prime } \sum_{\left\langle \left\langle i,j\right\rangle \right\rangle ,\sigma } {\hat{\mathcal{B}} }_{i,\sigma }^{\dagger } {\hat{\mathcal{B}} }_{j,\sigma } +\mathrm{H}.\mathrm{c}.\\
        &+J\sum_{\left\langle i,j\right\rangle } ({\hat{\mathbf{S}} }_i \cdot {\hat{\mathbf{S}} }_j - {\hat{n} }_i {\hat{n} }_j/4),
\end{aligned}
\end{eqnarray}
where ${\hat{\mathcal{B}} }_{i,\sigma }^{\dagger }$ (${\hat{\mathcal{B}} }_{i,\sigma }$) is the spinful hard-core boson creation (annihilation) operator with spin $\sigma = \uparrow, \downarrow$ on site $i=(x_i,y_i)$, ${\hat{\mathbf{S}} }_i =\frac{1}{2}\sum_{\alpha ,\alpha^{\prime } } {\hat{\mathcal{B}} }_{i,\alpha }^{\dagger } {\boldsymbol{\sigma}}_{\alpha \alpha^{\prime } } {\hat{\mathcal{B}} }_{i,\alpha^{\prime } }$ is the spin-$1/2$ operator and ${\hat{n} }_i =\sum_{\sigma } {\hat{\mathcal{B}} }_{i,\sigma }^{\dagger } {\hat{\mathcal{B}} }_{i,\sigma }$ is the charge density operator. 
The Hilbert space for each site is constrained by no-double occupancy.
Motivated by recent experiments~\cite{Qiao_2025}, where the hopping amplitudes decay as $t \propto r^{-3}$ and spin interactions decay as $J \propto r^{-6}$ with distance $r$, we therefore study the model that includes hopping up to the next-nearest-neighbor (NNN) sites while restricting spin and density interactions to the nearest-neighbor (NN) sites.
We set $t/J=3$ and focus on the doping regime $1/24 \le \delta \le 1/3$ on four-leg cylinder, complemented by selected doping levels on eight-leg cylinder. 
We tune the NNN hopping magnitude within the range $t^{\prime }\in \left\lbrack -0\ldotp 3t,0\ldotp 3t\right\rbrack $,
which broadly covers the range accessible by current experimental platforms~\cite{Qiao_2025}, as well as the extended parameter range enabled by the experimental scheme proposed later in our work [see Supplemental Material (SM) for more details~\cite{SM} ].

We consider a cylindrical geometry with open (periodic) boundary conditions along the $x$ ($y$) direction for four- and eight-leg cylinders. 
The length (width) of the lattice is denoted as $L_x$ ($L_y$), giving the total site number $N = L_x \times L_y$. 
The doping ratio $\delta$ is defined as $\delta = N_h / N$ ($N_h$ is the number of doped holes). 
We solve the ground state of the system by DMRG~\cite{dmrg_white_1992,dmrg_white_1993,dmrg_ulrich_2005,dmrg_ulrich_2011} calculations with $U(1)_{\mathrm{charge}} \times U(1)_{\mathrm{spin}}$ symmetries implemented, and keep the maximum bond dimensions up to $D=10000$ for $L_x \le 36$ and $D=48000$ for $L_x \le 80$ on four-leg systems. Besides, we perform calculations with $SU(2)_{\mathrm{spin}}$ symmetry implemented, and keep the maximum bond dimensions up to $D=20000$ symmetric multiplets (equivalent to $\sim72000$ $U(1)$ states) for $L_x = 12-24$ on eight-leg systems. All calculations ensure accurate results with the typical truncation error $\epsilon \sim{10}^{-6}$.

\begin{figure}
   \includegraphics[width=0.47\textwidth,angle=0]{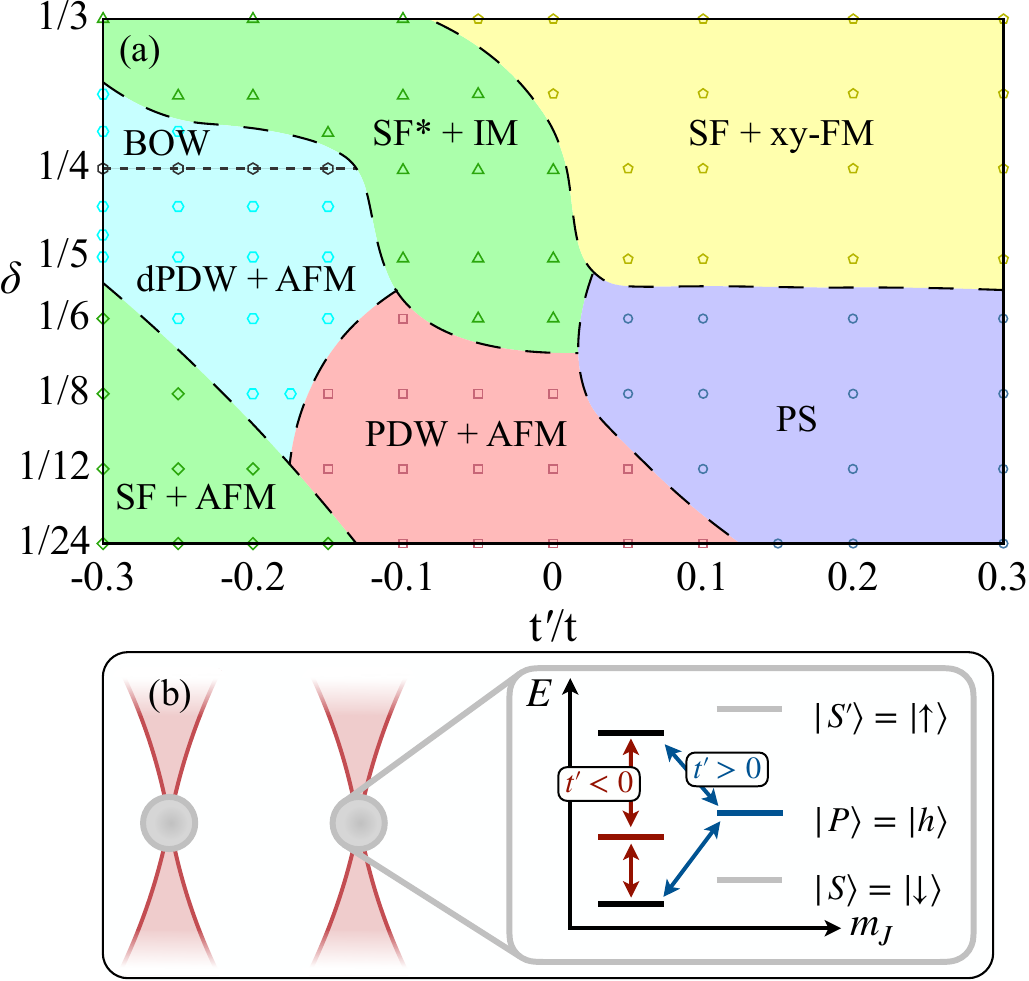}
   \caption{\label{Phase_Diagram}
   \textbf{Phase diagram and Rydberg experimental scheme.} (a) Phase diagram of the bosonic $t$-$t'$-$J$ model on four-leg cylinder. Within $-0.3 \le t' /t \le 0.3$ and $1/24\le \delta \le1/3$, we identify a phase separation (PS) and a SF+$xy$-ferromagnetic (FM) phase on the $t' \ge 0$ side; a SF+AFM phase and a SF*+incommensurate magnetism (IM) phase on the $t' \le 0$ side; a PDW+AFM phase and a dPDW+AFM phase sandwiched by SF+AFM and SF*+IM phases; a bond order wave (BOW) state at the special $\delta=1/4$. The symbols denote the calculated parameter points. 
   (b) The $t$-$t'$-$J$ model with hard-core bosonic holes can be implemented in three Rydberg levels. The tunneling term arises from dipole-dipole exchange interactions between $\ket{S}$ and $\ket{P}$ states. Without affecting the spin interactions, we have a freedom in choosing the magnetic sublevel of the $\ket{P}$ state (hole state) allowing one to implement both signs of tunneling $t'/t$ after a gauge transformation.
   }
\end{figure}

{\it Phase diagram.---} We summarize the phases of the four-leg system in Fig.~\ref{Phase_Diagram}(a).
At $t'=0$, we identify three phases with increased doping levels.
In the PDW+AFM phase at lower doping, the doped holes form tightly bound pairs that condense into a PDW state with $(\pi,\pi)$ spatial modulation on top of an AFM spin background~\cite{Zhang_2024}.
In the SF+$xy$-FM phase at larger doping ratios, the doped holes condense at the momentum $(0,0)$ with an in-plane $xy$-FM order~\cite{SM,Zhang_2024,Timothy_2024}. We find that these two phases are bridged by an intermediate phase denoted by SF*+IM,
in which the single holes are condensed at incommensurate momenta $(\pm k^*,0)$ and the Néel AFM order gives way to an IM order whose wave vector $2k^*$ is locked to the hole condensate. At $t' > 0$, we show that the ground state is a PS as an intermediate phase to replace the SF*+IM phase.
Here the doped holes prefer to cluster in localized regions, forming hole-rich FM phases, while other regions remain undoped AFM phases.
At $t' < 0$, the SF*+IM phase extends to a wider regime at higher doping, while the PDW+AFM phase at lower doping is turned into two new phases where the quasi-long-range AFM order is still maintained, one is the dPDW phase, characterized by the survival of local hole pairing without long-range coherence, either for individual bosons or for the pairs, resulting in only short-ranged pairing correlations.
At smaller doping, a large $|t'|$ can significantly lower the single-hole energy minima at momentum $(\pm\pi,0)$ and $(0,\pm\pi)$ to lead to a single-hole Bose-Einstein condensation (BEC), which is denoted by SF+AFM. 
Lastly, there is a BOW state at special $\delta=1/4$ (cf. Fig. S14). The phase boundaries in Fig.~\ref{Phase_Diagram}(a) are determined by examining the evolution of multiple physical quantities with respect to $t'$ and $\delta$, analyzing the distinctions among these quantities, and cross-checking them in several independent ways.

\begin{figure}
   \includegraphics[width=0.48\textwidth,angle=0]{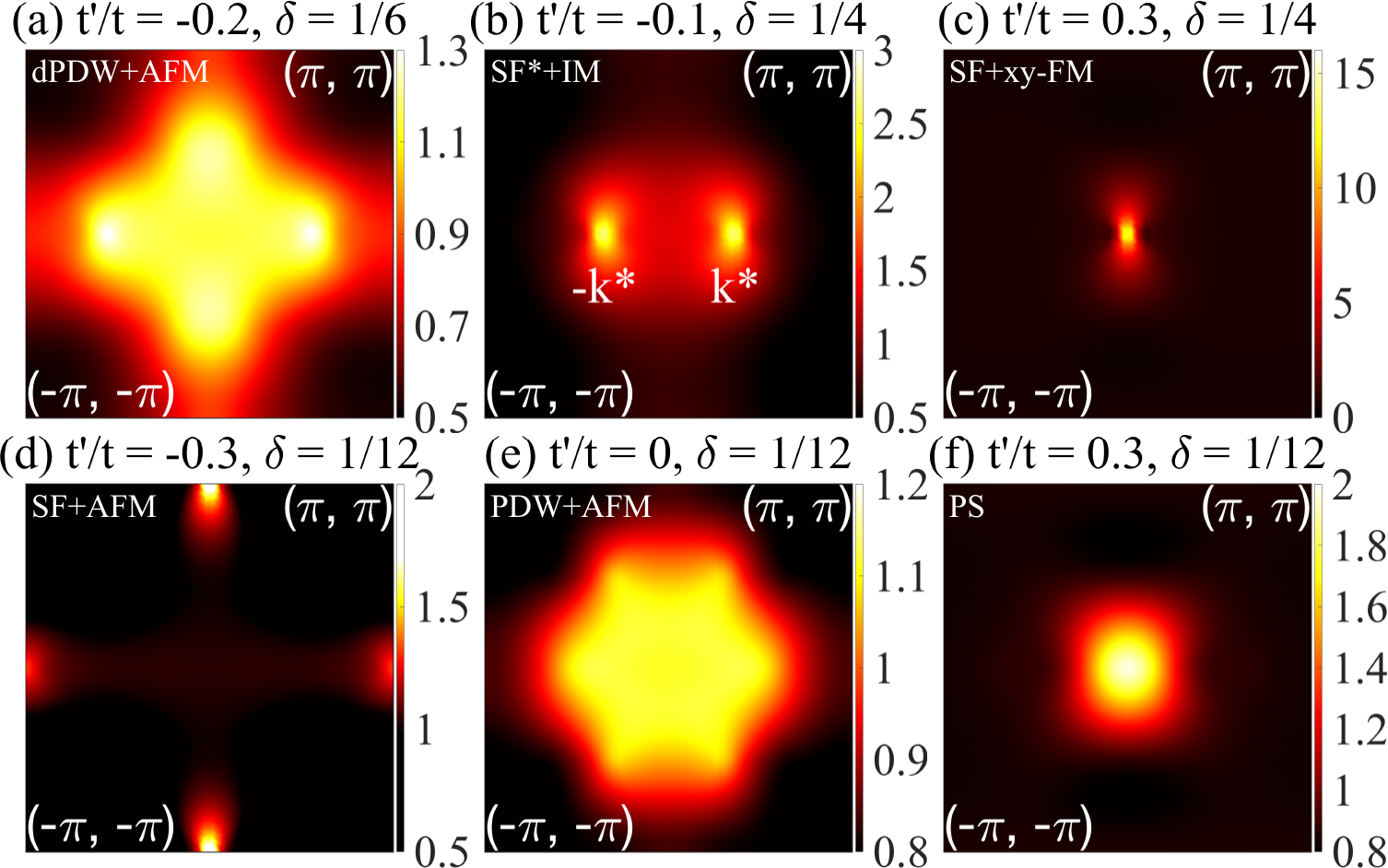}
   \caption{\label{Momentum}
   \textbf{Momentum distribution $n(\bf k)$.}
   (a) dPDW+AFM phase, (b) SF*+IM phase, (c) SF+FM phase, (d) SF+AFM phase, (e) PDW+AFM phase, and (f) PS. Here all the $n(\bf k)$ are obtained by taking the
   Fourier transformation for the all-to-all single-boson correlations.
   }
\end{figure}

{\it Momentum distribution.---}
In Fig.~\ref{Momentum}, we present the representative momentum distribution $n(\mathbf{k}) = (1/N) \sum_{i,j,\sigma} \langle {\hat{\mathcal{B}} }_{i,\sigma}^{\dagger } {\hat{\mathcal{B}} }_{j,\sigma} \rangle e^{i\mathbf{k}\cdot \left({\mathbf{r}}_i -{\mathbf{r}}_j \right)}$ in different phases. As evident from Figs.~\ref{Momentum}(b-d), $n(\bf k)$ exhibits sharp peaks signaling BEC of the single bosons in the ground states. 
Specifically, a SF of doped holes at momentum $(0,0)$ occurs in the spin FM background in Fig.~\ref{Momentum}(c) without pairing. Fig.~\ref{Momentum}(f) further shows a broadened peak at $(0,0)$ in the PS phase where the doped holes are still unpaired and concentrated in the stripe-like FM regions (cf. Fig. S1).
On the other hand, in the SF+AFM phase [Fig.~\ref{Momentum}(d)], bosonic holes are condensed at $(\pm\pi,0)$ and $(0,\pm\pi)$, which coexist with the background AFM order due to the dominant NNN hopping $t'<0$ (cf. Fig. S4). Fig.~\ref{Momentum}(b) features the most anomalous condensation of holes at emerging momenta $(\pm k^*,0)$ in the so-called SF*+IM phase. Here, the incommensurate $k^*$ depends on $\delta$ and $t'/t$ (cf. Fig. S7), signaling that the hole condensation has a density modulation.
The accompanied spin density wave shows a related wavevector $2k^*$ (see below), which is also identified in the wider eight-leg system (cf. Fig.~\ref{IM_8leg}).

By contrast, unlike the single-hole SF at large negative $t'$ and small doping (i.e., the SF+AFM phase), we find that holes are generally paired in an AFM background, where the single-hole momentum distribution shows incoherent or broad features without BEC as shown in Figs.~\ref{Momentum}(a) and (e), respectively. In the following, we further examine the corresponding spin structure factors.

\begin{figure}
   \includegraphics[width=0.463\textwidth,angle=0]{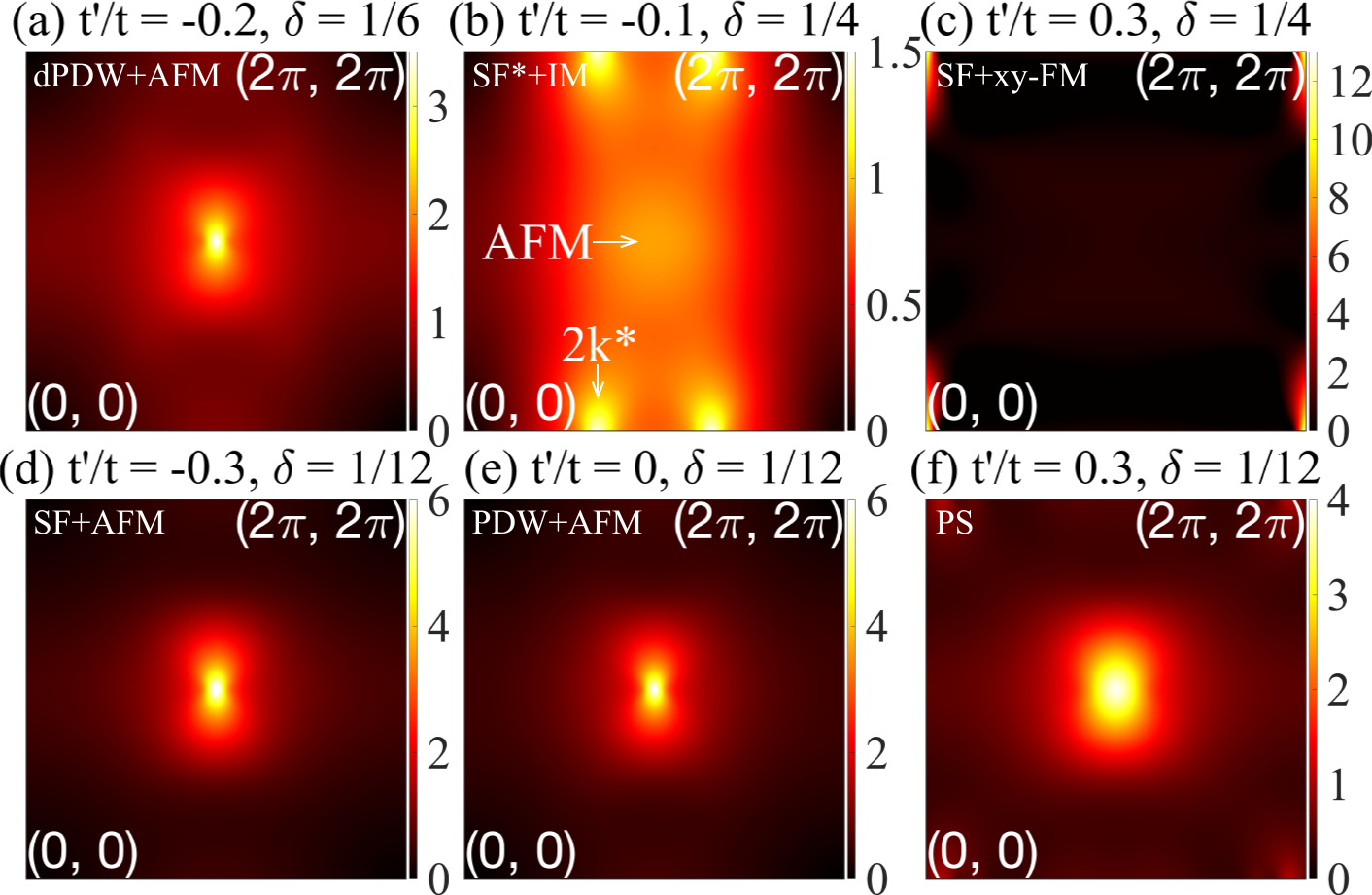}
   \caption{\label{Spin_Structure}
   \textbf{Spin structure factor $S(\bf k)$.}
   (a) dPDW+AFM phase, (b) SF*+IM phase, (c) SF+FM phase, (d) SF+AFM phase, (e) PDW+AFM phase, and (f) PS. Here all the $S(\bf k)$ are obtained by taking the
   Fourier transformation for the all-to-all spin correlations.
   }
\end{figure}

{\it Spin structure factor.---}
To characterize the magnetic properties of the system, we study the spin structure factor $S(\mathbf{k}) = (1/N) \sum_{i,j} \langle {{\hat{\mathbf{S}} }_i \cdot {\hat{\mathbf{S}} }_j} \rangle e^{i\mathbf{k}\cdot \left({\mathbf{r}}_i -{\mathbf{r}}_j \right)}$. Fig.~\ref{Spin_Structure} shows the typical features of $S(\mathbf{k})$ for the six distinct phases, which correspond to the $n(\mathbf{k})$ given in Fig.~\ref{Momentum} for the same parameters. For the dPDW+AFM, SF+AFM, and PDW+AFM phases, $S(\mathbf{k})$ indicates a commensurate AFM order at $(\pi,\pi)$ with quasi-long-range spin correlations in Figs.~\ref{Spin_Structure}(a), (d), and (e), respectively. In the PS phase, such an AFM peak becomes broadened in Fig.~\ref{Spin_Structure}(f) in which a weaker FM signature at $(0,0)$ starts to appear progressively (cf. Fig. S2). On the other hand, the AFM order also disappears in the SF*+IM phase, with a hole-induced incommensurate magnetism at momentum $(2k^*,0)$, as shown in Fig.~\ref{Spin_Structure}(b). Eventually, such IM peaks move to the FM ordered state at $(0,0)$ and symmetric points in Fig.~\ref{Spin_Structure}(c).

\begin{figure}
   \includegraphics[width=0.48\textwidth,angle=0]{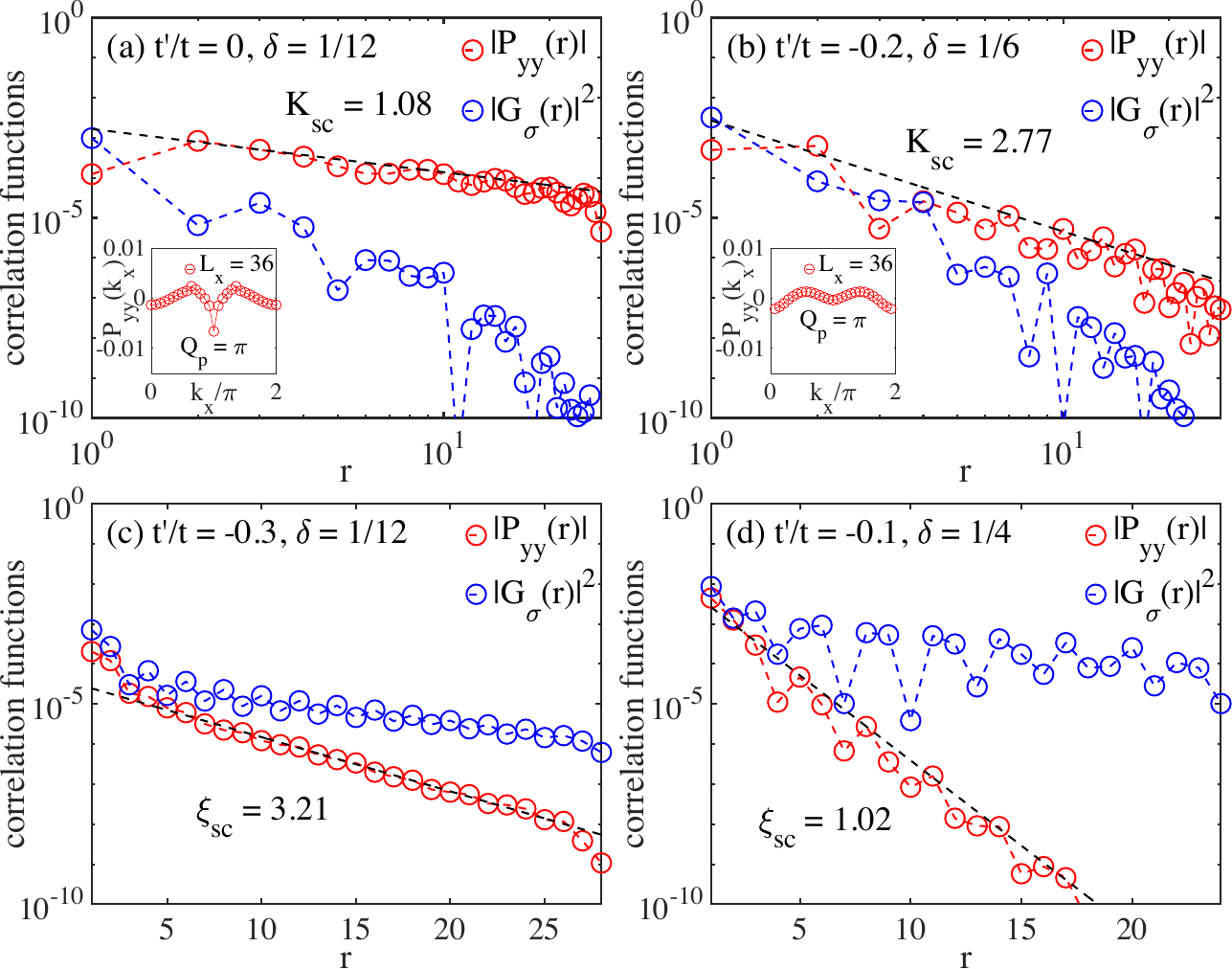}
   \caption{\label{Correlations}
   \textbf{Pairing and single-boson correlations.}
   (a) and (b) are the double-logarithmic plot of the pairing correlations $P_{yy}(r)$ and the product of two single-boson correlations $G_\sigma^2(r)$ in the PDW+AFM and dPDW+AFM phases, respectively. The power exponents $K_{\mathrm{sc}}$ are obtained by algebraic fitting with dash line. (c) and (d) are the semi-logarithmic plot of the $P_{yy}(r)$ and $G_\sigma^2(r)$ in the SF+AFM and SF*+IM phases, respectively. The correlation lengths $\xi_{\mathrm{sc}}$ are obtained by exponential fitting with dash line. The insets in (a) and (b) show the corresponding pairing structure factor $P_{yy}(k_x)$, where a singular peak or broad dome appears at $\mathbf{Q}_p=\pi$.
   }
\end{figure}

\begin{figure}
   \includegraphics[width=0.48\textwidth,angle=0]{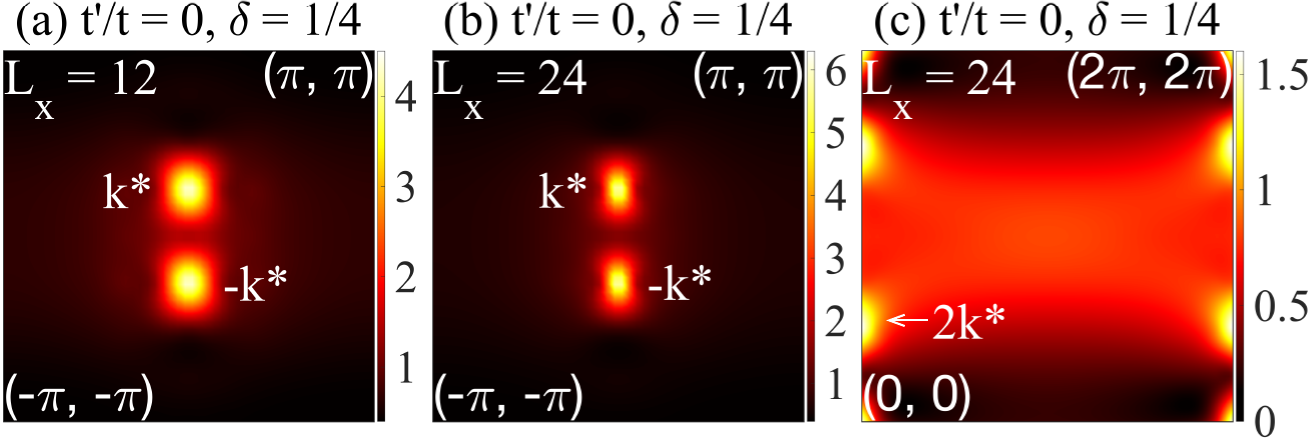}
   \caption{\label{IM_8leg}
   \textbf{Momentum distribution $n({\bf k})$ and spin structure factor $S({\bf k})$ in the SF*+IM phase on eight-leg systems.}
   (a) and (b) are $n(\bf k)$ at system length $L_{x}=12$ and $24$ for $t'/t=0$, $\delta=1/4$. (c) $S(\bf k)$ at $L_{x}=24$ for $t'/t=0$, $\delta=1/4$. $n(\bf k)$ and $S(\bf k)$ are obtained by taking the Fourier transformation for the all-to-all correlations.
   }
\end{figure}

{\it Correlation functions.---}
In Fig.~\ref{Correlations}, we further explore SC in different phases by examining the spin-singlet pairing correlations $P_{\alpha\beta }(\mathbf{r}) = \langle {\hat{\Delta}}_{\alpha }^{\dagger } ({\mathbf{r}}_0) {\hat{\Delta}}_{\beta }({\mathbf{r}}_0 +\mathbf{r}) \rangle$, where the spin-singlet pairing operator reads ${\hat{\Delta} }_{\alpha } \left(\mathbf{r}\right)=({\hat{\mathcal{B}} }_{\mathbf{r}\uparrow } {\hat{\mathcal{B}} }_{\mathbf{r}+{\mathbf{e}}_{\alpha } \downarrow } -{\hat{\mathcal{B}} }_{\mathbf{r}\downarrow } {\hat{\mathcal{B}} }_{\mathbf{r}+{\mathbf{e}}_{\alpha } \uparrow })/\sqrt{2}$ and ${\mathit{\mathbf{e}}}_{\alpha = x,y}$ denote the unit vectors along the $x$ and $y$ directions. In the PDW+AFM phase [Fig.~\ref{Correlations}(a)], we find that $P_{yy}(r)$ can be well fitted by the power-law behavior $P_{yy}(r) \sim r^{-K_{\mathrm{sc}} }$ with $K_{\mathrm{sc}} \simeq 1.08$, characterizing a quasi-long-range SC order. In the inset of Fig.~\ref{Correlations}(a), we also present the corresponding pairing structure factor $P_{yy}(k_x )=(1/N)\sum_{i,j} P_{yy} \left(r\right)e^{{ik}_x \cdot \left(x_i -x_j \right)}$, where a singular peak at $\mathbf{Q}_p=\pi$ confirms the instability towards a $2$-period PDW at zero temperature.
In contrast, $P_{yy}(r)$ decays faster with $K_{\mathrm{sc}} \simeq 2.77$ (in the power-law fitting) in the dPDW+AFM phase [Fig.~\ref{Correlations}(b)], and $P(k_x)$ also shows a broad dome reflecting the significantly weaker SC order. In order to further determine whether there is any residual hole-pairing in this phase, we also calculated the single-boson correlations $G_{\sigma}(r) = \langle  \hat{\mathcal{B}}_{x,y,\sigma }^{\dagger } \hat{\mathcal{B}}_{x+r,y,\sigma } \rangle$ and found that $P_{yy}(r)$ is still much stronger than the product of two single-boson correlations $G_\sigma^2(r)$ in the dPDW+AFM phase, demonstrating that the doped holes are still paired, although not as strong as in the PDW+AFM phase. 
For the SF+AFM and SF*+IM phases, which are distinct from the PDW+AFM and dPDW+AFM phases, we observe that $G_\sigma^2(r)$ is even stronger than $P_{yy}(r)$ as shown in Figs.~\ref{Correlations}(c) and (d), consistent with the essence of single-boson condensate.

{\it Robustness of the SF*+IM phase.---}
To confirm the robustness of the exotic SF*+IM phase, we further simulate the model on wider eight-leg cylinders. 
As shown in Fig.~\ref{IM_8leg}(a), the bosons condense at $(0, \pm k^*)$, and the condensation is further enhanced with increasing both system length $L_x$  [Fig.~\ref{IM_8leg}(b)] and system width $L_y$ (cf. Fig. S9). 
Unlike four-leg systems, here the peaks of $n(\bf k)$ are located along the $k_y$ direction.
While the condensation can spontaneously select $(0, \pm k^*)$ or $(\pm k^*, 0)$ in two-dimensional systems, a specific one may be chosen in finite-size systems due to geometry and boundary effects.
Correspondingly, the spin structure factor $S(\bf k)$ develops a consistent peak at $(0, 2k^*)$ [Fig.~\ref{IM_8leg}(c)].
With growing system width, the comparable power exponents $K_{\rm G}$ and $K_{\mathrm{s}}$ (cf. Fig. S8 and Fig. S11) strongly indicate that this SF*+IM phase may persist in two dimensions. 
We also notice the FM signature at $(0, 0)$ in Fig.~\ref{IM_8leg}(c), which originates from the $xy$-FM correlations, as this parameter point lies close to the SF+$xy$-FM phase on the eight-leg system. Inclusion of a finite $t'<0$ can suppress this FM signature.

{\it Experimental proposal.---}Recent experiments in Rydberg tweezer arrays realized a hard-core bosonic $t$-$t'$-$J$ model with $t'/t>0$~\cite{Qiao_2025,Lukas_PRL_2024}. In this scheme, three Rydberg levels $\{ \ket{S}, \ket{P},\ket{S'}\}$ of an atom are identified with the local $t$-$J$ Hilbert space, $\{ \ket{\!\!\downarrow}, \ket{\circ},\ket{\!\!\uparrow}\}$, see Fig.~\ref{Phase_Diagram}(b), such that the dipole-dipole interaction between pairs of atoms directly realizes a model similar to the Hamiltonian we study~\cite{Footnote_Rydberg}.

Here, we discuss an extension of the Rydberg scheme to access the full parameter regime, i.e. both $t'/t>0$ and $t'/t<0$: On the microscopic level, the tunneling terms~$t, t' \propto -C_3$ originate from resonant dipolar interactions with transition dipole matrix element~$C_3$ between $\ket{S}$ and~$\ket{P}$ states -- in contrast to Wannier function overlaps of ultracold atoms in optical lattices. To be explicit, the dipole matrix element~$C_3$ depends on the quantum numbers of the atomic pair state and has opposite sign for~$\Delta m =\pm 1$ and $\Delta m =0$, where~$\Delta m$ is the change of magnetic quantum number.
Hence, the global sign of tunneling~$t, t'$ can be set by the quantum number~$m_P$, while keeping~$m_{S}=m_{S'}$ fixed to leave the spin interaction unchanged.

On bipartite lattices we can always perform a local gauge transformation on one sublattice to obtain ferromagnetic nearest-neighbor tunnelings~$t \rightarrow -|t|$ regardless of the global sign of tunnelings. However, the next nearest-neighbor tunneling~$t'$ remains invariant under this transformation; hence implementing $t'/t > 0$ ($t'/t < 0$) for $\Delta m = \pm 1$ ($\Delta m = 0$), if the quantization axis is perpendicular to the atomic plane. The relative coupling strength~$|t'/t|$ depends on the interatomic distance and its relative angle~$\theta$ with respect to the quantization axis. For~$\theta=90^\circ$, the long-range dipolar interactions give rise to relatively large ratios~$|t'/t| = 2^{-3/2} \approx 0.35$ on the square lattice, and $|t'/t| = 3^{-3/2} \approx 0.19$ on the honeycomb lattice.
By exploiting the angular dependency of the dipolar interactions, the ratio~$|t'/t|$ can be tuned over a wider parameter range on the square lattice allowing one to access the full phase diagram in Fig.~\ref{Phase_Diagram}(a) [see SM for more details~\cite{SM}].

{\it Summary and discussion.---}
In this work, we present a comprehensive numerical study of the bosonic $t$-$t'$-$J$ model through large-scale DMRG simulations, and uncover a diverse class of  unconventional quantum phases beyond the standard paradigm of uniform condensation (cf. Fig.~\ref{Phase_Diagram}(a)). Specifically, at $t'=0$ and low doping, we reproduce the previously reported PDW+AFM phase, where tightly bound hole pairs exhibit long-range phase coherence, whereas individual holes remain incoherent, defying the conventional intuition that single bosons, with more favorable kinetic energy, should condense more readily than bound pairs.  As doping increases, pairing fades and single bosons regain coherence, condensing at incommensurate momenta $(\pm k^*,0)$, corresponding to the SF* phase. Notably, this incommensurate condensation persists even at $t'=0$, indicating that the emergent momentum shift originates from interaction effects rather than band structure. 
Upon further doping, such emergent momentum $k^*$ continuously evolves toward zero, eventually yielding a conventional SF+$xy$-FM phase (see Sec. D of SM~\cite{SM} for further details on this transition).
On the $t'>0$ side, dilute holes tend to cluster into larger aggregates, rather than forming coherent boson pairs as in the PDW phase. This spatially inhomogeneous configuration can be viewed as a natural interpolation via phase separation between the PDW+AFM phase and the uniform SF+$xy$-FM phase. On the $t'<0$ side, the SF*+IM phase identified at $t'=0$ persists over a broad doping range. At lower doping, we observe a quantum phase (dPDW) where holes remain paired, yet neither single bosons nor pairs exhibit long-range coherence. This phase closely resembles the pseudogap, characterized by preformed Cooper pairs that lack global phase coherence due to dilute charge carriers and strong fluctuations~\cite{Lee2006,Keimer2015}. Such a pseudogap phase has been observed previously in the cuprates~\cite{Lee2006,Keimer2015}, where it can be effectively described by the fermionic counterpart of the bosonic model studied here.

All these gapless \cite{SM} quantum states can be understood as distinct manifestations of interference frustration induced by the motion of doped holes in Mott antiferromagnets.  At $t'=0$, NN hopping is severely frustrated by the emergent $\mathbb{Z}_2$ frustration~\cite{Zhang_2024,Weng_Sheng_1996,Ting_Weng_1997}, leading to suppressed single-particle coherence. This frustration can be relieved through several distinct mechanisms, each giving rise to a corresponding saddle-point phase:
(i) tightly binding two holes into a coherent pair (PDW+AFM);  (ii) ferromagnetically polarizing the spin background (SF+$xy$-FM); or  (iii) recombining the doped hole with a local spin into an itinerant quasiparticle with an emergent momentum shift  previously identified in one-hole-doped two-leg fermionic ladders~\cite{Zhao2023} (SF*+IM). Upon introducing NNN hopping $t'$, the new kinetic channel enlarges the hole-pair size, leaving the $\mathbb{Z}_2$ frustration only partially canceled. This incomplete cancellation leads to strong phase fluctuations that destroy long-range coherence, while local pairing persists (dPDW+AFM).
Importantly, this scenario is further supported by replacing the $t$-hopping with spin-dependent $\sigma t$-hopping, leading to the bosonic $\sigma t$–$t'$–$J$ model~\cite{SM}. In this case, the doping-induced $\mathbb{Z}_2$ frustration is completely eliminated~\cite{Zhang_2024,lu2024sign}, resulting in the collapse of all exotic phases into a uniform SF phase, as demonstrated in Fig. S16.

Finally, given that recent Rydberg atom experiments~\cite{Qiao_2025} have already realized the bosonic $t$–$t'$–$J$ model with $t'>0$ and observed restricted hole motion reminiscent of the behavior found in the PS region of our phase diagram, we not only propose a concrete protocol to reverse the sign of $t'$ (i.e., $t'<0$) in such platforms, but also provide theoretical insights to guide and interpret future quantum simulations of doped Mott antiferromagnets.

\begin{acknowledgements}
{\it Acknowledgments.---}
We thank Hao-Kai Zhang, Antoine Browaeys, Mu Qiao, Romain Martin, Ya-Hui Zhang, Thomas Kiely, Feng Chen, Xin-Chi Zhou and Liujun Zou for stimulating discussions. 
X.~L. and S.~S.~G. were supported by the National Natural Science Foundation of China (No. 12534009 and No. 12274014), the Guangdong Provincial Quantum Science Strategic Initiative (No. GDZX2501006), the Special Project in Key Areas for Universities in Guangdong Province (No. 2023ZDZX3054), and the Dongguan Key Laboratory of Artificial Intelligence Design for Advanced Materials.
J.~X.~Z. was supported by the European Research Council (ERC) under the European Union’s Horizon 2020 research and innovation program (Grant Agreement No.\ 853116, acronym TRANSPORT). This research was supported in part by grant NSF PHY-2309135 to the Kavli Institute for Theoretical Physics (KITP).  
L.~H. was supported by the Simons Collaboration on Ultra-Quantum Matter, which is a grant from the Simons Foundation (651440).
D.~N.~S. was supported by the US National Science Foundation Grant No. PHY-2216774.
Z.~Y.~W. was supported by MOST of China (Grant No. 2021YFA1402101) and NSF of China (Grant No. 12347107).
The computational resources are supported by the SongShan Lake HPC Center (SSL-HPC) in Great Bay University.
\end{acknowledgements}

{\it Data availability.---}
The data are available from the authors upon reasonable request.

\twocolumngrid
\bibliography{refs}% Produces the bibliography via BibTeX.

\begin{thebibliography}{52}%
\makeatletter
\providecommand \@ifxundefined [1]{%
 \@ifx{#1\undefined}
}%
\providecommand \@ifnum [1]{%
 \ifnum #1\expandafter \@firstoftwo
 \else \expandafter \@secondoftwo
 \fi
}%
\providecommand \@ifx [1]{%
 \ifx #1\expandafter \@firstoftwo
 \else \expandafter \@secondoftwo
 \fi
}%
\providecommand \natexlab [1]{#1}%
\providecommand \enquote  [1]{``#1''}%
\providecommand \bibnamefont  [1]{#1}%
\providecommand \bibfnamefont [1]{#1}%
\providecommand \citenamefont [1]{#1}%
\providecommand \href@noop [0]{\@secondoftwo}%
\providecommand \href [0]{\begingroup \@sanitize@url \@href}%
\providecommand \@href[1]{\@@startlink{#1}\@@href}%
\providecommand \@@href[1]{\endgroup#1\@@endlink}%
\providecommand \@sanitize@url [0]{\catcode `\\12\catcode `\$12\catcode
  `\&12\catcode `\#12\catcode `\^12\catcode `\_12\catcode `\%12\relax}%
\providecommand \@@startlink[1]{}%
\providecommand \@@endlink[0]{}%
\providecommand \url  [0]{\begingroup\@sanitize@url \@url }%
\providecommand \@url [1]{\endgroup\@href {#1}{\urlprefix }}%
\providecommand \urlprefix  [0]{URL }%
\providecommand \Eprint [0]{\href }%
\providecommand \doibase [0]{https://doi.org/}%
\providecommand \selectlanguage [0]{\@gobble}%
\providecommand \bibinfo  [0]{\@secondoftwo}%
\providecommand \bibfield  [0]{\@secondoftwo}%
\providecommand \translation [1]{[#1]}%
\providecommand \BibitemOpen [0]{}%
\providecommand \bibitemStop [0]{}%
\providecommand \bibitemNoStop [0]{.\EOS\space}%
\providecommand \EOS [0]{\spacefactor3000\relax}%
\providecommand \BibitemShut  [1]{\csname bibitem#1\endcsname}%
\let\auto@bib@innerbib\@empty
%</preamble>
\bibitem [{\citenamefont {Lee}\ \emph {et~al.}(2006)\citenamefont {Lee},
  \citenamefont {Nagaosa},\ and\ \citenamefont {Wen}}]{Lee2006}%
  \BibitemOpen
  \bibfield  {author} {\bibinfo {author} {\bibfnamefont {P.~A.}\ \bibnamefont
  {Lee}}, \bibinfo {author} {\bibfnamefont {N.}~\bibnamefont {Nagaosa}},\ and\
  \bibinfo {author} {\bibfnamefont {X.-G.}\ \bibnamefont {Wen}},\ }\bibfield
  {title} {\bibinfo {title} {{Doping a Mott insulator: Physics of
  high-temperature superconductivity}},\ }\href
  {https://doi.org/10.1103/RevModPhys.78.17} {\bibfield  {journal} {\bibinfo
  {journal} {Rev. Mod. Phys.}\ }\textbf {\bibinfo {volume} {78}},\ \bibinfo
  {pages} {17} (\bibinfo {year} {2006})}\BibitemShut {NoStop}%
\bibitem [{\citenamefont {Keimer}\ \emph {et~al.}(2015)\citenamefont {Keimer},
  \citenamefont {Kivelson}, \citenamefont {Norman}, \citenamefont {Uchida},\
  and\ \citenamefont {Zaanen}}]{Keimer2015}%
  \BibitemOpen
  \bibfield  {author} {\bibinfo {author} {\bibfnamefont {B.}~\bibnamefont
  {Keimer}}, \bibinfo {author} {\bibfnamefont {S.~A.}\ \bibnamefont
  {Kivelson}}, \bibinfo {author} {\bibfnamefont {M.~R.}\ \bibnamefont
  {Norman}}, \bibinfo {author} {\bibfnamefont {S.}~\bibnamefont {Uchida}},\
  and\ \bibinfo {author} {\bibfnamefont {J.}~\bibnamefont {Zaanen}},\
  }\bibfield  {title} {\bibinfo {title} {{From quantum matter to
  high-temperature superconductivity in copper oxides}},\ }\href
  {https://doi.org/10.1038/nature14165} {\bibfield  {journal} {\bibinfo
  {journal} {Nature}\ }\textbf {\bibinfo {volume} {518}},\ \bibinfo {pages}
  {179} (\bibinfo {year} {2015})}\BibitemShut {NoStop}%
\bibitem [{\citenamefont {Proust}\ and\ \citenamefont
  {Taillefer}(2019)}]{Proust_ARCMP_2019}%
  \BibitemOpen
  \bibfield  {author} {\bibinfo {author} {\bibfnamefont {C.}~\bibnamefont
  {Proust}}\ and\ \bibinfo {author} {\bibfnamefont {L.}~\bibnamefont
  {Taillefer}},\ }\bibfield  {title} {\bibinfo {title} {{The Remarkable
  Underlying Ground States of Cuprate Superconductors}},\ }\href
  {https://doi.org/10.1146/annurev-conmatphys-031218-013210} {\bibfield
  {journal} {\bibinfo  {journal} {Annual Review of Condensed Matter Physics}\
  }\textbf {\bibinfo {volume} {10}},\ \bibinfo {pages} {409} (\bibinfo {year}
  {2019})}\BibitemShut {NoStop}%
\bibitem [{\citenamefont {Zhang}\ and\ \citenamefont {Rice}(1988)}]{Zhang1988}%
  \BibitemOpen
  \bibfield  {author} {\bibinfo {author} {\bibfnamefont {F.~C.}\ \bibnamefont
  {Zhang}}\ and\ \bibinfo {author} {\bibfnamefont {T.~M.}\ \bibnamefont
  {Rice}},\ }\bibfield  {title} {\bibinfo {title} {{Effective Hamiltonian for
  the superconducting Cu oxides}},\ }\href
  {https://doi.org/10.1103/PhysRevB.37.3759} {\bibfield  {journal} {\bibinfo
  {journal} {Phys. Rev. B}\ }\textbf {\bibinfo {volume} {37}},\ \bibinfo
  {pages} {3759} (\bibinfo {year} {1988})}\BibitemShut {NoStop}%
\bibitem [{\citenamefont {Ogata}\ and\ \citenamefont
  {Fukuyama}(2008)}]{Masao_RPP_2008}%
  \BibitemOpen
  \bibfield  {author} {\bibinfo {author} {\bibfnamefont {M.}~\bibnamefont
  {Ogata}}\ and\ \bibinfo {author} {\bibfnamefont {H.}~\bibnamefont
  {Fukuyama}},\ }\bibfield  {title} {\bibinfo {title} {{The $t$-$\mathit{J}$
  model for the oxide high-Tc superconductors}},\ }\href
  {https://doi.org/10.1088/0034-4885/71/3/036501} {\bibfield  {journal}
  {\bibinfo  {journal} {Reports on Progress in Physics}\ }\textbf {\bibinfo
  {volume} {71}},\ \bibinfo {pages} {036501} (\bibinfo {year}
  {2008})}\BibitemShut {NoStop}%
\bibitem [{\citenamefont {Jiang}\ and\ \citenamefont
  {Kivelson}(2021)}]{Jiang_PRL_2021}%
  \BibitemOpen
  \bibfield  {author} {\bibinfo {author} {\bibfnamefont {H.-C.}\ \bibnamefont
  {Jiang}}\ and\ \bibinfo {author} {\bibfnamefont {S.~A.}\ \bibnamefont
  {Kivelson}},\ }\bibfield  {title} {\bibinfo {title} {{High Temperature
  Superconductivity in a Lightly Doped Quantum Spin Liquid}},\ }\href
  {https://doi.org/10.1103/PhysRevLett.127.097002} {\bibfield  {journal}
  {\bibinfo  {journal} {Phys. Rev. Lett.}\ }\textbf {\bibinfo {volume} {127}},\
  \bibinfo {pages} {097002} (\bibinfo {year} {2021})}\BibitemShut {NoStop}%
\bibitem [{\citenamefont {Gong}\ \emph {et~al.}(2021)\citenamefont {Gong},
  \citenamefont {Zhu},\ and\ \citenamefont {Sheng}}]{Gong_PRL_2021}%
  \BibitemOpen
  \bibfield  {author} {\bibinfo {author} {\bibfnamefont {S.}~\bibnamefont
  {Gong}}, \bibinfo {author} {\bibfnamefont {W.}~\bibnamefont {Zhu}},\ and\
  \bibinfo {author} {\bibfnamefont {D.~N.}\ \bibnamefont {Sheng}},\ }\bibfield
  {title} {\bibinfo {title} {{Robust $d$-Wave Superconductivity in the
  Square-Lattice $t$-$\mathit{J}$ Model}},\ }\href
  {https://doi.org/10.1103/PhysRevLett.127.097003} {\bibfield  {journal}
  {\bibinfo  {journal} {Phys. Rev. Lett.}\ }\textbf {\bibinfo {volume} {127}},\
  \bibinfo {pages} {097003} (\bibinfo {year} {2021})}\BibitemShut {NoStop}%
\bibitem [{\citenamefont {Jiang}\ \emph {et~al.}(2021)\citenamefont {Jiang},
  \citenamefont {Scalapino},\ and\ \citenamefont {White}}]{White_PNAS_2021}%
  \BibitemOpen
  \bibfield  {author} {\bibinfo {author} {\bibfnamefont {S.}~\bibnamefont
  {Jiang}}, \bibinfo {author} {\bibfnamefont {D.~J.}\ \bibnamefont
  {Scalapino}},\ and\ \bibinfo {author} {\bibfnamefont {S.~R.}\ \bibnamefont
  {White}},\ }\bibfield  {title} {\bibinfo {title} {{Ground-state phase diagram
  of the $t$-$t^\prime$-$\mathit{J}$ model}},\ }\href
  {https://doi.org/10.1073/pnas.2109978118} {\bibfield  {journal} {\bibinfo
  {journal} {Proc. Natl. Acad. Sci. U.S.A.}\ }\textbf {\bibinfo {volume}
  {118}},\ \bibinfo {pages} {e2109978118} (\bibinfo {year} {2021})}\BibitemShut
  {NoStop}%
\bibitem [{\citenamefont {Jiang}\ \emph {et~al.}(2023)\citenamefont {Jiang},
  \citenamefont {Kivelson},\ and\ \citenamefont {Lee}}]{Jiang_PRB_2023}%
  \BibitemOpen
  \bibfield  {author} {\bibinfo {author} {\bibfnamefont {H.-C.}\ \bibnamefont
  {Jiang}}, \bibinfo {author} {\bibfnamefont {S.~A.}\ \bibnamefont
  {Kivelson}},\ and\ \bibinfo {author} {\bibfnamefont {D.-H.}\ \bibnamefont
  {Lee}},\ }\bibfield  {title} {\bibinfo {title} {{Superconducting valence bond
  fluid in lightly doped eight-leg $t$-$J$ cylinders}},\ }\href
  {https://doi.org/10.1103/PhysRevB.108.054505} {\bibfield  {journal} {\bibinfo
   {journal} {Phys. Rev. B}\ }\textbf {\bibinfo {volume} {108}},\ \bibinfo
  {pages} {054505} (\bibinfo {year} {2023})}\BibitemShut {NoStop}%
\bibitem [{\citenamefont {Lu}\ \emph {et~al.}(2024{\natexlab{a}})\citenamefont
  {Lu}, \citenamefont {Chen}, \citenamefont {Zhu}, \citenamefont {Sheng},\ and\
  \citenamefont {Gong}}]{lu2024emergent}%
  \BibitemOpen
  \bibfield  {author} {\bibinfo {author} {\bibfnamefont {X.}~\bibnamefont
  {Lu}}, \bibinfo {author} {\bibfnamefont {F.}~\bibnamefont {Chen}}, \bibinfo
  {author} {\bibfnamefont {W.}~\bibnamefont {Zhu}}, \bibinfo {author}
  {\bibfnamefont {D.~N.}\ \bibnamefont {Sheng}},\ and\ \bibinfo {author}
  {\bibfnamefont {S.-S.}\ \bibnamefont {Gong}},\ }\bibfield  {title} {\bibinfo
  {title} {{Emergent Superconductivity and Competing Charge Orders in
  Hole-Doped Square-Lattice $t$-$\mathit{J}$ Model}},\ }\href
  {https://doi.org/10.1103/PhysRevLett.132.066002} {\bibfield  {journal}
  {\bibinfo  {journal} {Phys. Rev. Lett.}\ }\textbf {\bibinfo {volume} {132}},\
  \bibinfo {pages} {066002} (\bibinfo {year} {2024}{\natexlab{a}})}\BibitemShut
  {NoStop}%
\bibitem [{\citenamefont {Chen}\ \emph {et~al.}(2025)\citenamefont {Chen},
  \citenamefont {Haldane},\ and\ \citenamefont {Sheng}}]{tJ_Feng_2023}%
  \BibitemOpen
  \bibfield  {author} {\bibinfo {author} {\bibfnamefont {F.}~\bibnamefont
  {Chen}}, \bibinfo {author} {\bibfnamefont {F.~D.~M.}\ \bibnamefont
  {Haldane}},\ and\ \bibinfo {author} {\bibfnamefont {D.~N.}\ \bibnamefont
  {Sheng}},\ }\bibfield  {title} {\bibinfo {title} {{Global phase diagram of
  D-wave superconductivity in the square-lattice $t$-$\mathit{J}$ Model}},\
  }\href {https://doi.org/10.1073/pnas.2420963122} {\bibfield  {journal}
  {\bibinfo  {journal} {Proc. Natl. Acad. Sci. U.S.A.}\ }\textbf {\bibinfo
  {volume} {122}},\ \bibinfo {pages} {e2420963122} (\bibinfo {year}
  {2025})}\BibitemShut {NoStop}%
\bibitem [{\citenamefont {Koepsell}\ \emph {et~al.}(2019)\citenamefont
  {Koepsell}, \citenamefont {Vijayan}, \citenamefont {Sompet}, \citenamefont
  {Grusdt}, \citenamefont {Hilker}, \citenamefont {Demler}, \citenamefont
  {Salomon}, \citenamefont {Bloch},\ and\ \citenamefont
  {Gross}}]{Koepsell_nature_2019}%
  \BibitemOpen
  \bibfield  {author} {\bibinfo {author} {\bibfnamefont {J.}~\bibnamefont
  {Koepsell}}, \bibinfo {author} {\bibfnamefont {J.}~\bibnamefont {Vijayan}},
  \bibinfo {author} {\bibfnamefont {P.}~\bibnamefont {Sompet}}, \bibinfo
  {author} {\bibfnamefont {F.}~\bibnamefont {Grusdt}}, \bibinfo {author}
  {\bibfnamefont {T.~A.}\ \bibnamefont {Hilker}}, \bibinfo {author}
  {\bibfnamefont {E.}~\bibnamefont {Demler}}, \bibinfo {author} {\bibfnamefont
  {G.}~\bibnamefont {Salomon}}, \bibinfo {author} {\bibfnamefont
  {I.}~\bibnamefont {Bloch}},\ and\ \bibinfo {author} {\bibfnamefont
  {C.}~\bibnamefont {Gross}},\ }\bibfield  {title} {\bibinfo {title} {{Imaging
  magnetic polarons in the doped Fermi–Hubbard model}},\ }\href
  {https://doi.org/10.1038/s41586-019-1463-1} {\bibfield  {journal} {\bibinfo
  {journal} {Nature}\ }\textbf {\bibinfo {volume} {572}},\ \bibinfo {pages}
  {358} (\bibinfo {year} {2019})}\BibitemShut {NoStop}%
\bibitem [{\citenamefont {Hartke}\ \emph {et~al.}(2020)\citenamefont {Hartke},
  \citenamefont {Oreg}, \citenamefont {Jia},\ and\ \citenamefont
  {Zwierlein}}]{Hartke_PRL_2020}%
  \BibitemOpen
  \bibfield  {author} {\bibinfo {author} {\bibfnamefont {T.}~\bibnamefont
  {Hartke}}, \bibinfo {author} {\bibfnamefont {B.}~\bibnamefont {Oreg}},
  \bibinfo {author} {\bibfnamefont {N.}~\bibnamefont {Jia}},\ and\ \bibinfo
  {author} {\bibfnamefont {M.}~\bibnamefont {Zwierlein}},\ }\bibfield  {title}
  {\bibinfo {title} {{Doublon-Hole Correlations and Fluctuation Thermometry in
  a Fermi-Hubbard Gas}},\ }\href
  {https://doi.org/10.1103/PhysRevLett.125.113601} {\bibfield  {journal}
  {\bibinfo  {journal} {Phys. Rev. Lett.}\ }\textbf {\bibinfo {volume} {125}},\
  \bibinfo {pages} {113601} (\bibinfo {year} {2020})}\BibitemShut {NoStop}%
\bibitem [{\citenamefont {Bohrdt}\ \emph {et~al.}(2021)\citenamefont {Bohrdt},
  \citenamefont {Homeier}, \citenamefont {Reinmoser}, \citenamefont {Demler},\
  and\ \citenamefont {Grusdt}}]{Bohrdt_AP2021}%
  \BibitemOpen
  \bibfield  {author} {\bibinfo {author} {\bibfnamefont {A.}~\bibnamefont
  {Bohrdt}}, \bibinfo {author} {\bibfnamefont {L.}~\bibnamefont {Homeier}},
  \bibinfo {author} {\bibfnamefont {C.}~\bibnamefont {Reinmoser}}, \bibinfo
  {author} {\bibfnamefont {E.}~\bibnamefont {Demler}},\ and\ \bibinfo {author}
  {\bibfnamefont {F.}~\bibnamefont {Grusdt}},\ }\bibfield  {title} {\bibinfo
  {title} {{Exploration of doped quantum magnets with ultracold atoms}},\
  }\href {https://doi.org/https://doi.org/10.1016/j.aop.2021.168651} {\bibfield
   {journal} {\bibinfo  {journal} {Annals of Physics}\ }\textbf {\bibinfo
  {volume} {435}},\ \bibinfo {pages} {168651} (\bibinfo {year}
  {2021})}\BibitemShut {NoStop}%
\bibitem [{\citenamefont {Gall}\ \emph {et~al.}(2021)\citenamefont {Gall},
  \citenamefont {Wurz}, \citenamefont {Samland}, \citenamefont {Chan},\ and\
  \citenamefont {Köhl}}]{Gall_nature_2021}%
  \BibitemOpen
  \bibfield  {author} {\bibinfo {author} {\bibfnamefont {M.}~\bibnamefont
  {Gall}}, \bibinfo {author} {\bibfnamefont {N.}~\bibnamefont {Wurz}}, \bibinfo
  {author} {\bibfnamefont {J.}~\bibnamefont {Samland}}, \bibinfo {author}
  {\bibfnamefont {C.~F.}\ \bibnamefont {Chan}},\ and\ \bibinfo {author}
  {\bibfnamefont {M.}~\bibnamefont {Köhl}},\ }\bibfield  {title} {\bibinfo
  {title} {{Competing magnetic orders in a bilayer Hubbard model with ultracold
  atoms}},\ }\href {https://doi.org/10.1038/s41586-020-03058-x} {\bibfield
  {journal} {\bibinfo  {journal} {Nature}\ }\textbf {\bibinfo {volume} {589}},\
  \bibinfo {pages} {40} (\bibinfo {year} {2021})}\BibitemShut {NoStop}%
\bibitem [{\citenamefont {Shao}\ \emph {et~al.}(2024)\citenamefont {Shao},
  \citenamefont {Wang}, \citenamefont {Zhu}, \citenamefont {Zhu}, \citenamefont
  {Sun}, \citenamefont {Chen}, \citenamefont {Zhang}, \citenamefont {Fan},
  \citenamefont {Deng}, \citenamefont {Yao}, \citenamefont {Chen},\ and\
  \citenamefont {Pan}}]{Pan_nature_2024}%
  \BibitemOpen
  \bibfield  {author} {\bibinfo {author} {\bibfnamefont {H.-J.}\ \bibnamefont
  {Shao}}, \bibinfo {author} {\bibfnamefont {Y.-X.}\ \bibnamefont {Wang}},
  \bibinfo {author} {\bibfnamefont {D.-Z.}\ \bibnamefont {Zhu}}, \bibinfo
  {author} {\bibfnamefont {Y.-S.}\ \bibnamefont {Zhu}}, \bibinfo {author}
  {\bibfnamefont {H.-N.}\ \bibnamefont {Sun}}, \bibinfo {author} {\bibfnamefont
  {S.-Y.}\ \bibnamefont {Chen}}, \bibinfo {author} {\bibfnamefont
  {C.}~\bibnamefont {Zhang}}, \bibinfo {author} {\bibfnamefont {Z.-J.}\
  \bibnamefont {Fan}}, \bibinfo {author} {\bibfnamefont {Y.}~\bibnamefont
  {Deng}}, \bibinfo {author} {\bibfnamefont {X.-C.}\ \bibnamefont {Yao}},
  \bibinfo {author} {\bibfnamefont {Y.-A.}\ \bibnamefont {Chen}},\ and\
  \bibinfo {author} {\bibfnamefont {J.-W.}\ \bibnamefont {Pan}},\ }\bibfield
  {title} {\bibinfo {title} {{Antiferromagnetic phase transition in a 3D
  fermionic Hubbard model}},\ }\href
  {https://doi.org/10.1038/s41586-024-07689-2} {\bibfield  {journal} {\bibinfo
  {journal} {Nature}\ }\textbf {\bibinfo {volume} {632}},\ \bibinfo {pages}
  {267} (\bibinfo {year} {2024})}\BibitemShut {NoStop}%
\bibitem [{\citenamefont {Prichard}\ \emph {et~al.}(2024)\citenamefont
  {Prichard}, \citenamefont {Spar}, \citenamefont {Morera}, \citenamefont
  {Demler}, \citenamefont {Yan},\ and\ \citenamefont
  {Bakr}}]{Prichard_nature_2024}%
  \BibitemOpen
  \bibfield  {author} {\bibinfo {author} {\bibfnamefont {M.~L.}\ \bibnamefont
  {Prichard}}, \bibinfo {author} {\bibfnamefont {B.~M.}\ \bibnamefont {Spar}},
  \bibinfo {author} {\bibfnamefont {I.}~\bibnamefont {Morera}}, \bibinfo
  {author} {\bibfnamefont {E.}~\bibnamefont {Demler}}, \bibinfo {author}
  {\bibfnamefont {Z.~Z.}\ \bibnamefont {Yan}},\ and\ \bibinfo {author}
  {\bibfnamefont {W.~S.}\ \bibnamefont {Bakr}},\ }\bibfield  {title} {\bibinfo
  {title} {{Directly imaging spin polarons in a kinetically frustrated Hubbard
  system}},\ }\href {https://doi.org/10.1038/s41586-024-07356-6} {\bibfield
  {journal} {\bibinfo  {journal} {Nature}\ }\textbf {\bibinfo {volume} {629}},\
  \bibinfo {pages} {323} (\bibinfo {year} {2024})}\BibitemShut {NoStop}%
\bibitem [{\citenamefont {Bohrdt}\ \emph {et~al.}(2024)\citenamefont {Bohrdt},
  \citenamefont {Wei}, \citenamefont {Adler}, \citenamefont {Srakaew},
  \citenamefont {Agrawal}, \citenamefont {Weckesser}, \citenamefont {Bloch},
  \citenamefont {Grusdt},\ and\ \citenamefont {Zeiher}}]{Annabelle_2024}%
  \BibitemOpen
  \bibfield  {author} {\bibinfo {author} {\bibfnamefont {A.}~\bibnamefont
  {Bohrdt}}, \bibinfo {author} {\bibfnamefont {D.}~\bibnamefont {Wei}},
  \bibinfo {author} {\bibfnamefont {D.}~\bibnamefont {Adler}}, \bibinfo
  {author} {\bibfnamefont {K.}~\bibnamefont {Srakaew}}, \bibinfo {author}
  {\bibfnamefont {S.}~\bibnamefont {Agrawal}}, \bibinfo {author} {\bibfnamefont
  {P.}~\bibnamefont {Weckesser}}, \bibinfo {author} {\bibfnamefont
  {I.}~\bibnamefont {Bloch}}, \bibinfo {author} {\bibfnamefont
  {F.}~\bibnamefont {Grusdt}},\ and\ \bibinfo {author} {\bibfnamefont
  {J.}~\bibnamefont {Zeiher}},\ }\href@noop {} {\bibinfo {title} {{Microscopy
  of bosonic charge carriers in staggered magnetic fields}}} (\bibinfo {year}
  {2024}),\ \Eprint {https://arxiv.org/abs/2410.19500} {arXiv:2410.19500}
  \BibitemShut {NoStop}%
\bibitem [{\citenamefont {Bakr}\ \emph {et~al.}(2025)\citenamefont {Bakr},
  \citenamefont {Ba},\ and\ \citenamefont {Prichard}}]{bakr2025}%
  \BibitemOpen
  \bibfield  {author} {\bibinfo {author} {\bibfnamefont {W.~S.}\ \bibnamefont
  {Bakr}}, \bibinfo {author} {\bibfnamefont {Z.}~\bibnamefont {Ba}},\ and\
  \bibinfo {author} {\bibfnamefont {M.~L.}\ \bibnamefont {Prichard}},\
  }\href@noop {} {\bibinfo {title} {{Microscopy of Ultracold Fermions in
  Optical Lattices}}} (\bibinfo {year} {2025}),\ \Eprint
  {https://arxiv.org/abs/2507.04042} {arXiv:2507.04042} \BibitemShut {NoStop}%
\bibitem [{\citenamefont {Xu}\ \emph {et~al.}(2025)\citenamefont {Xu},
  \citenamefont {Kendrick}, \citenamefont {Kale}, \citenamefont {Gang},
  \citenamefont {Feng}, \citenamefont {Zhang}, \citenamefont {Young},
  \citenamefont {Lebrat},\ and\ \citenamefont {Greiner}}]{Muqing_nature_2025}%
  \BibitemOpen
  \bibfield  {author} {\bibinfo {author} {\bibfnamefont {M.}~\bibnamefont
  {Xu}}, \bibinfo {author} {\bibfnamefont {L.~H.}\ \bibnamefont {Kendrick}},
  \bibinfo {author} {\bibfnamefont {A.}~\bibnamefont {Kale}}, \bibinfo {author}
  {\bibfnamefont {Y.}~\bibnamefont {Gang}}, \bibinfo {author} {\bibfnamefont
  {C.}~\bibnamefont {Feng}}, \bibinfo {author} {\bibfnamefont {S.}~\bibnamefont
  {Zhang}}, \bibinfo {author} {\bibfnamefont {A.~W.}\ \bibnamefont {Young}},
  \bibinfo {author} {\bibfnamefont {M.}~\bibnamefont {Lebrat}},\ and\ \bibinfo
  {author} {\bibfnamefont {M.}~\bibnamefont {Greiner}},\ }\bibfield  {title}
  {\bibinfo {title} {{A neutral-atom Hubbard quantum simulator in the cryogenic
  regime}},\ }\href {https://doi.org/10.1038/s41586-025-09112-w} {\bibfield
  {journal} {\bibinfo  {journal} {Nature}\ }\textbf {\bibinfo {volume} {642}},\
  \bibinfo {pages} {909} (\bibinfo {year} {2025})}\BibitemShut {NoStop}%
\bibitem [{\citenamefont {Qiao}\ \emph {et~al.}(2025)\citenamefont {Qiao},
  \citenamefont {Emperauger}, \citenamefont {Chen}, \citenamefont {Homeier},
  \citenamefont {Hollerith}, \citenamefont {Bornet}, \citenamefont {Martin},
  \citenamefont {Gély}, \citenamefont {Klein}, \citenamefont {Barredo},
  \citenamefont {Geier}, \citenamefont {Chiu}, \citenamefont {Grusdt},
  \citenamefont {Bohrdt}, \citenamefont {Lahaye},\ and\ \citenamefont
  {Browaeys}}]{Qiao_2025}%
  \BibitemOpen
  \bibfield  {author} {\bibinfo {author} {\bibfnamefont {M.}~\bibnamefont
  {Qiao}}, \bibinfo {author} {\bibfnamefont {G.}~\bibnamefont {Emperauger}},
  \bibinfo {author} {\bibfnamefont {C.}~\bibnamefont {Chen}}, \bibinfo {author}
  {\bibfnamefont {L.}~\bibnamefont {Homeier}}, \bibinfo {author} {\bibfnamefont
  {S.}~\bibnamefont {Hollerith}}, \bibinfo {author} {\bibfnamefont
  {G.}~\bibnamefont {Bornet}}, \bibinfo {author} {\bibfnamefont
  {R.}~\bibnamefont {Martin}}, \bibinfo {author} {\bibfnamefont
  {B.}~\bibnamefont {Gély}}, \bibinfo {author} {\bibfnamefont
  {L.}~\bibnamefont {Klein}}, \bibinfo {author} {\bibfnamefont
  {D.}~\bibnamefont {Barredo}}, \bibinfo {author} {\bibfnamefont
  {S.}~\bibnamefont {Geier}}, \bibinfo {author} {\bibfnamefont {N.-C.}\
  \bibnamefont {Chiu}}, \bibinfo {author} {\bibfnamefont {F.}~\bibnamefont
  {Grusdt}}, \bibinfo {author} {\bibfnamefont {A.}~\bibnamefont {Bohrdt}},
  \bibinfo {author} {\bibfnamefont {T.}~\bibnamefont {Lahaye}},\ and\ \bibinfo
  {author} {\bibfnamefont {A.}~\bibnamefont {Browaeys}},\ }\bibfield  {title}
  {\bibinfo {title} {{Realization of a doped quantum antiferromagnet in a
  Rydberg tweezer array}},\ }\href {https://doi.org/10.1038/s41586-025-09377-1}
  {\bibfield  {journal} {\bibinfo  {journal} {Nature}\ }\textbf {\bibinfo
  {volume} {644}},\ \bibinfo {pages} {889} (\bibinfo {year}
  {2025})}\BibitemShut {NoStop}%
\bibitem [{\citenamefont {Homeier}\ \emph {et~al.}(2024)\citenamefont
  {Homeier}, \citenamefont {Harris}, \citenamefont {Blatz}, \citenamefont
  {Geier}, \citenamefont {Hollerith}, \citenamefont {Schollw\"ock},
  \citenamefont {Grusdt},\ and\ \citenamefont {Bohrdt}}]{Lukas_PRL_2024}%
  \BibitemOpen
  \bibfield  {author} {\bibinfo {author} {\bibfnamefont {L.}~\bibnamefont
  {Homeier}}, \bibinfo {author} {\bibfnamefont {T.~J.}\ \bibnamefont {Harris}},
  \bibinfo {author} {\bibfnamefont {T.}~\bibnamefont {Blatz}}, \bibinfo
  {author} {\bibfnamefont {S.}~\bibnamefont {Geier}}, \bibinfo {author}
  {\bibfnamefont {S.}~\bibnamefont {Hollerith}}, \bibinfo {author}
  {\bibfnamefont {U.}~\bibnamefont {Schollw\"ock}}, \bibinfo {author}
  {\bibfnamefont {F.}~\bibnamefont {Grusdt}},\ and\ \bibinfo {author}
  {\bibfnamefont {A.}~\bibnamefont {Bohrdt}},\ }\bibfield  {title} {\bibinfo
  {title} {{Antiferromagnetic Bosonic $t$-$J$ Models and Their Quantum
  Simulation in Tweezer Arrays}},\ }\href
  {https://doi.org/10.1103/PhysRevLett.132.230401} {\bibfield  {journal}
  {\bibinfo  {journal} {Phys. Rev. Lett.}\ }\textbf {\bibinfo {volume} {132}},\
  \bibinfo {pages} {230401} (\bibinfo {year} {2024})}\BibitemShut {NoStop}%
\bibitem [{\citenamefont {Jepsen}\ \emph {et~al.}(2021)\citenamefont {Jepsen},
  \citenamefont {Ho}, \citenamefont {Amato-Grill}, \citenamefont {Dimitrova},
  \citenamefont {Demler},\ and\ \citenamefont {Ketterle}}]{Wolfgang_PRX_2021}%
  \BibitemOpen
  \bibfield  {author} {\bibinfo {author} {\bibfnamefont {P.~N.}\ \bibnamefont
  {Jepsen}}, \bibinfo {author} {\bibfnamefont {W.~W.}\ \bibnamefont {Ho}},
  \bibinfo {author} {\bibfnamefont {J.}~\bibnamefont {Amato-Grill}}, \bibinfo
  {author} {\bibfnamefont {I.}~\bibnamefont {Dimitrova}}, \bibinfo {author}
  {\bibfnamefont {E.}~\bibnamefont {Demler}},\ and\ \bibinfo {author}
  {\bibfnamefont {W.}~\bibnamefont {Ketterle}},\ }\bibfield  {title} {\bibinfo
  {title} {{Transverse Spin Dynamics in the Anisotropic Heisenberg Model
  Realized with Ultracold Atoms}},\ }\href
  {https://doi.org/10.1103/PhysRevX.11.041054} {\bibfield  {journal} {\bibinfo
  {journal} {Phys. Rev. X}\ }\textbf {\bibinfo {volume} {11}},\ \bibinfo
  {pages} {041054} (\bibinfo {year} {2021})}\BibitemShut {NoStop}%
\bibitem [{\citenamefont {Sun}\ \emph {et~al.}(2021)\citenamefont {Sun},
  \citenamefont {Yang}, \citenamefont {Wang}, \citenamefont {Zhou},
  \citenamefont {Su}, \citenamefont {Dai}, \citenamefont {Yuan},\ and\
  \citenamefont {Pan}}]{Sun_np_2021}%
  \BibitemOpen
  \bibfield  {author} {\bibinfo {author} {\bibfnamefont {H.}~\bibnamefont
  {Sun}}, \bibinfo {author} {\bibfnamefont {B.}~\bibnamefont {Yang}}, \bibinfo
  {author} {\bibfnamefont {H.-Y.}\ \bibnamefont {Wang}}, \bibinfo {author}
  {\bibfnamefont {Z.-Y.}\ \bibnamefont {Zhou}}, \bibinfo {author}
  {\bibfnamefont {G.-X.}\ \bibnamefont {Su}}, \bibinfo {author} {\bibfnamefont
  {H.-N.}\ \bibnamefont {Dai}}, \bibinfo {author} {\bibfnamefont {Z.-S.}\
  \bibnamefont {Yuan}},\ and\ \bibinfo {author} {\bibfnamefont {J.-W.}\
  \bibnamefont {Pan}},\ }\bibfield  {title} {\bibinfo {title} {{Realization of
  a bosonic antiferromagnet}},\ }\href
  {https://doi.org/10.1038/s41567-021-01277-1} {\bibfield  {journal} {\bibinfo
  {journal} {Nature Physics}\ }\textbf {\bibinfo {volume} {17}},\ \bibinfo
  {pages} {990} (\bibinfo {year} {2021})}\BibitemShut {NoStop}%
\bibitem [{\citenamefont {Duan}\ \emph {et~al.}(2003)\citenamefont {Duan},
  \citenamefont {Demler},\ and\ \citenamefont {Lukin}}]{Duan_PRL_2003}%
  \BibitemOpen
  \bibfield  {author} {\bibinfo {author} {\bibfnamefont {L.-M.}\ \bibnamefont
  {Duan}}, \bibinfo {author} {\bibfnamefont {E.}~\bibnamefont {Demler}},\ and\
  \bibinfo {author} {\bibfnamefont {M.~D.}\ \bibnamefont {Lukin}},\ }\bibfield
  {title} {\bibinfo {title} {{Controlling Spin Exchange Interactions of
  Ultracold Atoms in Optical Lattices}},\ }\href
  {https://doi.org/10.1103/PhysRevLett.91.090402} {\bibfield  {journal}
  {\bibinfo  {journal} {Phys. Rev. Lett.}\ }\textbf {\bibinfo {volume} {91}},\
  \bibinfo {pages} {090402} (\bibinfo {year} {2003})}\BibitemShut {NoStop}%
\bibitem [{\citenamefont {Harris}\ \emph {et~al.}(2024)\citenamefont {Harris},
  \citenamefont {Schollwöck}, \citenamefont {Bohrdt},\ and\ \citenamefont
  {Grusdt}}]{Timothy_2024}%
  \BibitemOpen
  \bibfield  {author} {\bibinfo {author} {\bibfnamefont {T.~J.}\ \bibnamefont
  {Harris}}, \bibinfo {author} {\bibfnamefont {U.}~\bibnamefont {Schollwöck}},
  \bibinfo {author} {\bibfnamefont {A.}~\bibnamefont {Bohrdt}},\ and\ \bibinfo
  {author} {\bibfnamefont {F.}~\bibnamefont {Grusdt}},\ }\href@noop {}
  {\bibinfo {title} {{Kinetic magnetism and stripe order in the
  antiferromagnetic bosonic $t$-$J$ model}}} (\bibinfo {year} {2024}),\ \Eprint
  {https://arxiv.org/abs/2410.00904} {arXiv:2410.00904} \BibitemShut {NoStop}%
\bibitem [{\citenamefont {Zhang}(2022)}]{YHZhang2022}%
  \BibitemOpen
  \bibfield  {author} {\bibinfo {author} {\bibfnamefont {Y.-H.}\ \bibnamefont
  {Zhang}},\ }\bibfield  {title} {\bibinfo {title} {{Doping a Mott insulator
  with excitons in a moir\'e bilayer: Fractional superfluid, neutral Fermi
  surface, and Mott transition}},\ }\href
  {https://doi.org/10.1103/PhysRevB.106.195120} {\bibfield  {journal} {\bibinfo
   {journal} {Phys. Rev. B}\ }\textbf {\bibinfo {volume} {106}},\ \bibinfo
  {pages} {195120} (\bibinfo {year} {2022})}\BibitemShut {NoStop}%
\bibitem [{\citenamefont {Yang}\ and\ \citenamefont {Zhang}(2024)}]{Yang2024}%
  \BibitemOpen
  \bibfield  {author} {\bibinfo {author} {\bibfnamefont {H.}~\bibnamefont
  {Yang}}\ and\ \bibinfo {author} {\bibfnamefont {Y.-H.}\ \bibnamefont
  {Zhang}},\ }\bibfield  {title} {\bibinfo {title} {{Exciton- and light-induced
  ferromagnetism from doping a moir\'e Mott insulator}},\ }\href
  {https://doi.org/10.1103/PhysRevB.110.L041115} {\bibfield  {journal}
  {\bibinfo  {journal} {Phys. Rev. B}\ }\textbf {\bibinfo {volume} {110}},\
  \bibinfo {pages} {L041115} (\bibinfo {year} {2024})}\BibitemShut {NoStop}%
\bibitem [{\citenamefont {Zhang}\ \emph {et~al.}(2025)\citenamefont {Zhang},
  \citenamefont {Zhang}, \citenamefont {Xu}, \citenamefont {Jiang},\ and\
  \citenamefont {Weng}}]{Zhang_2024}%
  \BibitemOpen
  \bibfield  {author} {\bibinfo {author} {\bibfnamefont {H.-K.}\ \bibnamefont
  {Zhang}}, \bibinfo {author} {\bibfnamefont {J.-X.}\ \bibnamefont {Zhang}},
  \bibinfo {author} {\bibfnamefont {J.-S.}\ \bibnamefont {Xu}}, \bibinfo
  {author} {\bibfnamefont {H.-C.}\ \bibnamefont {Jiang}},\ and\ \bibinfo
  {author} {\bibfnamefont {Z.-Y.}\ \bibnamefont {Weng}},\ }\bibfield  {title}
  {\bibinfo {title} {{Quantum-interference-induced pairing in bosonic doped
  antiferromagnets}},\ }\href {https://doi.org/10.1103/1mc9-xqhy} {\bibfield
  {journal} {\bibinfo  {journal} {Phys. Rev. B}\ }\textbf {\bibinfo {volume}
  {112}},\ \bibinfo {pages} {224511} (\bibinfo {year} {2025})}\BibitemShut
  {NoStop}%
\bibitem [{SM()}]{SM}%
  \BibitemOpen
  \href@noop {} {\bibinfo {title} {See supplemental materials for more
  supporting data.}}\BibitemShut {Stop}%
\bibitem [{\citenamefont {White}(1992)}]{dmrg_white_1992}%
  \BibitemOpen
  \bibfield  {author} {\bibinfo {author} {\bibfnamefont {S.~R.}\ \bibnamefont
  {White}},\ }\bibfield  {title} {\bibinfo {title} {{Density matrix formulation
  for quantum renormalization groups}},\ }\href
  {https://doi.org/10.1103/PhysRevLett.69.2863} {\bibfield  {journal} {\bibinfo
   {journal} {Phys. Rev. Lett.}\ }\textbf {\bibinfo {volume} {69}},\ \bibinfo
  {pages} {2863} (\bibinfo {year} {1992})}\BibitemShut {NoStop}%
\bibitem [{\citenamefont {White}(1993)}]{dmrg_white_1993}%
  \BibitemOpen
  \bibfield  {author} {\bibinfo {author} {\bibfnamefont {S.~R.}\ \bibnamefont
  {White}},\ }\bibfield  {title} {\bibinfo {title} {{Density-matrix algorithms
  for quantum renormalization groups}},\ }\href
  {https://doi.org/10.1103/PhysRevB.48.10345} {\bibfield  {journal} {\bibinfo
  {journal} {Phys. Rev. B}\ }\textbf {\bibinfo {volume} {48}},\ \bibinfo
  {pages} {10345} (\bibinfo {year} {1993})}\BibitemShut {NoStop}%
\bibitem [{\citenamefont {Schollw\"ock}(2005)}]{dmrg_ulrich_2005}%
  \BibitemOpen
  \bibfield  {author} {\bibinfo {author} {\bibfnamefont {U.}~\bibnamefont
  {Schollw\"ock}},\ }\bibfield  {title} {\bibinfo {title} {{The density-matrix
  renormalization group}},\ }\href {https://doi.org/10.1103/RevModPhys.77.259}
  {\bibfield  {journal} {\bibinfo  {journal} {Rev. Mod. Phys.}\ }\textbf
  {\bibinfo {volume} {77}},\ \bibinfo {pages} {259} (\bibinfo {year}
  {2005})}\BibitemShut {NoStop}%
\bibitem [{\citenamefont {Schollw\"ock}(2011)}]{dmrg_ulrich_2011}%
  \BibitemOpen
  \bibfield  {author} {\bibinfo {author} {\bibfnamefont {U.}~\bibnamefont
  {Schollw\"ock}},\ }\bibfield  {title} {\bibinfo {title} {{The density-matrix
  renormalization group in the age of matrix product states}},\ }\href
  {https://doi.org/https://doi.org/10.1016/j.aop.2010.09.012} {\bibfield
  {journal} {\bibinfo  {journal} {Annals of Physics}\ }\textbf {\bibinfo
  {volume} {326}},\ \bibinfo {pages} {96} (\bibinfo {year} {2011})}\BibitemShut
  {NoStop}%
\bibitem [{Foo()}]{Footnote_Rydberg}%
  \BibitemOpen
  \href@noop {} {\bibinfo {title} {{We note that the microscopic interactions
  depend on the set of Rydberg states and on external fields. For the recent
  implementation in $^{87}$Rb, the spin interactions realize an anisotropic XXZ
  model and the density-density interaction can depend on the orietation with
  respect to the quantization axis; the spin-density interaction is neglible
  }}}\BibitemShut {NoStop}%
\bibitem [{\citenamefont {Sheng}\ \emph {et~al.}(1996)\citenamefont {Sheng},
  \citenamefont {Chen},\ and\ \citenamefont {Weng}}]{Weng_Sheng_1996}%
  \BibitemOpen
  \bibfield  {author} {\bibinfo {author} {\bibfnamefont {D.~N.}\ \bibnamefont
  {Sheng}}, \bibinfo {author} {\bibfnamefont {Y.~C.}\ \bibnamefont {Chen}},\
  and\ \bibinfo {author} {\bibfnamefont {Z.~Y.}\ \bibnamefont {Weng}},\
  }\bibfield  {title} {\bibinfo {title} {{Phase String Effect in a Doped
  Antiferromagnet}},\ }\href {https://doi.org/10.1103/PhysRevLett.77.5102}
  {\bibfield  {journal} {\bibinfo  {journal} {Phys. Rev. Lett.}\ }\textbf
  {\bibinfo {volume} {77}},\ \bibinfo {pages} {5102} (\bibinfo {year}
  {1996})}\BibitemShut {NoStop}%
\bibitem [{\citenamefont {Weng}\ \emph {et~al.}(1997)\citenamefont {Weng},
  \citenamefont {Sheng}, \citenamefont {Chen},\ and\ \citenamefont
  {Ting}}]{Ting_Weng_1997}%
  \BibitemOpen
  \bibfield  {author} {\bibinfo {author} {\bibfnamefont {Z.~Y.}\ \bibnamefont
  {Weng}}, \bibinfo {author} {\bibfnamefont {D.~N.}\ \bibnamefont {Sheng}},
  \bibinfo {author} {\bibfnamefont {Y.-C.}\ \bibnamefont {Chen}},\ and\
  \bibinfo {author} {\bibfnamefont {C.~S.}\ \bibnamefont {Ting}},\ }\bibfield
  {title} {\bibinfo {title} {{Phase string effect in the $t$-$\mathit{J}$
  model: General theory}},\ }\href {https://doi.org/10.1103/PhysRevB.55.3894}
  {\bibfield  {journal} {\bibinfo  {journal} {Phys. Rev. B}\ }\textbf {\bibinfo
  {volume} {55}},\ \bibinfo {pages} {3894} (\bibinfo {year}
  {1997})}\BibitemShut {NoStop}%
\bibitem [{\citenamefont {Zhao}\ \emph {et~al.}(2023)\citenamefont {Zhao},
  \citenamefont {Chen}, \citenamefont {Sun},\ and\ \citenamefont
  {Weng}}]{Zhao2023}%
  \BibitemOpen
  \bibfield  {author} {\bibinfo {author} {\bibfnamefont {J.-Y.}\ \bibnamefont
  {Zhao}}, \bibinfo {author} {\bibfnamefont {S.~A.}\ \bibnamefont {Chen}},
  \bibinfo {author} {\bibfnamefont {R.-Y.}\ \bibnamefont {Sun}},\ and\ \bibinfo
  {author} {\bibfnamefont {Z.-Y.}\ \bibnamefont {Weng}},\ }\bibfield  {title}
  {\bibinfo {title} {{Continuous transition from a Landau quasiparticle to a
  neutral spinon}},\ }\href {https://doi.org/10.1103/PhysRevB.107.085112}
  {\bibfield  {journal} {\bibinfo  {journal} {Phys. Rev. B}\ }\textbf {\bibinfo
  {volume} {107}},\ \bibinfo {pages} {085112} (\bibinfo {year}
  {2023})}\BibitemShut {NoStop}%
\bibitem [{\citenamefont {Lu}\ \emph {et~al.}(2024{\natexlab{b}})\citenamefont
  {Lu}, \citenamefont {Zhang}, \citenamefont {Gong}, \citenamefont {Sheng},\
  and\ \citenamefont {Weng}}]{lu2024sign}%
  \BibitemOpen
  \bibfield  {author} {\bibinfo {author} {\bibfnamefont {X.}~\bibnamefont
  {Lu}}, \bibinfo {author} {\bibfnamefont {J.-X.}\ \bibnamefont {Zhang}},
  \bibinfo {author} {\bibfnamefont {S.-S.}\ \bibnamefont {Gong}}, \bibinfo
  {author} {\bibfnamefont {D.~N.}\ \bibnamefont {Sheng}},\ and\ \bibinfo
  {author} {\bibfnamefont {Z.-Y.}\ \bibnamefont {Weng}},\ }\bibfield  {title}
  {\bibinfo {title} {{Sign structure of the $t$-$t^\prime$-$\mathit{J}$ model
  and its physical consequences}},\ }\href
  {https://doi.org/10.1103/PhysRevB.110.165127} {\bibfield  {journal} {\bibinfo
   {journal} {Phys. Rev. B}\ }\textbf {\bibinfo {volume} {110}},\ \bibinfo
  {pages} {165127} (\bibinfo {year} {2024}{\natexlab{b}})}\BibitemShut
  {NoStop}%
\bibitem [{\citenamefont {Zhang}\ \emph {et~al.}(2024)\citenamefont {Zhang},
  \citenamefont {Chen}, \citenamefont {Zhang},\ and\ \citenamefont
  {Weng}}]{zhang2024_hg}%
  \BibitemOpen
  \bibfield  {author} {\bibinfo {author} {\bibfnamefont {J.-X.}\ \bibnamefont
  {Zhang}}, \bibinfo {author} {\bibfnamefont {C.}~\bibnamefont {Chen}},
  \bibinfo {author} {\bibfnamefont {J.-H.}\ \bibnamefont {Zhang}},\ and\
  \bibinfo {author} {\bibfnamefont {Z.-Y.}\ \bibnamefont {Weng}},\ }\bibfield
  {title} {\bibinfo {title} {Hourglasslike spin excitation in a doped mott
  insulator},\ }\href {https://doi.org/10.1103/PhysRevResearch.6.013109}
  {\bibfield  {journal} {\bibinfo  {journal} {Phys. Rev. Res.}\ }\textbf
  {\bibinfo {volume} {6}},\ \bibinfo {pages} {013109} (\bibinfo {year}
  {2024})}\BibitemShut {NoStop}%
\bibitem [{\citenamefont {Agterberg}\ \emph {et~al.}(2020)\citenamefont
  {Agterberg}, \citenamefont {Davis}, \citenamefont {Edkins}, \citenamefont
  {Fradkin}, \citenamefont {Van~Harlingen}, \citenamefont {Kivelson},
  \citenamefont {Lee}, \citenamefont {Radzihovsky}, \citenamefont {Tranquada},\
  and\ \citenamefont {Wang}}]{Agterberg_PDW_2020}%
  \BibitemOpen
  \bibfield  {author} {\bibinfo {author} {\bibfnamefont {D.~F.}\ \bibnamefont
  {Agterberg}}, \bibinfo {author} {\bibfnamefont {J.~S.}\ \bibnamefont
  {Davis}}, \bibinfo {author} {\bibfnamefont {S.~D.}\ \bibnamefont {Edkins}},
  \bibinfo {author} {\bibfnamefont {E.}~\bibnamefont {Fradkin}}, \bibinfo
  {author} {\bibfnamefont {D.~J.}\ \bibnamefont {Van~Harlingen}}, \bibinfo
  {author} {\bibfnamefont {S.~A.}\ \bibnamefont {Kivelson}}, \bibinfo {author}
  {\bibfnamefont {P.~A.}\ \bibnamefont {Lee}}, \bibinfo {author} {\bibfnamefont
  {L.}~\bibnamefont {Radzihovsky}}, \bibinfo {author} {\bibfnamefont {J.~M.}\
  \bibnamefont {Tranquada}},\ and\ \bibinfo {author} {\bibfnamefont
  {Y.}~\bibnamefont {Wang}},\ }\bibfield  {title} {\bibinfo {title} {{The
  Physics of Pair-Density Waves: Cuprate Superconductors and Beyond}},\ }\href
  {https://doi.org/https://doi.org/10.1146/annurev-conmatphys-031119-050711}
  {\bibfield  {journal} {\bibinfo  {journal} {Annual Review of Condensed Matter
  Physics}\ }\textbf {\bibinfo {volume} {11}},\ \bibinfo {pages} {231}
  (\bibinfo {year} {2020})}\BibitemShut {NoStop}%
\bibitem [{\citenamefont {Jiang}\ \emph {et~al.}(2020)\citenamefont {Jiang},
  \citenamefont {Zaanen}, \citenamefont {Devereaux},\ and\ \citenamefont
  {Jiang}}]{Jiang_PRR_2020}%
  \BibitemOpen
  \bibfield  {author} {\bibinfo {author} {\bibfnamefont {Y.-F.}\ \bibnamefont
  {Jiang}}, \bibinfo {author} {\bibfnamefont {J.}~\bibnamefont {Zaanen}},
  \bibinfo {author} {\bibfnamefont {T.~P.}\ \bibnamefont {Devereaux}},\ and\
  \bibinfo {author} {\bibfnamefont {H.-C.}\ \bibnamefont {Jiang}},\ }\bibfield
  {title} {\bibinfo {title} {{Ground state phase diagram of the doped {Hubbard}
  model on the four-leg cylinder}},\ }\href
  {https://doi.org/10.1103/PhysRevResearch.2.033073} {\bibfield  {journal}
  {\bibinfo  {journal} {Phys. Rev. Res.}\ }\textbf {\bibinfo {volume} {2}},\
  \bibinfo {pages} {033073} (\bibinfo {year} {2020})}\BibitemShut {NoStop}%
\bibitem [{\citenamefont {Lu}\ \emph {et~al.}(2023)\citenamefont {Lu},
  \citenamefont {Qu}, \citenamefont {Qi}, \citenamefont {Li},\ and\
  \citenamefont {Gong}}]{Lu_PRB_2023}%
  \BibitemOpen
  \bibfield  {author} {\bibinfo {author} {\bibfnamefont {X.}~\bibnamefont
  {Lu}}, \bibinfo {author} {\bibfnamefont {D.-W.}\ \bibnamefont {Qu}}, \bibinfo
  {author} {\bibfnamefont {Y.}~\bibnamefont {Qi}}, \bibinfo {author}
  {\bibfnamefont {W.}~\bibnamefont {Li}},\ and\ \bibinfo {author}
  {\bibfnamefont {S.-S.}\ \bibnamefont {Gong}},\ }\bibfield  {title} {\bibinfo
  {title} {{Ground-state phase diagram of the extended two-leg $t$-$J$
  ladder}},\ }\href {https://doi.org/10.1103/PhysRevB.107.125114} {\bibfield
  {journal} {\bibinfo  {journal} {Phys. Rev. B}\ }\textbf {\bibinfo {volume}
  {107}},\ \bibinfo {pages} {125114} (\bibinfo {year} {2023})}\BibitemShut
  {NoStop}%
\bibitem [{\citenamefont {Lu}\ \emph {et~al.}(2025)\citenamefont {Lu},
  \citenamefont {Guo}, \citenamefont {Chen}, \citenamefont {Sheng},\ and\
  \citenamefont {Gong}}]{lu_epc_2024}%
  \BibitemOpen
  \bibfield  {author} {\bibinfo {author} {\bibfnamefont {X.}~\bibnamefont
  {Lu}}, \bibinfo {author} {\bibfnamefont {H.}~\bibnamefont {Guo}}, \bibinfo
  {author} {\bibfnamefont {W.-Q.}\ \bibnamefont {Chen}}, \bibinfo {author}
  {\bibfnamefont {D.~N.}\ \bibnamefont {Sheng}},\ and\ \bibinfo {author}
  {\bibfnamefont {S.-S.}\ \bibnamefont {Gong}},\ }\bibfield  {title} {\bibinfo
  {title} {{Tuning competition between charge order and superconductivity in
  the square-lattice $t$-$t^\prime$-$\mathit{J}$ model}},\ }\href
  {https://doi.org/10.1103/PhysRevB.111.035139} {\bibfield  {journal} {\bibinfo
   {journal} {Phys. Rev. B}\ }\textbf {\bibinfo {volume} {111}},\ \bibinfo
  {pages} {035139} (\bibinfo {year} {2025})}\BibitemShut {NoStop}%
\bibitem [{\citenamefont {Roux}\ \emph {et~al.}(2007)\citenamefont {Roux},
  \citenamefont {Orignac}, \citenamefont {White},\ and\ \citenamefont
  {Poilblanc}}]{Roux_PRB_2007}%
  \BibitemOpen
  \bibfield  {author} {\bibinfo {author} {\bibfnamefont {G.}~\bibnamefont
  {Roux}}, \bibinfo {author} {\bibfnamefont {E.}~\bibnamefont {Orignac}},
  \bibinfo {author} {\bibfnamefont {S.~R.}\ \bibnamefont {White}},\ and\
  \bibinfo {author} {\bibfnamefont {D.}~\bibnamefont {Poilblanc}},\ }\bibfield
  {title} {\bibinfo {title} {{Diamagnetism of doped two-leg ladders and probing
  the nature of their commensurate phases}},\ }\href
  {https://doi.org/10.1103/PhysRevB.76.195105} {\bibfield  {journal} {\bibinfo
  {journal} {Phys. Rev. B}\ }\textbf {\bibinfo {volume} {76}},\ \bibinfo
  {pages} {195105} (\bibinfo {year} {2007})}\BibitemShut {NoStop}%
\bibitem [{\citenamefont {Calabrese}\ and\ \citenamefont
  {Cardy}(2004)}]{Calabrese2004}%
  \BibitemOpen
  \bibfield  {author} {\bibinfo {author} {\bibfnamefont {P.}~\bibnamefont
  {Calabrese}}\ and\ \bibinfo {author} {\bibfnamefont {J.}~\bibnamefont
  {Cardy}},\ }\bibfield  {title} {\bibinfo {title} {{Entanglement entropy and
  quantum field theory}},\ }\href
  {https://doi.org/10.1088/1742-5468/2004/06/p06002} {\bibfield  {journal}
  {\bibinfo  {journal} {Journal of Statistical Mechanics: Theory and
  Experiment}\ }\textbf {\bibinfo {volume} {2004}},\ \bibinfo {pages} {P06002}
  (\bibinfo {year} {2004})}\BibitemShut {NoStop}%
\bibitem [{\citenamefont {Fagotti}\ and\ \citenamefont
  {Calabrese}(2011)}]{Calabrese2011}%
  \BibitemOpen
  \bibfield  {author} {\bibinfo {author} {\bibfnamefont {M.}~\bibnamefont
  {Fagotti}}\ and\ \bibinfo {author} {\bibfnamefont {P.}~\bibnamefont
  {Calabrese}},\ }\bibfield  {title} {\bibinfo {title} {{Universal parity
  effects in the entanglement entropy of XX chains with open boundary
  conditions}},\ }\href {https://doi.org/10.1088/1742-5468/2011/01/p01017}
  {\bibfield  {journal} {\bibinfo  {journal} {Journal of Statistical Mechanics:
  Theory and Experiment}\ }\textbf {\bibinfo {volume} {2011}},\ \bibinfo
  {pages} {P01017} (\bibinfo {year} {2011})}\BibitemShut {NoStop}%
\bibitem [{\citenamefont {Sandvik}(2010)}]{Sandvik2010}%
  \BibitemOpen
  \bibfield  {author} {\bibinfo {author} {\bibfnamefont {A.~W.}\ \bibnamefont
  {Sandvik}},\ }\bibfield  {title} {\bibinfo {title} {{Computational Studies of
  Quantum Spin Systems}},\ }\href {https://doi.org/10.1063/1.3518900}
  {\bibfield  {journal} {\bibinfo  {journal} {AIP Conference Proceedings}\
  }\textbf {\bibinfo {volume} {1297}},\ \bibinfo {pages} {135} (\bibinfo {year}
  {2010})}\BibitemShut {NoStop}%
\bibitem [{\citenamefont {Altland}\ and\ \citenamefont
  {Simons}(2010)}]{Altland2010}%
  \BibitemOpen
  \bibfield  {author} {\bibinfo {author} {\bibfnamefont {A.}~\bibnamefont
  {Altland}}\ and\ \bibinfo {author} {\bibfnamefont {B.~D.}\ \bibnamefont
  {Simons}},\ }\href {https://doi.org/https://doi.org/10.1017/CBO9780511789984}
  {\emph {\bibinfo {title} {{Condensed Matter Field Theory}}}}\ (\bibinfo
  {publisher} {Cambridge University Press},\ \bibinfo {year}
  {2010})\BibitemShut {NoStop}%
\bibitem [{\citenamefont {Auerbach}(2012)}]{Auerbach2012}%
  \BibitemOpen
  \bibfield  {author} {\bibinfo {author} {\bibfnamefont {A.}~\bibnamefont
  {Auerbach}},\ }\href
  {https://doi.org/https://doi.org/10.1007/978-1-4612-0869-3} {\emph {\bibinfo
  {title} {{Interacting Electrons and Quantum Magnetism}}}}\ (\bibinfo
  {publisher} {Springer New York},\ \bibinfo {year} {2012})\BibitemShut
  {NoStop}%
\bibitem [{\citenamefont {Wu}\ \emph {et~al.}(2008)\citenamefont {Wu},
  \citenamefont {Weng},\ and\ \citenamefont {Zaanen}}]{Wu2008}%
  \BibitemOpen
  \bibfield  {author} {\bibinfo {author} {\bibfnamefont {K.}~\bibnamefont
  {Wu}}, \bibinfo {author} {\bibfnamefont {Z.~Y.}\ \bibnamefont {Weng}},\ and\
  \bibinfo {author} {\bibfnamefont {J.}~\bibnamefont {Zaanen}},\ }\bibfield
  {title} {\bibinfo {title} {{Sign structure of the $t$-$\mathit{J}$ model}},\
  }\href {https://doi.org/10.1103/PhysRevB.77.155102} {\bibfield  {journal}
  {\bibinfo  {journal} {Phys. Rev. B}\ }\textbf {\bibinfo {volume} {77}},\
  \bibinfo {pages} {155102} (\bibinfo {year} {2008})}\BibitemShut {NoStop}%
\bibitem [{\citenamefont {Weber}\ \emph {et~al.}(2017)\citenamefont {Weber},
  \citenamefont {Tresp}, \citenamefont {Menke}, \citenamefont {Urvoy},
  \citenamefont {Firstenberg}, \citenamefont {B{\"u}chler},\ and\ \citenamefont
  {Hofferberth}}]{Weber2017}%
  \BibitemOpen
  \bibfield  {author} {\bibinfo {author} {\bibfnamefont {S.}~\bibnamefont
  {Weber}}, \bibinfo {author} {\bibfnamefont {C.}~\bibnamefont {Tresp}},
  \bibinfo {author} {\bibfnamefont {H.}~\bibnamefont {Menke}}, \bibinfo
  {author} {\bibfnamefont {A.}~\bibnamefont {Urvoy}}, \bibinfo {author}
  {\bibfnamefont {O.}~\bibnamefont {Firstenberg}}, \bibinfo {author}
  {\bibfnamefont {H.~P.}\ \bibnamefont {B{\"u}chler}},\ and\ \bibinfo {author}
  {\bibfnamefont {S.}~\bibnamefont {Hofferberth}},\ }\bibfield  {title}
  {\bibinfo {title} {{Tutorial: Calculation of Rydberg interaction
  potentials}},\ }\href {https://doi.org/10.1088/1361-6455/aa743a} {\bibfield
  {journal} {\bibinfo  {journal} {Journal of Physics B: Atomic, Molecular and
  Optical Physics}\ }\textbf {\bibinfo {volume} {50}},\ \bibinfo {pages}
  {133001} (\bibinfo {year} {2017})}\BibitemShut {NoStop}%
\end{thebibliography}%
\clearpage
\appendix
\widetext
\begin{center}
	\textbf{\large Supplemental Materials for: “Competing and Intertwined Orders in Boson-Doped Mott Antiferromagnets”}
\end{center}
\vspace{1mm}
\renewcommand\thefigure{\thesection S\arabic{figure}}
\renewcommand\theequation{\thesection S\arabic{equation}}
\setcounter{figure}{0} 
\setcounter{equation}{0}

In the Supplemental Materials, we provide more results to support the conclusions we have discussed in the main text. 
In Sec. A, we show the doping evolution of the charge density profile $n(x)$ as well as the spin structure factor $S(\bf k)$ in the PS regime. 
In Sec. B, we present more details of the SF+$xy$-FM phase. 
In Sec. C, we discuss the origin of the condensation pattern of momentum distribution $n(\bf k)$ in the SF+AFM phase and demonstrate various correlation functions.
In Sec. D, we provide the structure factors in the SF*+IM phase, analyze the mechanism of incommensurate magnetism, and present more complementary results on the eight-leg cylinders.
In Sec. E, and Sec. F, we show more details of the PDW+AFM and dPDW+AFM phases, respectively, and compare their similarities and differences.
In Sec. G, we present more details about the BOW state at $\delta=1/4$. 
In Sec. H and Sec. I, we respectively provide a detailed derivation of the intrinsic $\mathbb{Z}_{2}$ Berry phase in the bosonic $t$-$t'$-$J$ model as well as the effective bosonic $\sigma t$-$t'$-$J$ model from the spinful hard-core Bose-Hubbard model at large-$U$ limit. 
In Sec. J, we show the phase diagram of the bosonic $\sigma t$-$t'$-$J$ model and related physical quantities. 
In the last Sec. K, we elaborate on the detailed experimental implementation of the bosonic $t$-$t'$-$J$ model on Rydberg tweezer platforms.

\section{\label{sec_PS}A. Doping evolution of the charge density profiles and spin structure factors in the PS regime}
\begin{figure}[h]
   \includegraphics[width=0.6\textwidth,angle=0]{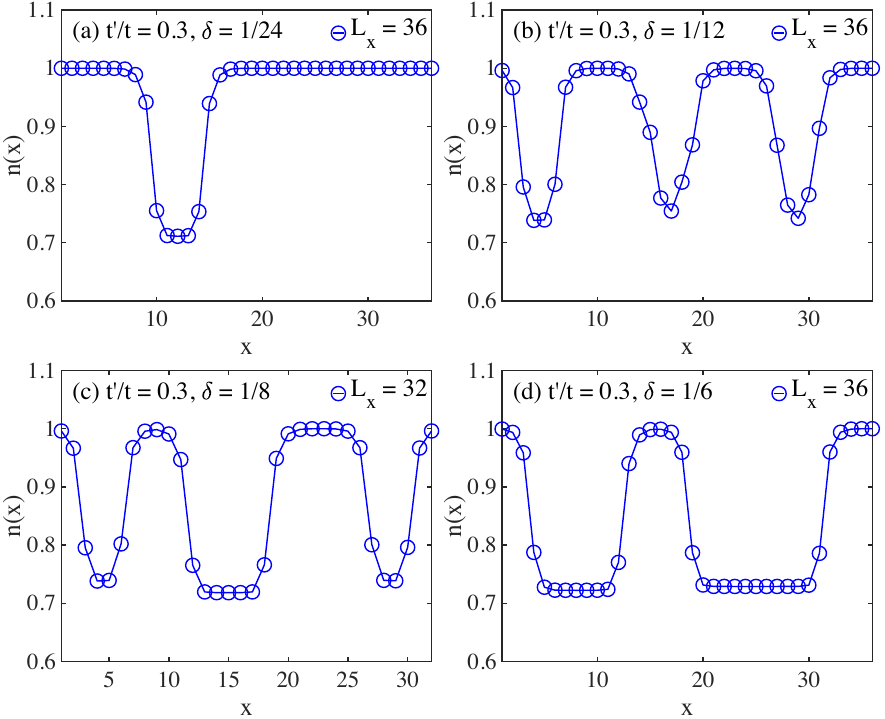}
   \caption{\label{SM_PS_CDW}
   \textbf{Charge density profiles $n(x)$ in the PS regime.}
   $n(x)$ with fixed $t'/t=0.3$ at (a) $\delta =1/24$, (b) $\delta =1/12$, (c) $\delta =1/8$, and (d) $\delta =1/6$.
   }
\end{figure}

\begin{figure}[h]
   \includegraphics[width=0.45\textwidth,angle=0]{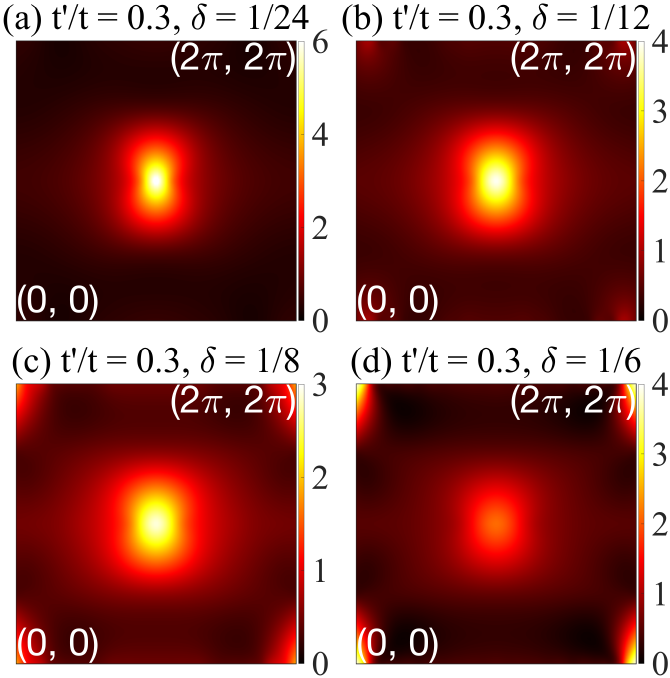}
   \caption{\label{SM_PS_Sk}
   \textbf{Spin structure factors $S(\bf k)$ in the PS regime.}
   $S(\bf k)$ with fixed $t'/t=0.3$ at (a) $\delta =1/24$, (b) $\delta =1/12$, (c) $\delta =1/8$, and (d) $\delta =1/6$. Here all the $S(\bf k)$ are obtained by taking the Fourier transformation for the all-to-all correlations.
   }
\end{figure}

In the $t'>0$ side of the phase diagram, we find the system manifests as stripe-like PS at low doping levels. In Fig.~\ref{SM_PS_CDW}, we show the evolution of the charge density profiles $n(x) = \sum_{y=1}^{L_y} \langle {\hat{n} }_{x,y} \rangle / L_y$, as the doping ratio increases from $\delta=1/24$ to $1/6$ with fixed $t'/t=0.3$ across the PS regime. It is evident that the doped holes do not distribute uniformly throughout the system. Instead, they tend to cluster together, leading to the formation of hole-rich regions and undoped regions. Moreover, the propagation of the doped holes remains confined within the hole-rich regions, further underscoring the spatial segregation induced by the phase separation. On the other hand, we also need to state that the magnetic background in the hole-rich regions and undoped regions is distinct. The undoped regions remain robust AFM background, while the hole-rich regions behave as FM tendencies. To illustrate this point more clearly, we present the corresponding evolution of the spin structure factors $S(\bf k)$ in the PS regime, as shown in Fig.~\ref{SM_PS_Sk}. At lower doping levels, the global $S(\bf k)$ does not exhibit clear FM signatures as the hole-rich regions remain narrow. However, as the doping level increases, these regions broaden, and $S(\bf k)$ gradually develops an FM peak at $(0, 0)$. Concurrently, AFM peak at $(\pi, \pi)$ is gradually suppressed, reflecting the shrinkage of undoped regions. If the doping level continues to increase, the undoped regions will eventually disappear completely, and the system transitions into a uniform SF+$xy$-FM phase, characterized by a single dominant peak at $(0, 0)$ in the $S(\bf k)$, which will be further discussed in the next section.

\section{\label{SM_FM}B. More details in the uniform SF+$xy$-FM phase}

In this section, we show the charge density profiles $n(x)$, spin correlations $F(r)=F_{+-}(r)+F_{zz}(r)$, single-boson correlations $G(r)=\sum_{\sigma}G_{\sigma}(r)$, and charge density correlations $D(r) = \langle {\hat{n} }_{x,y} {\hat{n} }_{x+r,y} \rangle - \langle {\hat{n} }_{x,y} \rangle \langle {\hat{n} }_{x+r,y} \rangle$
in the SF+$xy$-FM phase at $t'/t=0.3$ across doping levels from $\delta=1/5$ to $1/3$ to complement the discussion in the main text. In contrast with the PS, the $n(x)$ here are quite flat, doped holes are uniformly distributed throughout the bulk of systems within the SF+$xy$-FM phase, as shown in Figs.~\ref{SM_FM}(a1-c1). Besides, both the $F(r)$ [Figs.~\ref{SM_FM}(a2-c2)] and $G(r)$ [Figs.~\ref{SM_FM}(a3-c3)] also exhibit nearly long-range behavior, which is consistent with the sharp peaks of $S(\bf k)$ and $n(\bf k)$ in Fig. 3(c) and Fig. 2(c), respectively. 
These features collectively indicate the gapless nature of the SF+$xy$-FM phase.
Here, we need to emphasize that the SF+$xy$-FM phase spontaneously breaks the $SU(2)_{\mathrm{spin}}$ symmetry. As shown in Figs.~\ref{SM_FM}(a2-c2), one can find that although both in-plane spin correlation $F_{+-}(r)=\langle {\hat{{S}}^{+}_{x,y}} {\hat{{S}}}^{-}_{x+r,y} +\mathrm{H}.\mathrm{c}.\rangle/2$ and longitudinal spin correlation $F_{zz}(r)=\langle {\hat{{S}}^{z}_{x,y}} {\hat{{S}}}^{z}_{x+r,y} \rangle$ exhibit FM-type behavior, the latter is much weaker than the former. The ground state of the SF+$xy$-FM phase always lies in the sector with $S^{z}_{\mathrm{total}} =0$, which we have verified for both cases with and without enforcing the $U(1)_{\mathrm{spin}}$ symmetry. This indicates that the FM order is dominated by in-plane spin alignment.
For the $D(r)$ [Figs.~\ref{SM_FM}(a4-c4)], we find that it decays much faster than $F(r)$ and $G(r)$, which implies that there are no other intertwined charge orders in the SF+$xy$-FM phase. Here, we do not present the corresponding pairing correlations $P_{yy}(r)$ because they are extremely weak (near vanishing), and there is no singlet binding in the SF+$xy$-FM phase.

\begin{figure}[h]
   \includegraphics[width=0.9\textwidth,angle=0]{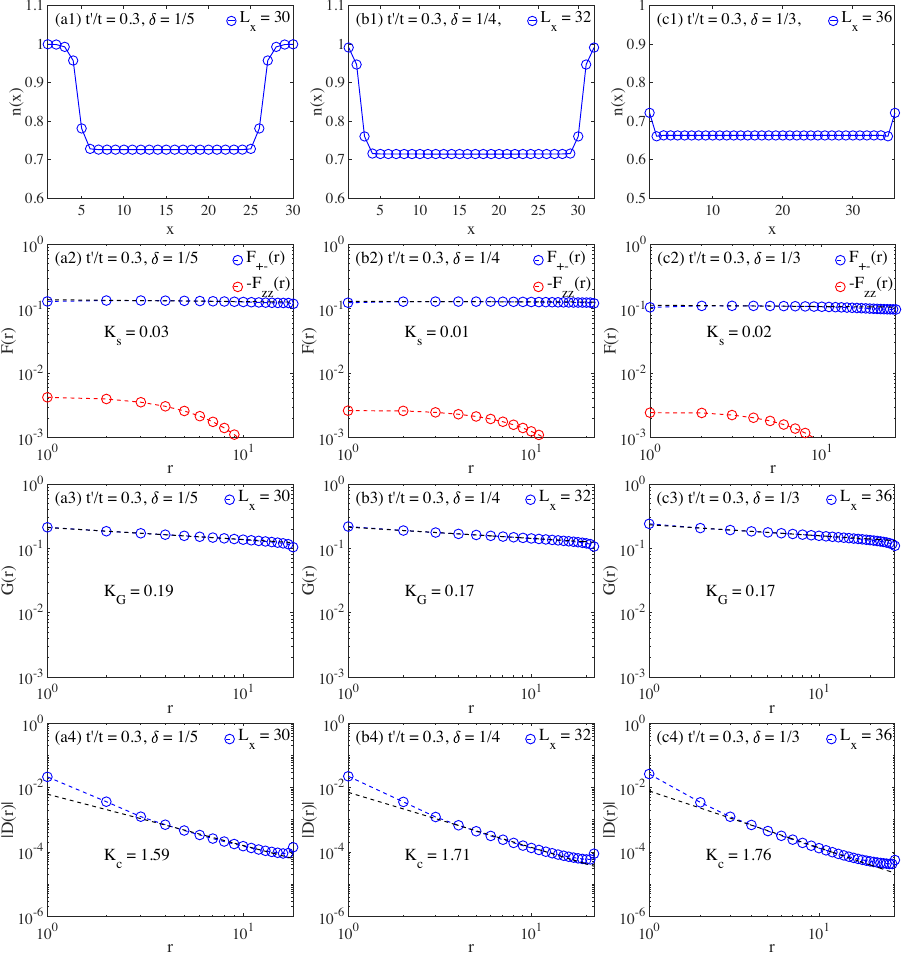}
   \caption{\label{SM_FM}
   \textbf{Charge density profiles and correlation functions in the SF+$xy$-FM phase.} (a1-a4) are respectively the charge density profile $n(x)$, double-logarithmic plot of the spin correlation $F(r)$, double-logarithmic plot of the single-boson correlation $G(r)$, and double-logarithmic plot of the charge density correlation $D(r)$ at $\delta=1/5$ with fixed $t'/t=0.3$. The power exponents $K_{\mathrm{s}}$, $K_{\mathrm{G}}$, and $K_{\mathrm{c}}$ are obtained by algebraic fitting with dash line. (b1-b4) and (c1-c4) are the similar plots at $\delta=1/4$ and $\delta=1/3$, respectively.
   }
\end{figure}

\section{\label{sec_SF_AFM}C. NNN hopping dominated SF+AFM phase}
\begin{figure}[h]
   \includegraphics[width=0.45\textwidth,angle=0]{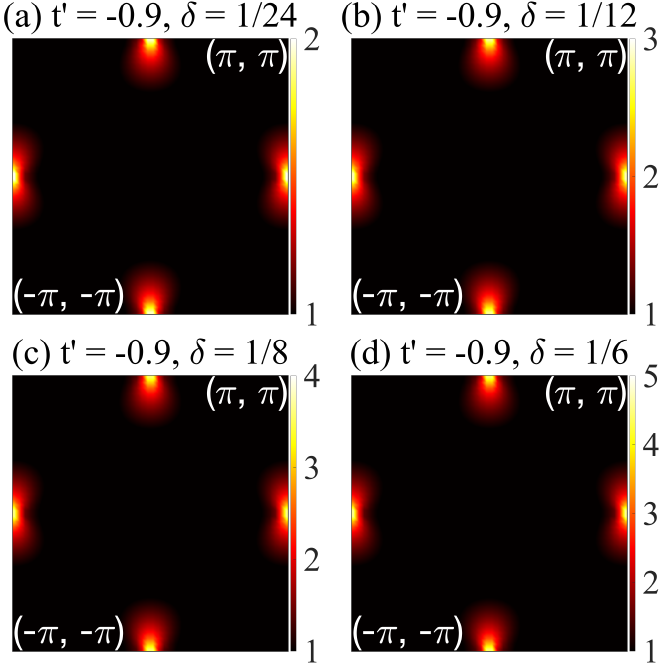}
   \caption{\label{SM_SFAFM_nk}
   \textbf{Momentum distributions $n(\bf k)$ in the SF+AFM phase with $t=0$.}
   $n(\bf k)$ with fixed $t'=-0.9$ at (a) $\delta =1/24$, (b) $\delta =1/12$, (c) $\delta =1/8$, and (d) $\delta =1/6$. Here all the $n(\bf k)$ are obtained by taking the Fourier transformation for the all-to-all correlations.
   }
\end{figure}

In the main text, we have shown the momentum distribution $n(\bf k)$ for the SF+AFM phase in Fig. 2(d), where one can see that the bosons are condensed at four symmetric momenta $(\pm\pi,0)$ and $(0,\pm\pi)$. To elucidate the origin of this condensation pattern, we demonstrate that it arises from the dominant NNN hopping processes at low doping levels.
As shown in Fig.~\ref{SM_SFAFM_nk}, we present analogous $n(\bf k)$ across doping levels 
$\delta=1/24-1/6$, calculated by suppressing NN hopping ($t=0$) while retaining a finite NNN hopping amplitude ($t'=-0.9$). Notably, the four-fold symmetric condensation peaks persist at $(\pm\pi,0)$ and $(0,\pm\pi)$, confirming that the doped bosons preferentially undergo diagonal hopping along the square lattice. This observation underscores the pivotal role of NNN hopping in shaping the low-doping condensate structure. Physically, this behavior arises because at very low doping density, the spin exchange energy dominates over the kinetic energy of hole hopping. As a result, the spin background strongly favors AFM order. However, such an AFM background suppresses the motion of individual holes via the NN hopping channel, effectively confining hole motion to the NNN paths. Consequently, the $t$–$t'$–$J$ model at small doping with finite $t$ exhibits the same phase behavior as the $t'$–$J$ model. In contrast, the situation is different for the $\sigma t$–$t'$–$J$ model, where the NN $\sigma t$ hopping term actually favors local AFM correlations. Thus, it is compatible with the AFM spin background. This explains the numerical observation that the $\sigma t$–$t'$–$J$ model at low doping with finite $t$ still displays the same behavior as the $\sigma t$–$J$ model, indicating that the dominant hopping channel remains the NN one.

\begin{figure}[h]
   \includegraphics[width=0.6\textwidth,angle=0]{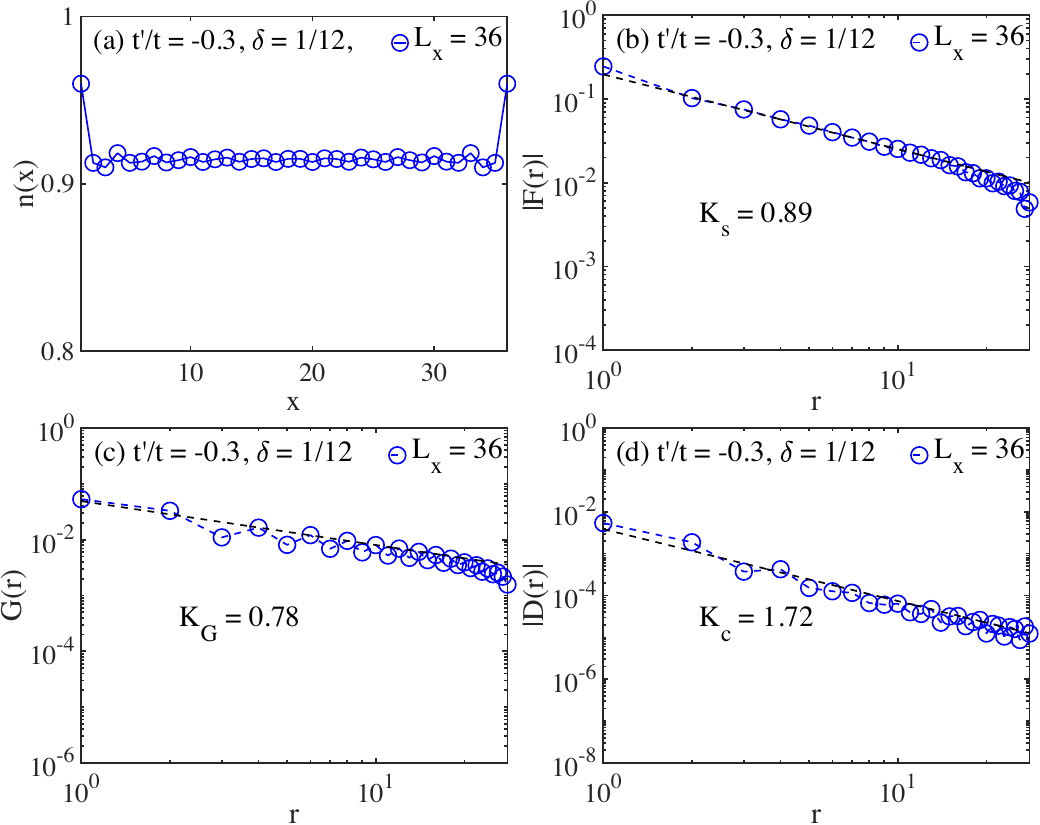}
   \caption{\label{SM_SFAFM_CDWcorre}
   \textbf{Charge density profile and correlation functions in the SF+AFM phase.} (a) Charge density profile $n(x)$ at $t'/t=-0.3, \delta=1/12$. (b-c) are respectively the double-logarithmic plot of spin correlation $F(r)$, single-boson correlation $G(r)$, and charge density correlation $D(r)$ with the same parameters of (a). The power exponents $K_{\mathrm{s}}$, $K_{\mathrm{G}}$, and $K_{\mathrm{c}}$ are obtained by algebraic fitting with dash line.
   }
\end{figure}

In Fig.~\ref{SM_SFAFM_CDWcorre}, we also show the charge density profile $n(x)$ and other correlation functions to further substantiate our findings. Firstly, we still observe the flat $n(x)$ [or near vanishing charge density oscillation] within the SF+AFM phase, as shown in Fig.~\ref{SM_SFAFM_CDWcorre}(a), which is similar to the SF+$xy$-FM phase as discussed in the previous section. Moreover, both the spin correlation $F(r)$ [Fig.~\ref{SM_SFAFM_CDWcorre}(b)] and single-boson correlation $G(r)$ [Fig.~\ref{SM_SFAFM_CDWcorre}(c)] show strong quasi-long-range order, characterized by $K_{\mathrm{s}}<1$ and $K_{\mathrm{G}}<1$, respectively. This confirms that both the strong magnetic correlations and the single-hole condensation remain robust and gapless. Notably, in contrast to the suppression of pairing correlation $P_{yy}(r)$ observed in Fig. 4 (c), here the charge density correlation $D(r)$ does not suppress completely. Instead, $D(r)$ exhibits a superb power-law behavior with $K_{\mathrm{c}}<2$, as shown in Fig.~\ref{SM_SFAFM_CDWcorre}(d).

\section{\label{SM_IM}D. Incommensurate magnetism in the SF*+IM phase}

\begin{figure}
   \includegraphics[width=0.72\textwidth,angle=0]{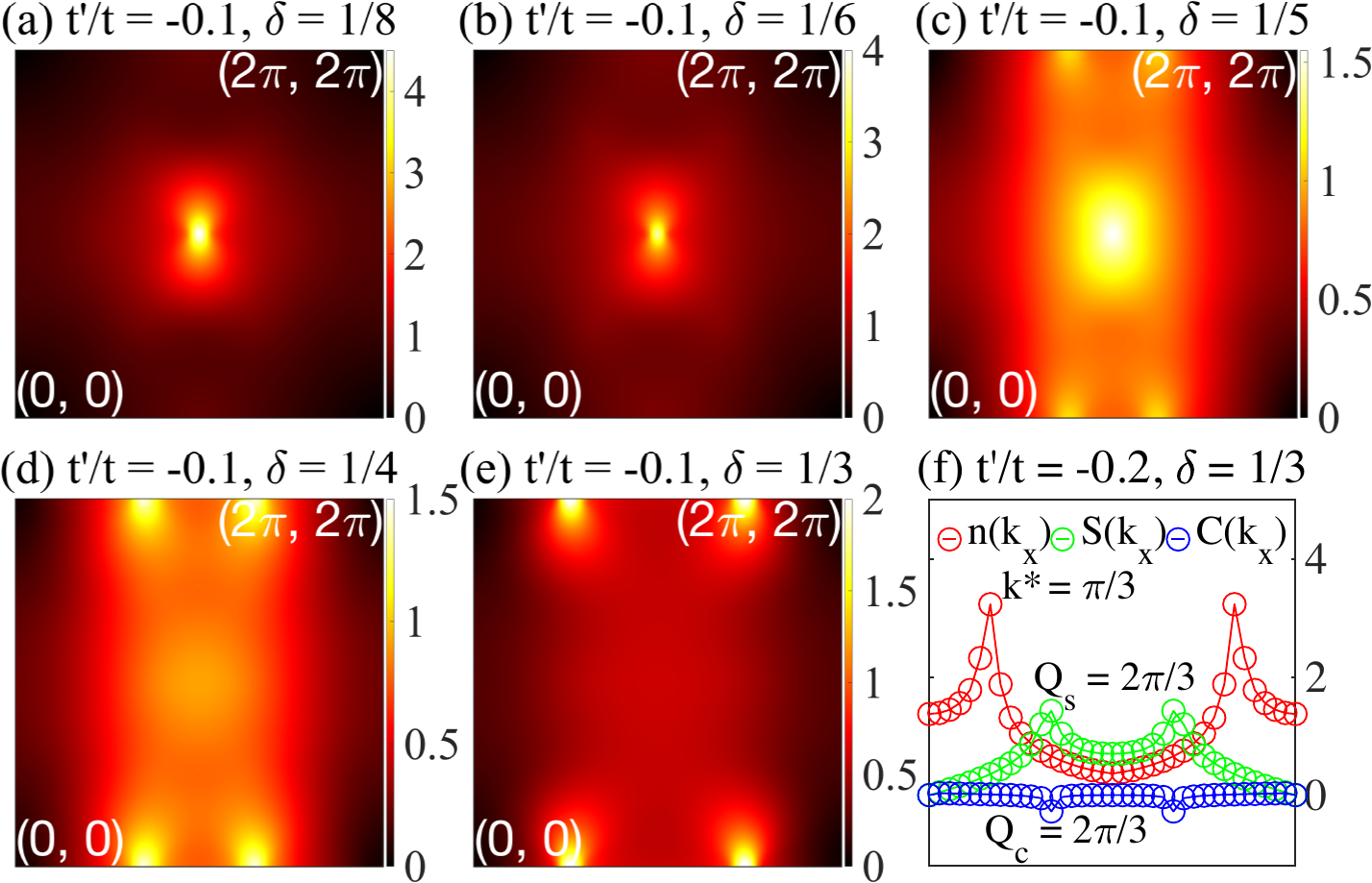}
   \caption{\label{Itinerant_magnetism}
   \textbf{Spin structure factors $S(\bf k)$ in the PDW+AFM and SF*+IM phases on four-leg systems.}
   (a-e) are the evolution of $S(\bf k)$ in the PDW+AFM and SF*+IM phases with fixed $t'/t=-0.1$. (f) $n(\bf k)$, $S(\bf k)$, and CDW structure factor $C(\bf k)$ in the SF*+IM phase with $t'/t=-0.2$. Here all the structure factors are obtained by taking the Fourier transformation for the all-to-all correlations.
   }
\end{figure}

In this section, we present more details about the incommensurate magnetism as discussed in the main text. On the $t'<0$ side, the system consistently exhibits quasi-long-range N\'eel AFM order except in the SF*+IM phase, where the original N\'eel AFM order is suppressed. To illustrate this point clearly, we show the doping evolution of the spin structure factors $S(\bf k)$ from $\delta=1/8$ to $1/3$ with fixed $t'/t=-0.1$ across the PDW+AFM and SF*+IM phases in Fig.~\ref{Itinerant_magnetism}. One can clearly find that the original N\'eel order in the PDW+AFM phase [Figs.~\ref{Itinerant_magnetism}(a-b)] is gradually suppressed. Once the system transitions into the SF*+IM phase [Figs.~\ref{Itinerant_magnetism}(c-e)], a new magnetic order emerges at $(\mathbf{Q}_s,0)$, simultaneously. To understand the origin mechanism of such IM order, we also study the momentum distribution $n(\mathbf{k})$ and charge density wave (CDW) structure factor $C(\mathbf{k})=(1/N)\sum_i \langle {\hat{n} }_i -(1-\delta)\rangle e^{i\mathbf{k}\cdot {\mathbf{r}}_i }$ as shown in Fig.~\ref{Itinerant_magnetism}(f), we find a universal relation within the SF*+IM phase, i.e. $\mathbf{Q}_c = \mathbf{Q}_s = 2k^*$. 

\begin{figure}
   \includegraphics[width=0.6\textwidth,angle=0]{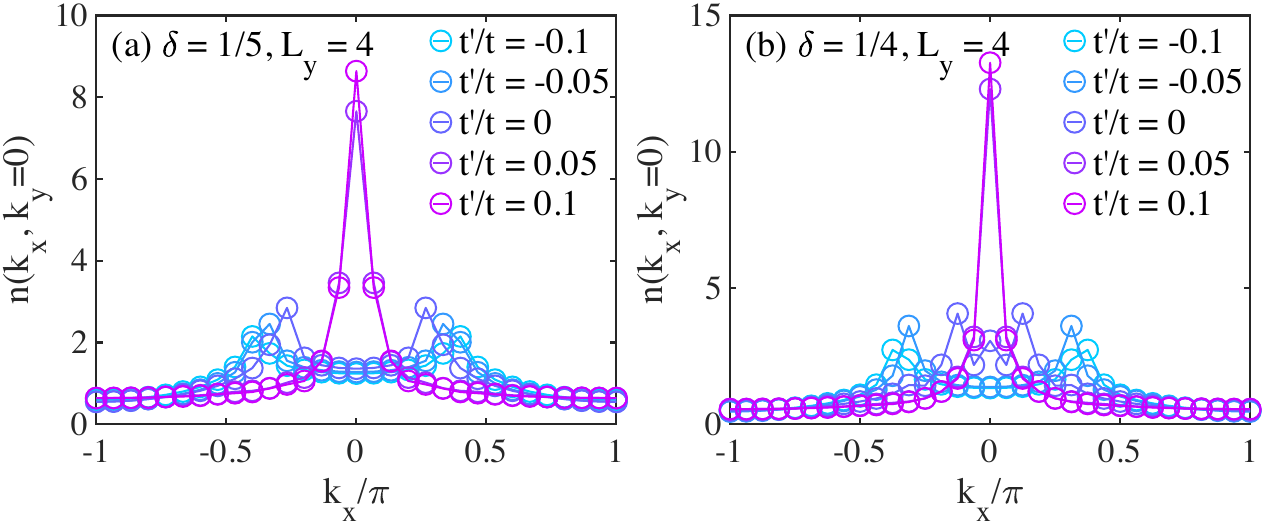}
   \caption{\label{R_KIM}
   \textbf{Momentum distribution $n(k_x, k_y=0)$ in the SF*+IM and SF+$xy$-FM phases on four-leg systems.}
   (a) $n(k_x, k_y=0)$ at $\delta=1/5$ with varied $t'/t$. 
   (b) $n(k_x, k_y=0)$ at $\delta=1/4$ with varied $t'/t$.
   Here all the structure factors are obtained by taking the Fourier transformation for the all-to-all correlations.
   }
\end{figure}

In Fig. 2(b), we have shown that the SF*+IM phase features the most anomalous condensation of holes at $(\pm k^*,0)$. In Fig.~\ref{R_KIM}, we further present the dependence of the $k^*$ on $\delta$ and $t'/t$. Clearly, when we fix $\delta$ and vary only the $t'/t$, the position of $k^*$ shifts continuously. As the system moves toward the $t'/t < 0$ side, the separation between the two $k^*$ peaks increases; conversely, as it moves toward the $t'/t > 0$ side, the peaks become higher and sharper. When the two peaks eventually merge at $(0,0)$, the system transitions into the SF+$xy$-FM phase. In addition, within the SF*+IM phase, increasing the doping level leads to progressively higher $k^*$ peaks, indicating an enhanced condensation at $\pm k^*$. Although it remains challenging to establish an explicit functional relationship between $k^*$ and other physical parameters prior to a full understanding of the microscopic mechanism underlying the SF*+IM phase, our numerical results nevertheless unambiguously demonstrate that $k^*$ is closely related to both $\delta$ and $t'/t$.
Previous studies of the single-hole-doped Mott antiferromagnets have shown that hole motion leads to emergent frustration, resulting in nontrivial momentum transfer and spin current accompanying the doped hole~\cite{Zhao2023}. This behavior bears strong resemblance to the emergence of the incommensurate momentum $k^*$ observed here. Concurrently, the AFM order at $(\pi,\pi)$ --- a remnant of the local moments from the parent Mott insulator --- gives way to IM with ordering vector $2k^*$, which emerges from the recombination of low-lying bosonic modes with condensate momenta $\pm k^*$. This suggests a binary nature of spin structure, with both localized and itinerant components present in doped Mott antiferromagnets~\cite{zhang2024_hg}.

\begin{figure}[h]
   \includegraphics[width=0.6\textwidth,angle=0]{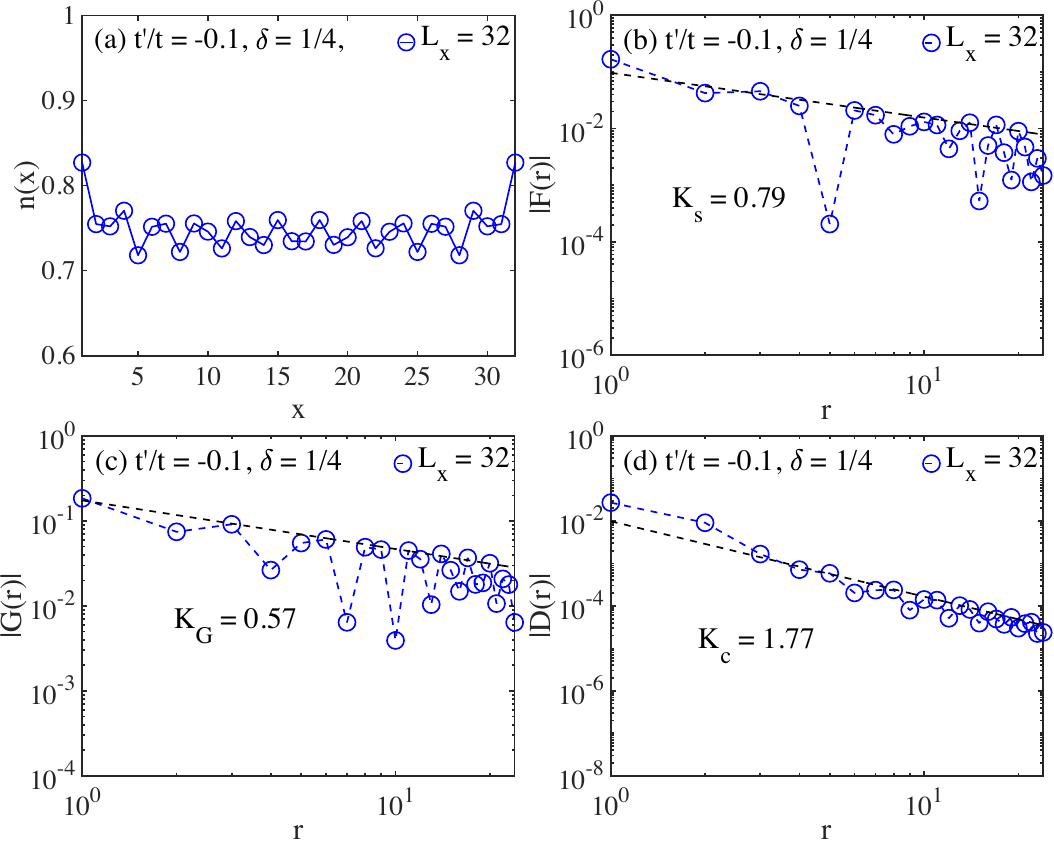}
   \caption{\label{SM_SFIM_CDWcorre}
   \textbf{Charge density profile and correlation functions in the SF*+IM phase on four-leg systems.} (a) Charge density profile $n(x)$ at $t'/t=-0.1, \delta=1/4$. (b-d) are respectively the double-logarithmic plot of spin correlation $F(r)$, single-boson correlation $G(r)$, and charge density correlation $D(r)$ with the same parameters of (a). The power exponents $K_{\mathrm{s}}$, $K_{\mathrm{G}}$, and $K_{\mathrm{c}}$ are obtained by algebraic fitting with dash line.
   }
\end{figure}

In Fig.~\ref{SM_SFIM_CDWcorre}, we further present the representative charge density profile $n(x)$ and other correlation functions to consolidate our conclusions in the SF*+IM phase on four-leg cylinders. Different from the SF+$xy$-FM and SF+AFM phases, here the $n(x)$ shows obvious non-uniform charge density modulation, as shown in Fig.~\ref{SM_SFIM_CDWcorre}(a), a direct consequence of the single-boson condensation at finite momentum discussed earlier. While for the spin correlation $F(r)$ [Fig.~\ref{SM_SFIM_CDWcorre}(b)], single-boson correlation $G(r)$ [Fig.~\ref{SM_SFIM_CDWcorre}(c)], and charge density correlation $D(r)$ [Fig.~\ref{SM_SFIM_CDWcorre}(d)], we find that they are highly similar to those in the SF+$xy$-FM and SF+AFM phases and likewise indicate gapless behavior.  All three correlation functions exhibit quite strong quasi-long-range order with power exponents $K_{\mathrm{s}}<1$, $K_{\mathrm{G}}<1$, and $K_{\mathrm{c}}<2$, respectively.

\begin{figure}[h]
   \includegraphics[width=0.45\textwidth,angle=0]{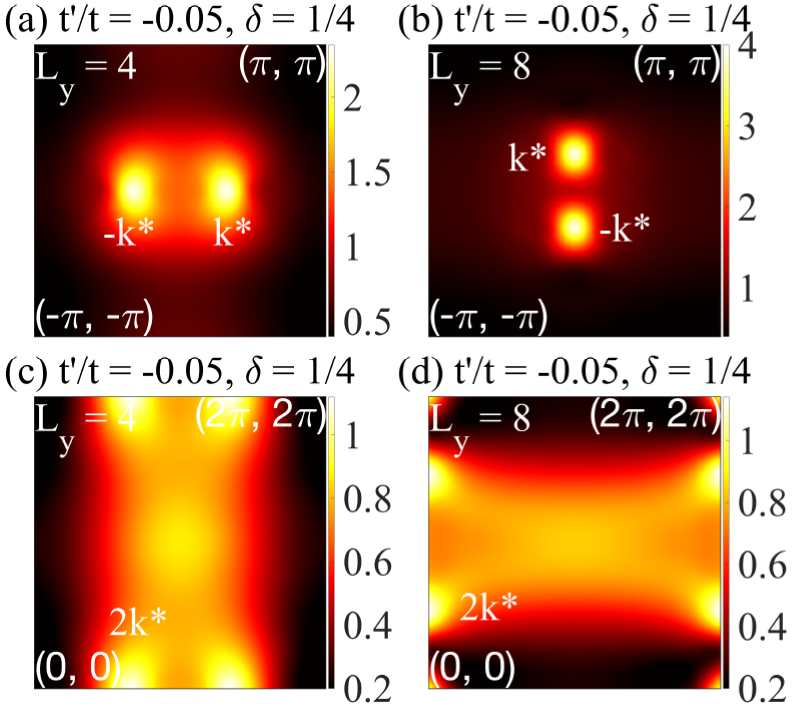}
   \caption{\label{R_SFIM8}
  \textbf{Momentum distribution $n({\bf k})$ and spin structure factor $S({\bf k})$ for fixed $L_{x}=12$, $t'/t=-0.05$, and $\delta=1/4$ on four- and eight-leg systems.} (a) and (b) show $n(\mathbf{k})$ for the four- and eight-leg systems, respectively. (c) and (d) show $S(\mathbf{k})$ for the four- and eight-leg systems, respectively. Here all the $n({\bf k})$ and $S({\bf k})$ are obtained by taking the Fourier transformation for the all-to-all correlations.
   }
\end{figure}

\begin{figure}
   \includegraphics[width=0.8\textwidth,angle=0]{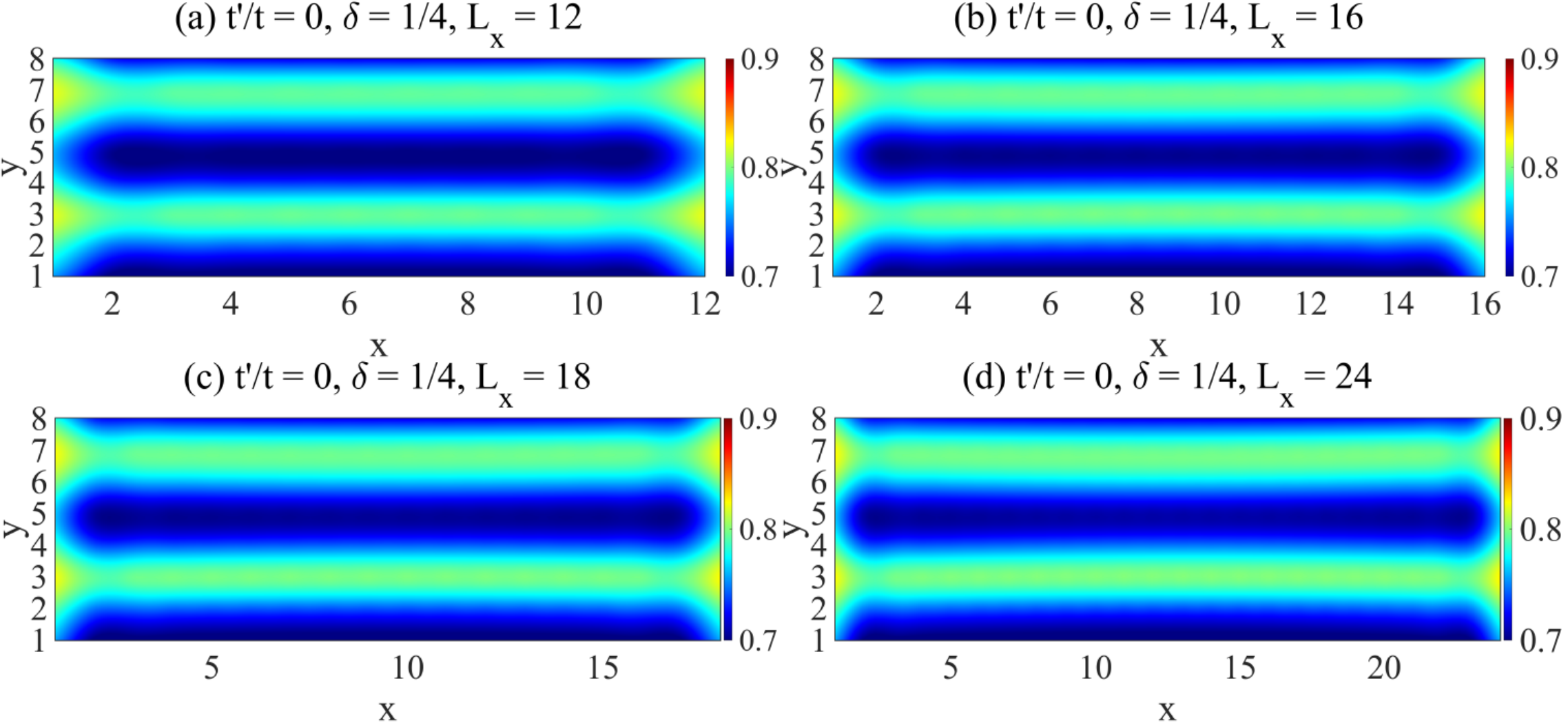}
   \caption{\label{SM_8legcdw}
   \textbf{Charge density distributions $n(x,y)$ in the SF*+IM phase on eight-leg systems at $t'/t=0$, $\delta=1/4$.}
   (a) $L_{x}=12$, (b) $L_{x}=16$, (c) $L_{x}=18$, and (d) $L_{x}=24$.
   }
\end{figure}

\begin{figure}
   \includegraphics[width=0.6\textwidth,angle=0]{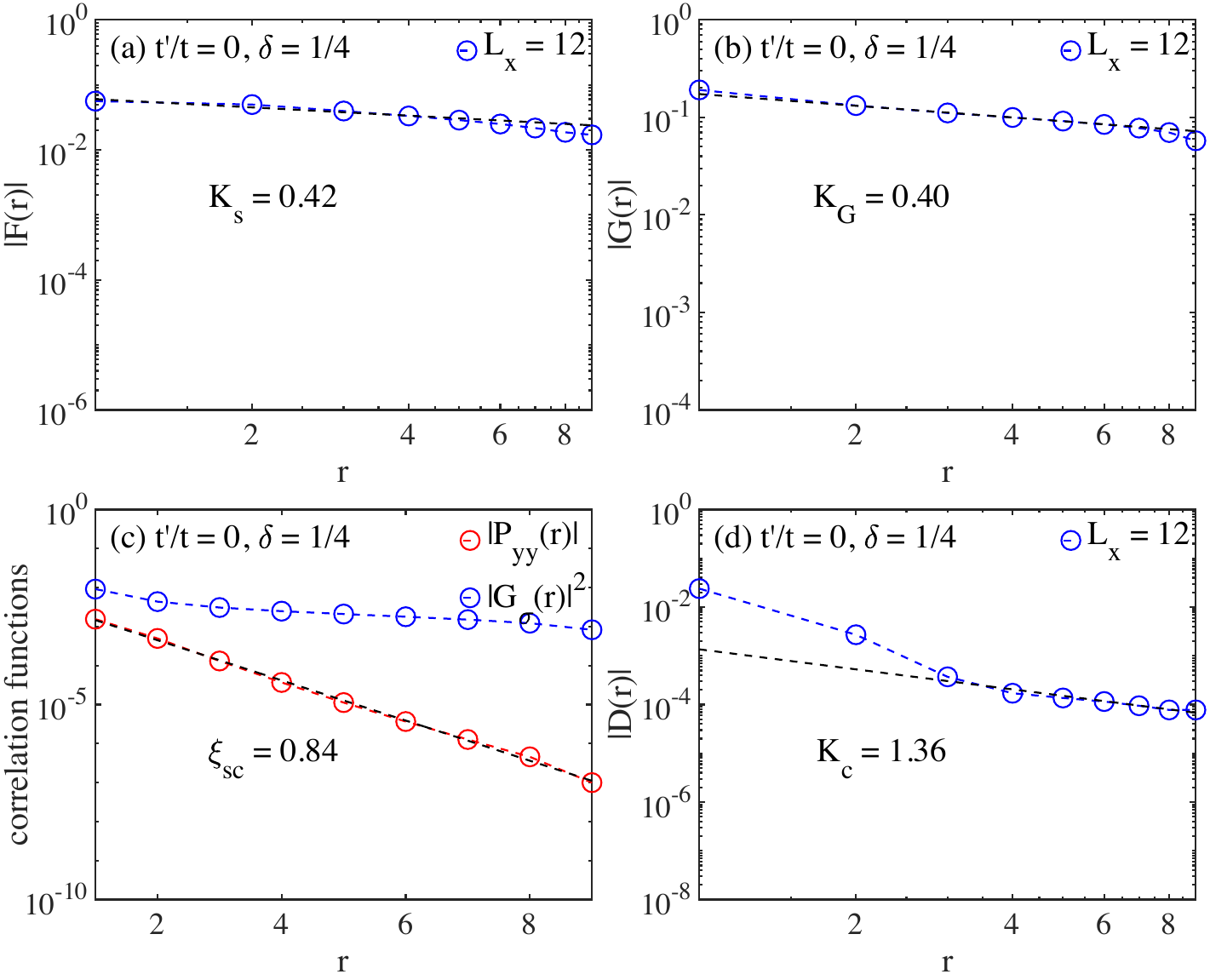}
   \caption{\label{SM_Corre8leg}
   \textbf{Correlation functions in the SF*+IM phase on eight-leg systems.} (a) and (b) are respectively the double-logarithmic plot of spin correlation $F(r)$ and single-boson correlation $G(r)$. (c) is the semi-logarithmic plot of the pairing correlations $P_{yy}(r)$ and $G_\sigma^2(r)$. (d) is the double-logarithmic plot of the charge density correlation $D(r)$. The power exponents $K_{\mathrm{s}}$, $K_{\mathrm{G}}$, and $K_{\mathrm{c}}$ are obtained by algebraic fitting with dash line. The correlation lengths $\xi_{\mathrm{sc}}$ is obtained by exponential fitting with dash line.
   }
\end{figure}

In Fig. 5, we have demonstrated that the exotic SF*+IM phase is highly robust, with the condensation further enhanced as the system length $L_x$ increases on wider eight-leg systems. To facilitate a more direct comparison between different system widths $L_y$ and to demonstrate the robustness of the SF*+IM phase toward the 2D limit, in Fig.~\ref{R_SFIM8} we show the evolution of $n(\mathbf{k})$ and $S(\mathbf{k})$ with respect to $L_y$, while fixing the system length to $L_x = 12$ and the model parameters to $t'/t = -0.05$ and $\delta = 1/4$. We find that the peak height of $n(\mathbf{k})$ on the eight-leg system [Fig.~\ref{R_SFIM8}(b)] is approximately twice that on the four-leg system [Fig.~\ref{R_SFIM8}(a)], and the IM order remains very stable (the peak at $2k^*$ on the eight-leg system [Fig.~\ref{R_SFIM8}(d)] is even slightly higher than that of the four-leg system [Fig.~\ref{R_SFIM8}(c)]). These results indicate that the SF*+IM phase remains robust as the system approaches the 2D limit, suggesting that it is likely to persist in true 2D systems. 
Next, we provide further details regarding the corresponding 2D charge density distributions $n(x, y)=\langle {\hat{n} }_{x,y} \rangle$ and correlation functions. Firstly, we observe that the orientation of oscillations in $n(x, y)$ is closely related to the splitting direction of the sub-peaks of $n(\bf k)$. Specifically, the real-space oscillation of $n(x, y)$ orients along the $x$ or $y$ direction when the sub-peaks of $n(\bf k)$ split along the $k_x$ or $k_y$ direction, respectively, as shown in Fig.~\ref{SM_SFIM_CDWcorre}(a) on four-leg systems and Fig.~\ref{SM_8legcdw} on eight-leg systems. Besides, to exclude possible finite size effects, we performed calculations on eight-leg cylinders with different system lengths $L_{x}=12, 16, 18, 24$ [Fig.~\ref{SM_8legcdw}]. Our results show that the $n(x, y)$ is always uniform in the $x$ direction but modulates solely along the $y$ direction with a period of $4$, and the relation $\mathbf{Q}_c = \mathbf{Q}_s = 2k^*$ is still satisfied in this case. Moreover, in order to further characterize and compare the quasi-long-range superfluidity and magnetic properties of the SF* state on eight-leg systems, we also calculate the real-space correlation functions, as shown in Fig.~\ref{SM_Corre8leg}. Obviously, the single-boson correlation $G(r)$ exhibits superb power-law behavior with $K_{\mathrm{G}}\simeq0.40$ [Fig.~\ref{SM_Corre8leg}(b)], and the squared single-boson correlation $G_\sigma^2(r)$ is stronger than the pairing correlation $P_{yy}(r)$ [Fig.~\ref{SM_Corre8leg}(c)], which confirms the SF* nature and single-boson condensation. Notably, the power exponent $K_{\mathrm{s}}\simeq0.42$ extracted from the spin correlation $F(r)$ [Fig.~\ref{SM_Corre8leg}(a)] is even smaller than that in the SF*+IM phase on four-leg systems [Fig.~\ref{SM_SFIM_CDWcorre}(b)], suggesting that the emergent IM order may persist in the 2D limit for the SF* state. Lastly, as for the charge density correlation $D(r)$ [Fig.~\ref{SM_Corre8leg}(d)], it remains weaker than $G(r)$ and $F(r)$, indicating that there is no other competing charge order alongside the SF* and IM orders, consistent with the behavior observed in the four-leg case.

\section{\label{SM_PDW_sec}E. More details in the PDW+AFM phase}

\begin{figure}
   \includegraphics[width=0.9\textwidth,angle=0]{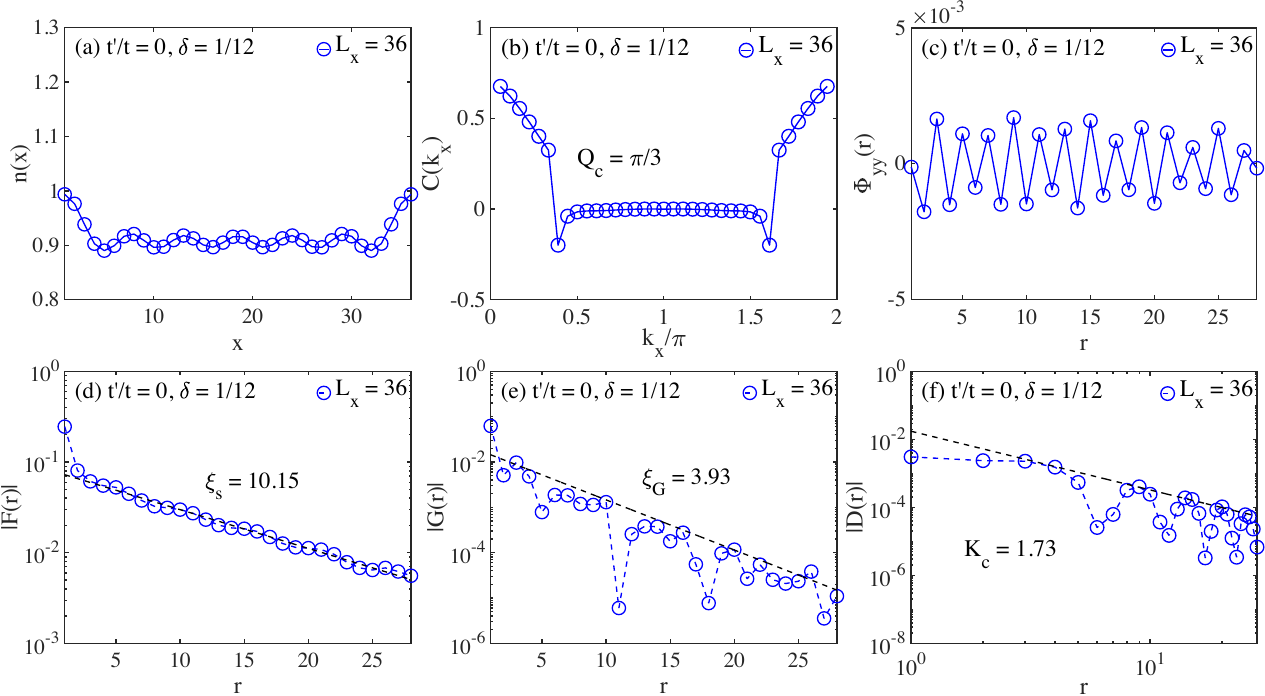}
   \caption{\label{SM_PDW}
   \textbf{PDW+AFM state at $t'/t=0$, $\delta=1/12$.}
   (a) Charge density profile $n(x)$. (b) CDW structure factor $C(\bf k)$ of (a). (c) Pairing oscillation $\Phi_{yy}(r)$, (d) and (e) are respectively the semi-logarithmic plot of spin correlation $F(r)$ and single-boson correlation $G(r)$, (f) double-logarithmic plot of charge density correlation $D(r)$. The correlation lengths $\xi_{\mathrm{s}}$ and $\xi_{\mathrm{G}}$ are obtained by exponential fitting with dash lines. The power exponent $K_{\mathrm{c}}$ is obtained by algebraic fitting with dash line.
   }
\end{figure}

The PDW is a SC state that carries a finite center-of-mass momentum, with a SC order parameter that varies periodically in real space such that its spatial average vanishes.~\cite{Agterberg_PDW_2020}. In the main text, we have shown in the phase diagram that there is a PDW+AFM phase at low doping levels. In this section, we present more details about this phase. Firstly, we find the charge density profiles are half-filled, indicating charge density oscillation with a period of $\lambda_{c} =1/2\delta$, as shown in Fig.~\ref{SM_PDW}(a). To further clarify this behavior, we also calculate the corresponding CDW structure factor $C(\bf k)$ in Fig.~\ref{SM_PDW}(b), which reveals a prominent peak at the ordering momentum $\mathbf{Q}_{c}=4\pi\delta$, confirming the periodicity of the charge density modulation. In general, a half-filled CDW is always superconducting~\cite{Jiang_PRR_2020}. According to the definition of PDW, we can factorize its pairing correlations $P_{yy}(r)$ as $P_{{yy}} \left(r\right)=r^{-K_{\mathrm{sc}} } *\Phi_{{yy}} \left(r\right)$, where $\Phi_{{yy}}\left(r\right) \sim\mathrm{cos}\left(\mathbf{Q}_{p}\cdot \mathbf{r}+\varphi \right)$ is the spatial oscillation term that captures the positive and negative sign variations of the $P_{yy}(r)$. In Fig. 4(a), we have fitted the $K_{\mathrm{sc}}\simeq1.08$, then we can extract the pairing oscillations $\Phi_{yy}(r)$ in Fig.~\ref{SM_PDW}(c), where one can find that $\Phi_{yy}(r)$ exhibits periodic sign reversals with a characteristic wave vector $\mathbf{Q}_{p}=\pi$, confirming the oscillatory behavior of the PDW and the fact that its spatial average vanishes. Besides, we also calculate other correlation functions as shown in Figs.~\ref{SM_PDW}(d-f). For the single-boson correlations $G(r)$ [Fig.~\ref{SM_PDW}(e)], we find that they decay exponentially with a very short correlation length $\xi_{\mathrm{G}}\simeq3.93$, implying gapped single-boson excitations. On the other hand, we also find that the charge density correlations $D(r)$ [Fig.~\ref{SM_PDW}(f)] still decay algebraically with $K_{\mathrm{c}}\simeq1.73>K_{\mathrm{sc}}\simeq1.08$, which is highly similar to the Luther-Emery liquid in Fermi systems~\cite{Lu_PRB_2023,Jiang_PRR_2020}, but here the spin correlations $F(r)$ [Fig.~\ref{SM_PDW}(d)] with $\xi_{\mathrm{s}}\simeq10.15$ (or with $K_{\mathrm{s}}\simeq0.76$ by algebraic fitting) are much stronger than their fermionic counterparts. This behavior is consistent with the pronounced AFM peak observed in Fig. 3(e) and indicates that the spin sector remains gapless. 

\section{\label{SM_dPDW}F. More details in the dPDW+AFM phase}

\begin{figure}
   \includegraphics[width=0.9\textwidth,angle=0]{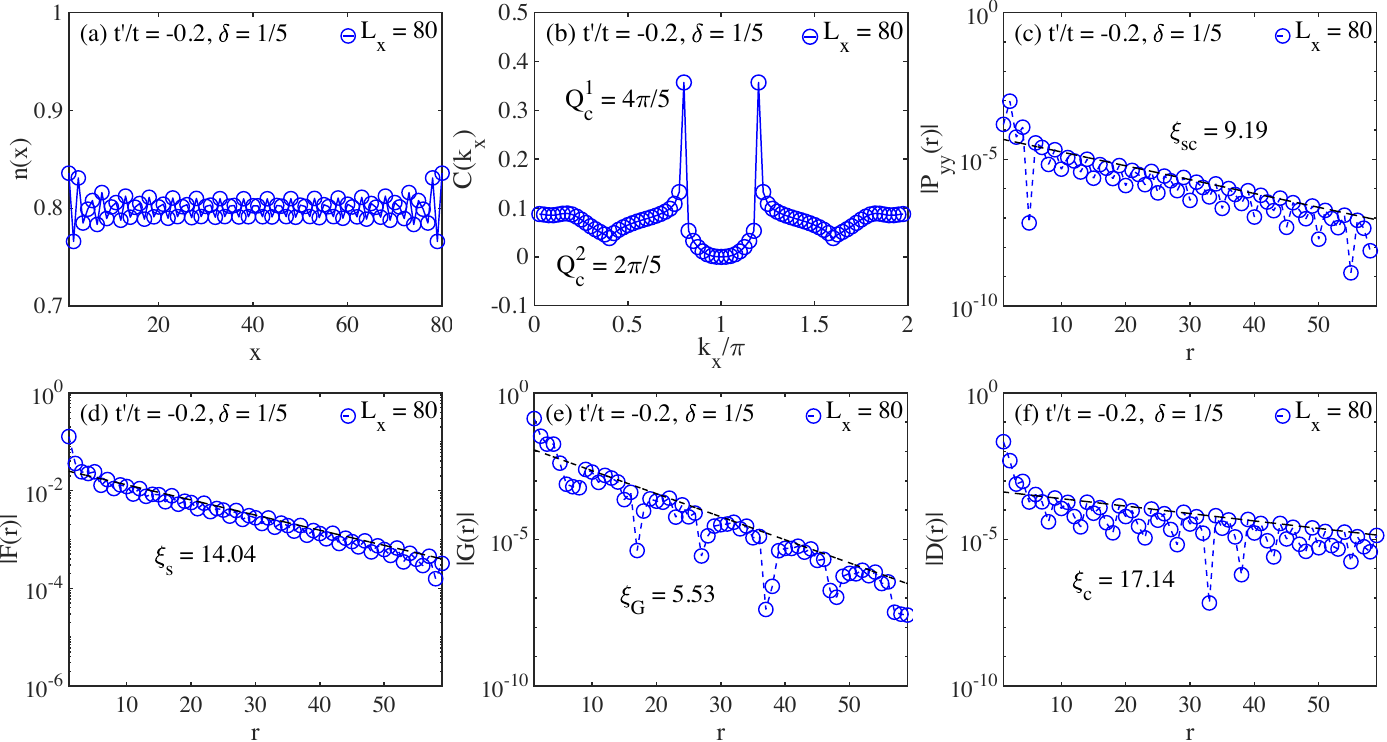}
   \caption{\label{SM_DPDW}
   \textbf{dPDW+AFM state at $t'/t=-0.2$, $\delta=1/5$.}
   (a) Charge density profile $n(x)$. (b) CDW structure factor $C(\bf k)$ of (a). (c-f) are respectively the semi-logarithmic plot of pairing correlation $P_{yy}(r)$, spin correlation $F(r)$, single-boson correlation $G(r)$, and charge density correlation $D(r)$. The correlation lengths $\xi_{\mathrm{sc}}$, $\xi_{\mathrm{s}}$, $\xi_{\mathrm{G}}$, and $\xi_{\mathrm{c}}$ are obtained by exponential fitting with dash line.
   }
\end{figure}

For the dPDW+AFM phase, it shares similar properties with the PDW+AFM phase, but also has a lot of differences. We have already shown in the main text that the SC in the dPDW+AFM phase is very weak, but hole pairing still exists. In this section, we further show more details of this phase. In Fig.~\ref{SM_DPDW}(a), we provide a representative charge density profile $n(x)$ with longer $L_x=80$ in the dPDW+AFM phase, where the periodicity is not as clear as that of Fig.~\ref{SM_PDW}(a), which is due to the joint modulation of the two charge density periods. In Fig.~\ref{SM_DPDW}(b), we show the corresponding CDW structure factor $C(\bf k)$ of Fig.~\ref{SM_DPDW}(a). In contrast with the PDW+AFM phase, we find that there are two characteristic momenta in the dPDW+AFM phase, i.e. $\mathbf{Q}_{c}^1=4\pi\delta$ and $\mathbf{Q}_{c}^2=2\pi\delta$, referring to charge density modulation with $\lambda_{c}^1 =1/2\delta$ and $\lambda_{c}^2 =1/\delta$. 
Besides, the correlation functions also show some differences. In Figs.~\ref{SM_DPDW}(c-f), we respectively show the pairing correlation $P_{yy}(r)$ [Fig.~\ref{SM_DPDW}(c)], spin correlation $F(r)$ [Fig.~\ref{SM_DPDW}(d)], single-boson correlation $G(r)$ [Fig.~\ref{SM_DPDW}(e)], and charge density correlation $D(r)$ [Fig.~\ref{SM_DPDW}(f)] in the dPDW+AFM phase. Here, we present a unified semi-logarithmic plot of these four correlation functions to facilitate a direct comparison of their correlation lengths. Notably, the $G(r)$ exhibits the shortest correlation length ($\xi_{\mathrm{G}}\simeq5.53$) among the four, maintaining its characteristic exponential decay. Meanwhile, the $F(r)$ continues to demonstrate robust AFM correlations with a long correlation length ($\xi_{\mathrm{s}}\simeq14.04$). Interestingly, different from the PDW+AFM phase as discussed in the previous section, here the $D(r)$ (with $\xi_{\mathrm{c}}\simeq17.14$) is much stronger than the $P_{yy}(r)$ (with $\xi_{\mathrm{sc}}\simeq9.19$). This charge-density-dominated non-SC phase with hole pairing naturally evokes comparisons to the stripe phase observed in Fermi systems~\cite{Gong_PRL_2021,lu2024emergent,lu_epc_2024}.

\section{\label{SM_BOW}G. Bond order wave at $\delta =1/4$ doping}

\begin{figure}
   \includegraphics[width=0.9\textwidth,angle=0]{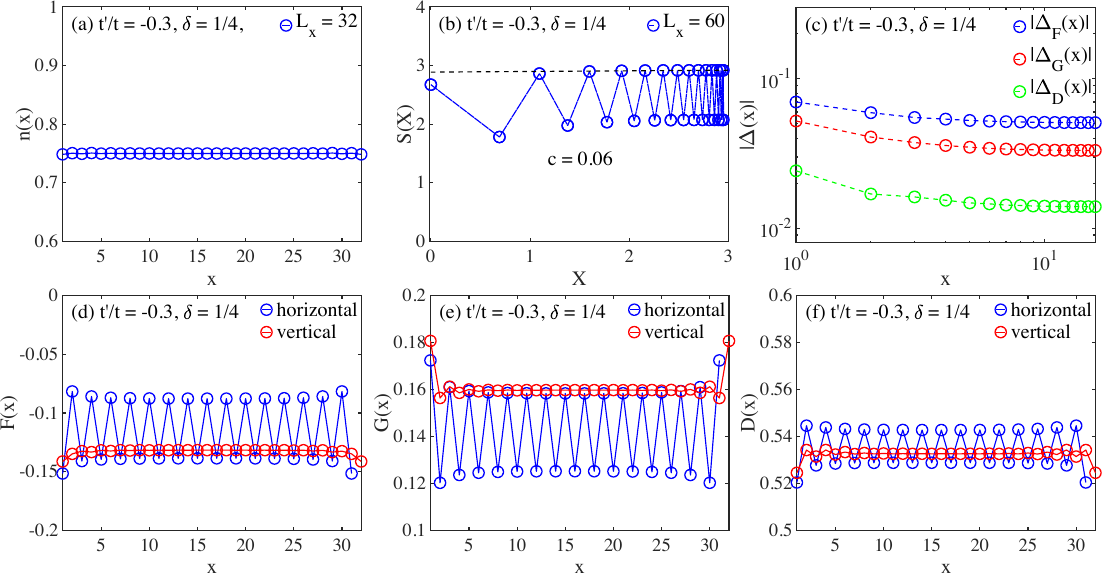}
   \caption{\label{SM_BOW}
   \textbf{BOW state at $t'/t=-0.3$, $\delta=1/4$.}
   (a) Charge density profile $n(x)$, (b) fitting of central charge $c$, (c) double-logarithmic plot of bond order parameters $\Delta(r)$, and (d-f) are the bond energy profiles for NN spin correlation $F(x)$, NN single-boson correlation $G(x)$, and NN charge density correlation $D(x)$.
   }
\end{figure}

In Fig. 1(a) at special $\delta=1/4$, we identify a BOW state, which means the charge density distribution is uniform within the system, while the NN bond energy exhibits a long-range periodic oscillation~\cite{Lu_PRB_2023,Roux_PRB_2007}. In Fig.~\ref{SM_BOW}, we show the BOW characteristics for the representative case of $t'/t=-0.3$, $\delta=1/4$. We first calculate the charge density profile $n(x)$, as shown in Fig.~\ref{SM_BOW}(a), which exhibits no spatial modulation and maintains an exact value of $n(x)=1-\delta=3/4$ throughout the system. In Figs.~\ref{SM_BOW}(d-f), we calculate the bond energies $F(x)=\langle {\hat{\bf S}}_i \cdot {\hat{\bf S}}_j \rangle$, $G(x)=\sum_{\sigma} \langle \hat{\mathcal{B}}^{\dagger}_{i,\sigma} \hat{\mathcal{B}}_{j,\sigma} \rangle$, and $D(x)=\langle \hat{n}_i \hat{n}_j\rangle$ for all the NN horizontal and vertical bonds, where $x$ denotes the site number along $x$ direction in one chain. One can find that all vertical bond energies are uniform, while the horizontal bond energies exhibit strong oscillations with a period of $\lambda=2$. To characterize the fact that translational symmetry is broken, we also calculate the dimer order parameters $\Delta_F(x) = \langle {\hat{\bf S}}_x \cdot {\hat{\bf S}}_{x+1} \rangle - \langle {\hat{\bf S}}_{x+1} \cdot {\hat{\bf S}}_{x+2} \rangle$, $\Delta_G(x) = \sum_{\sigma} \langle \hat{\mathcal{B}}^{\dagger}_{x,\sigma} \hat{\mathcal{B}}_{x+1,\sigma} \rangle - \sum_{\sigma} \langle \hat{\mathcal{B}}^{\dagger}_{x+1, \sigma} \hat{\mathcal{B}}_{x+2, \sigma} \rangle$, and $\Delta_D(x) = \langle \hat{n}_x \hat{n}_{x+1} \rangle - \langle \hat{n}_{x+1} \hat{n}_{x+2} \rangle$. As shown in Fig.~\ref{SM_BOW}(c), it is evident that all the dimer order parameters clearly do not exhibit algebraic decay but instead converge to constant values at long distances. Besides, we also try to fit the central charge $c$ of the system from the entanglement entropy $S\left(x\right)=-\mathrm{Tr}\left\lbrack \hat{\rho}_x \mathrm{ln}\hat{\rho}_x \right\rbrack$, where $\hat{\rho}_x$ is the reduced density matrix of the subsystem with rung number $x$. According to the formula~\cite{Calabrese2004,Calabrese2011}
\begin{equation}\label{entropy2}
S(x)=\frac{c}{6}\mathrm{log}[\frac{L_x }{\pi }\mathrm{sin}(\frac{\pi x}{L_x })] +\tilde{S},
\end{equation}
where we define $X=\mathrm{log}\left\lbrack \left(L_{x}/\pi \right)\mathrm{sin}\left(\pi x/L_{x}\right)\right\rbrack$ as the conformal distance, $\tilde{S}$ is a model-dependent parameter. In Fig.~\ref{SM_BOW}(b), we fit the central charge with $c\sim0$ on a longer cylinder with $L_x=60$, indicating a fully gapped BOW state.

\section{H. Derivation of the intrinsic $\mathbb{Z}_{2}$ Berry phase}

\begin{figure}[h]
   \includegraphics[width=0.85\textwidth,angle=0]{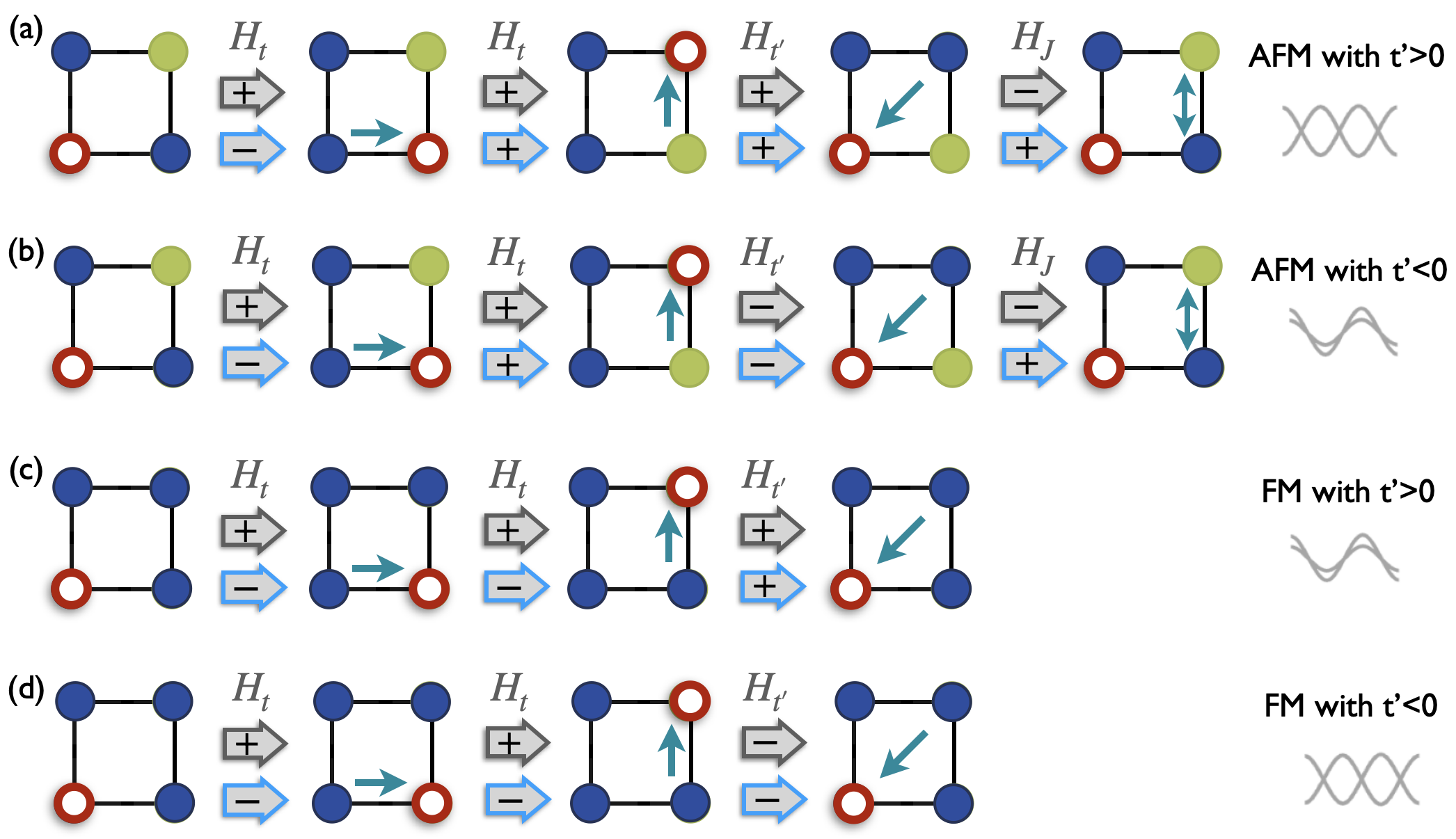}
   \caption{\textbf{Schematic Illustration of Interference Frustration in the Presence of NNN Hopping} Red and blue arrows indicate the frustrated processes before and after the basis transformation, respectively. The $\pm$ sign on the arrows denotes the $\mathbb{Z}_2$ sign generated by the corresponding evolution step. Green and blue circles denote spins in states $\left| \uparrow \right\rangle$ and $\left| \downarrow \right\rangle$; red hollow circles represent doped holes $\left|0\right\rangle$. Destructive (constructive) interference around triangular loops with local AFM correlations for $t'>0$ ($t'<0$); reversed interference pattern for FM correlations.
}
   \label{squ}

\end{figure}

In this section, we analyze the quantum frustration --- i.e., the sign structure --- of the models discussed in the main text by performing explicit power-series expansions of the partition function. Here, ``quantum frustration'' (or equivalently, the ``sign structure'') refers to the arrangement of sign or phase factors that appear in the path-integral–like expansion of the partition function. Our goal is to identify two distinct origins of frustration: a $\mathbb{Z}_2$-type phase frustration induced by NN hopping in the presence of doped holes, and a geometric frustration associated with NNN hopping. We further examine how the interplay between these two frustrations affects magnetic behaviors.

The partition function of a generic quantum system, $Z = \operatorname{Tr} e^{-\beta H}$, can be expanded as a sum over closed paths in imaginary-time evolution~\cite{Sandvik2010}:
\begin{equation}\label{eq:Z_expansion}
    Z=\sum_{n=0}^{\infty} \sum_{\left\{\alpha\right\}_n} \frac{\beta^{n}}{n !} \prod_{k=0}^{n-1} \left\langle\alpha_{k+1}\right| (-H) \left| \alpha_{k}\right\rangle,
\end{equation} 
where ${ \ket{\alpha_k} }$ is a sequence of basis states (e.g., the real-space Fock basis) satisfying the periodic boundary condition $\ket{\alpha_n} = \ket{\alpha_0}$. Each non-zero matrix element $\left\langle \alpha_{k+1} | (-H) | \alpha_k \right\rangle$ corresponds to a single evolution step. Decomposing each element into its magnitude and phase, the partition function can be written in a path-integral form as
\begin{equation}\label{eq:Z_tau_W}
    Z=\sum_C \tau_C W[C],
\end{equation}
where $W[C]$ is a non-negative weight and $\tau_C$ is the accumulated phase factor (or sign) associated with a path $C$.

This amplitude–phase structure also applies to other expansion schemes, such as the auxiliary-field quantum Monte Carlo method. The phase factor $\tau_C$ captures the quantum interference among different evolution paths and reflects the “quantumness” of the system. When $\tau_C$ is strictly positive (i.e., the sign problem is absent), the system effectively reduces to a classical one and becomes amenable to Monte Carlo simulations. However, when $\tau_C$ fluctuates strongly in sign or phase, the system exhibits quantum frustration, commonly referred to as the “sign problem.” Beyond its computational significance, this frustration phase has profound physical implications and can even be interpreted as a discrete version of the Berry phase action in continuous path-integral formulations~\cite{Altland2010,Auerbach2012}. It represents an adiabatic phase accumulated along a closed evolution loop, independent of the evolution speed, in contrast to the dynamical phase arising from real-time evolution.

It is important to clarify the basis dependence of the sign problem. Under purely phase-type basis transformations, the sign structure remains invariant. This is because the partition function involves matched conjugate pairs of basis vectors along each path. Specifically, for an $n$-step evolution path $C$, the accumulated phase $\tau_C = e^{i\Theta_C}$ can be expressed as
\begin{equation}
  \Theta_C = \sum_{k=0}^{n-1} \opr{Im} \ln \left\langle \alpha_{k+1} \right| (-H) \left| \alpha_{k} \right\rangle,
\end{equation}
with the periodic condition $\ket{\alpha_0} = \ket{\alpha_n}$ imposed. Since one can always shift the Hamiltonian to make diagonal elements real and positive, they contribute no net phase. Moreover, a phase rotation of the basis state $\left| \alpha_k \right\rangle \rightarrow e^{i\theta} \left| \alpha_k \right\rangle$ induces a transformation that cancels out in the conjugate pair:
\begin{equation}
    e^{i\theta}\left| \alpha_{k} \right\rangle \left\langle \alpha_{k} \right| e^{-i\theta} = \left| \alpha_{k} \right\rangle \left\langle \alpha_{k}\right|.
\end{equation}
Therefore, while such a transformation may affect individual matrix elements, the total phase $\Theta_C$ accumulated over a closed path remains unchanged.

We now analyze the quantum frustration --- i.e., the sign structure --- of the bosonic $t$–$t'$–$J$ model on bipartite lattices. The Hamiltonian given in Eq.~(1) of the main text can be rewritten as
\begin{equation}
    H_{t\text{-}J} \rightarrow -t\left(P_{o \up}+P_{o \dn}\right) -t' T_o +\frac{J}{2} T_{\up \dn} -\frac{J}{2} V_{\up \dn},
\end{equation}
where
\begin{eqnarray}
    P_{o \sigma} &=&\sum_{\avg{ij}} \hat{\mathcal{B}}_{i \sigma}^{\dagger} \hat{\mathcal{B}}_{j \sigma} + \text{H.c.}, \\
    T_{o}&=&\sum_{\langle\langle i j\rangle\rangle\sigma} \hat{\mathcal{B}}_{i \sigma}^{\dagger} \hat{\mathcal{B}}_{j \sigma}+\text { H.c. }, \\
    T_{\up \dn} &=&\sum_{\langle i j\rangle} \hat{\mathcal{B}}_{i \up}^{\dagger} \hat{\mathcal{B}}_{i \dn} \hat{\mathcal{B}}_{j \dn}^{\dagger} \hat{\mathcal{B}}_{j \up} + \text{H.c.}, \\
    V_{\up \dn} &=&\sum_{\avg{ij}} \left(\hat{n}_{i \up} \hat{n}_{j \dn}+\hat{n}_{i \dn} \hat{n}_{j \up}\right).
\end{eqnarray}
Here we omit to write down the no-double-occupancy projector $\mathcal{P}_s$ explicitly and absorb it into each single-particle operator for simplicity (the same below). The terms $P_{o \sigma}$ and $T_o$ represent NN and NNN hopping (exchange) between holes and spins, respectively; $T_{\uparrow \downarrow}$ corresponds to spin-flip processes (or equivalently, NN exchange between $\uparrow$ and $\downarrow$ spins), and $V_{\uparrow \downarrow}$ denotes the NN interaction energy between anti-parallel spins. The corresponding partition function can be expanded into a power series in $\beta$ as~\cite{Wu2008}:
\begin{equation}\label{eq:expand_tj}
\begin{aligned}
    Z_{t\text{-} t'\text{-}J}&= \opr{Tr} e^{-\beta  H_{t\text{-} t'\text{-}J}} \\
    &= \sum_{n=0}^{\infty}  \frac{\beta^{n}}{n !} \opr{Tr}\left[\sum \cdots \left(t P_{o \up}\right) \cdots \left(-\frac{J}{2} T_{\up \dn}\right) \cdots\left(t P_{o \dn}\right) \cdots  \left(t' T_{o }\right) \cdots \left(\frac{J}{2} V_{\up \dn} \right) \cdots\right]_{n} \\
    &= \sum_{n=0}^{\infty} \frac{\beta^{n}}{n !} \opr{Tr}\left[\sum (-1)^{N_{\uparrow \downarrow}} \left[\operatorname{sgn}\left(t^{\prime}\right)\right]^{N_{t^{\prime}}^{h}} \cdots\left(t P_{o \up}\right) \cdots \left(\frac{J}{2} T_{\up \dn}\right) \cdots\left(t P_{o \dn}\right) \cdots  \left(|t'| T_{o }\right) \cdots \left(\frac{J}{2} V_{\up \dn}\right) \cdots\right]_{n}.
\end{aligned}
\end{equation}
The formal notation $[\sum \cdots]_n$ indicates the summation over all length-$n$ process combination of $P_{o\up}$, $P_{o\dn}$, $T_{o}$, $T_{\up \dn}$, and $V_{\up \dn}$. Here $N_{\up\dn}$ denotes the total number of NN spin flip, and $N_{t^{\prime}}^{h}$ represents total number of NNN exchanges between holes and spins. We remark that the remaining part in Eq.\,\eqref{eq:expand_tj} except $(-1)^{\uparrow \downarrow}\left[\operatorname{sgn}\left(t^{\prime}\right)\right]^{N_{t^{\prime}}^{h}}$ is always non-negative because each matrix element in $P_{o\up}$, $P_{o\dn}$, $T_o$, $T_{\up \dn}$, and $V_{\up \dn}$ is non-negative.

By further expanding $P_{o\up}$, $P_{o\dn}$, $T_o$, $T_{\up \dn}$, and $V_{\up \dn}$ into elementary local terms and writing the trace as the sum of expectations over the complete Fock basis (or equivalently by inserting complete bases between each ``time slice''), the partition function can be expressed by a huge summation of real numbers with each number indexed by a discrete evolution of hole-spin configurations, where in each step, one of the five events $T_{o\up}$, $T_{o\dn}$, $T_{\up \dn}$, and $V_{\up \dn}$ occurs. Note that due to the trace operation, the initial and final hole-spin configurations in the evolution should be the same, so in each possible evolution path $C$, the motion of holes and spins must form closed loops. Namely, the partition function can be expressed by a summation that runs over all possible closed evolution paths $C$, i.e.,
\begin{equation}\label{eq:partition_sign_tj}
    Z_{t\text{-}t'\text{-}J} = \sum_{C} \tau_{C} W_{t\text{-}t'\text{-}J} [C],
\end{equation}
where
\begin{equation}\label{eq:sign_tj}
    \tau_{C} = (-1)^{N_{\up\dn}}\left[\operatorname{sgn}\left(t^{\prime}\right)\right]^{N_{t^{\prime}}^{h}},
\end{equation}
is the sign structure of the bosonic $t$-$t'$-$J$ model and $W_{t\text{-}t'\text{-}J}[C]\geq 0$ denotes a non-negative weight corresponding to the evolution path $C$ with its explicit form omitted. Note that the summation over evolution length $n$ in Eq.\,\eqref{eq:expand_tj} has been included in the summation over evolution path $C$ in Eq.\,\eqref{eq:partition_sign_tj}. the two components in the sign factor $\tau_C$ defined in Eq.\eqref{eq:sign_tj} correspond to distinct origin of quantum frustration: the factor $(-1)^{N_{\uparrow \downarrow}}$ encodes frustration induced by spin-flip processes, while $\left[\operatorname{sgn}(t')\right]^{N_{t'}^h}$ captures the geometric frustration associated with NNN hopping.

Furthermore, as discussed above, phase-only transformations of the basis do not alter the quantum frustration (i.e., the sign problem), since they leave the accumulated phase $\tau_C$ invariant for all closed evolution paths. To better illustrate the physical origin of doped-induced frustration in the bosonic $t$–$t'$-$J$ model, we therefore apply a Marshall transformation:
\begin{equation}\label{MS}
    \hat{\mathcal{B}}_{i \sigma} \rightarrow \sigma \hat{\mathcal{B}}_{i \sigma}, \quad i\in A,
\end{equation}
under which the spin operators transform as
\begin{equation}
    \hat{S}_i^{ \pm} \rightarrow-\hat{S}_i^{ \pm}, \quad \hat{S}_i^z \rightarrow \hat{S}_i^z, \quad i\in A.
\end{equation}
As a result, the bosonic $t$–$t'$–$J$ Hamiltonian transforms into the following form:
\begin{equation}
    H_{t\text{-}J} \rightarrow -t\left(P_{o \up}-P_{o \dn}\right) -t' T_o -\frac{J}{2} \left(T_{\up \dn} + V_{\up \dn}\right).
\end{equation}
Then, similar to Eq.~\eqref{eq:expand_tj}, the series expansion of the transformed Hamiltonian reads:
\begin{equation}\label{eq:expand_tj2}
    Z_{t\text{-} t'\text{-}J}= \sum_{n=0}^{\infty} \frac{\beta^{n}}{n !} \opr{Tr}\left[\sum (-1)^{N_{ \dn}^h} \left[\operatorname{sgn}\left(t^{\prime}\right)\right]^{N_{t^{\prime}}^{h}} \cdots\left(t P_{o \up}\right) \cdots \left(\frac{J}{2} T_{\up \dn}\right) \cdots\left(t P_{o \dn}\right) \cdots  \left(|t'| T_{o }\right) \cdots \left(\frac{J}{2} V_{\up \dn}\right) \cdots\right]_{n},
\end{equation}
the sign structure (or frustration structure) given in Eq.~\eqref{eq:sign_tj} is modified as follows:

\begin{equation}\label{eq:sign_tj}
    \tau_{C} = (-1)^{N_{\dn}^h}\left[\operatorname{sgn}\left(t^{\prime}\right)\right]^{N_{t^{\prime}}^{h}}.
\end{equation}
Importantly, the two terms in Eq.~\eqref{eq:sign_tj} represent two distinct sources of frustration. The factor $(-1)^{N_{\downarrow}^h}$ can be interpreted as a dopant-induced frustration, originating from the Hilbert space constraint $n_i \neq 1$. This reflects the intrinsic interplay between charge and spin degrees of freedom in a doped Mott insulator. In contrast, the factor $\left[\operatorname{sgn}(t')\right]^{N_{t'}^h}$ corresponds to geometric frustration, which arises from the fact that NNN hopping effectively renders the bipartite lattice non-bipartite. The interplay between these two types of frustration can lead to nontrivial consequences, particularly in the system’s magnetic behavior. A representative example is illustrated in Fig.~\ref{squ}, where red and blue arrows indicate the frustrated processes before and after the basis transformation, respectively. Green and blue circles denote spins in states $\left| \uparrow \right\rangle$ and $\left| \downarrow \right\rangle$, while red hollow circles represent doped holes $\left|0\right\rangle$. It is evident that when holes move around a small triangular loop with local AFM correlations, the interference is destructive (constructive) for $t'>0$ ($t'<0$). Conversely, with local FM correlations, the interference pattern is reversed. As a result, $t'<0$ ($t'>0$) favors AFM (FM) order on the square lattice due to the corresponding frustration.

It is important to note that, although the specific steps responsible for frustration differ before and after the Marshall basis transformation, the accumulated frustration --- i.e., the many-body Berry phase acquired after completing a closed world-line loop --- remains invariant, as exemplified in Fig.~\ref{squ}. In the main text, we focus on the sign structure after applying the Marshall transformation. This choice facilitates a smoother connection to the undoped limit and highlights the role of doped holes. At half-filling, where the model reduces to the Heisenberg model (for which the Marshall transformation was originally introduced), the sign structure $(-1)^{N_{\downarrow}^h}$ naturally vanishes due to the absence of holes. Meanwhile, the spin-flip–induced factor $(-1)^{N_{\uparrow\downarrow}}$ remains formally present, though it ultimately cancels due to the temporal periodic boundary condition. This observation implies that, at low doping, it is both natural and advantageous to focus on the sign structure $(-1)^{N_{\downarrow}^h}$, which treats doped holes as essential degrees of freedom and captures the core physics of dopant-induced quantum frustration.

\section{\label{SM_sigmatJ}I. Spinful Hard-Core Bose-Hubbard model at large-$U$ limit}
In this section, we derive the effective low-energy model of the spinful hard-core Bose-Hubbard model in the strongly interacting limit $U \gg t$, and show that it is equivalent—up to an on-site unitary (phase) transformation on bipartite lattices—to the bosonic $\sigma t$–$J$ model studied in this work. This derivation parallels the well-known mapping from the fermionic Hubbard model to the standard fermionic $t$–$J$ model via second-order perturbation theory. For comparison, we also present the corresponding derivation for the conventional (soft-core) spinful Bose-Hubbard model and highlight its qualitative differences from the bosonic $\sigma t$–$J$ model.

We begin with the one-band spinful hard-core Bose-Hubbard Hamiltonian:
\begin{equation}\label{eq:spinful_hard_core_bose_hubbard}
    H_{t\text{-}U} = H_t + H_U = - t \sum_{\avg{ij} \sigma} \left( \hat{\mathcal{B}}^{\dagger}_{i\sigma} \hat{\mathcal{B}}_{j\sigma} + \text{H.c.}\right) + \sum_{i,\sigma} \frac{1}{2}U \hat{n}_{i\sigma}(\hat{n}_{i\sigma}-1) 
\end{equation}
where $\hat{\mathcal{B}}_{i\sigma}^\dagger$ ($\hat{\mathcal{B}}_{i\sigma}$) creates (annihilates) a spin-$\sigma$ hard-core boson at site $i$, with $\sigma \in { \uparrow, \downarrow }$. These operators obey the commutation relations:
\begin{equation}
\begin{aligned}
    &[\hat{\mathcal{B}}_{i\sigma}, \hat{\mathcal{B}}_{j\sigma'}^\dagger] = 0, ~~[\hat{\mathcal{B}}_{i\sigma}, \hat{\mathcal{B}}_{j\sigma'}] = 0, \quad \text{if } i\neq j \text{ or } \sigma\neq\sigma',\\
    &\{\hat{\mathcal{B}}_{i\sigma}, \hat{\mathcal{B}}_{j\sigma'}^\dagger \} = 1, \quad \text{if } i = j \text{ and }  \sigma=\sigma',
\end{aligned}
\end{equation}
or equivalently,
\begin{equation}
    [\hat{\mathcal{B}}_{i\sigma}, \hat{\mathcal{B}}^\dagger_{j\sigma'}] = \delta_{ij}\delta_{\sigma\sigma'} \left(1 - 2 \hat{\mathcal{B}}^\dagger_{j\sigma'} \hat{\mathcal{B}}_{i\sigma} \right)  ,~~[\hat{\mathcal{B}}_{i\sigma}, \hat{\mathcal{B}}                            _{j\sigma'}] = 0,
\end{equation}
where $\delta_{ij}$ denotes the Kronecker delta function and the identity operator is omitted for simplicity and denoted as $1$. $\hat{n}_{i\sigma}=\hat{\mathcal{B}}_{i\sigma}^\dagger \hat{\mathcal{B}}_{i\sigma}$ is the particle density operator of spin $\sigma$ at site $i$. We use $\hat{n}_{i}=\sum_\sigma \hat{n}_{i\sigma}$ to denote the total particle density at site $i$. $t\geq 0$ is the hopping integral between the nearest-neighbor (NN) sites $\avg{ij}$ and $U\geq 0$ is the on-site Hubbard repulsion.

Assuming the system is below half-filling, in the large-$U$ limit $U/t \rightarrow \infty$, the ground-state manifold becomes highly degenerate, consisting of all no-double-occupancy configurations. Denote the corresponding projection operator as $\mathcal{P}_s$. According to Brillouin-Wigner perturbation theory, the effective Hamiltonian projected to this subspace can be expanded up to second order as:
\begin{equation}
\begin{aligned}
    & H_{\text{eff}}^{(0)} = \mathcal{P}_s H_U \mathcal{P}_s = 0, \\
    & H_{\text{eff}}^{(1)} = \mathcal{P}_s H_t \mathcal{P}_s, \\
    & H_{\text{eff}}^{(2)} = \mathcal{P}_s H_t \frac{1-\mathcal{P}_s}{0-H_U} H_t \mathcal{P}_s = -\frac{1}{U} \mathcal{P}_s H_t (1-\mathcal{P}_s) H_t \mathcal{P}_s \\
    &\quad\quad= -\frac{t^2}{U}\sum_{\avg{ij}\avg{kl}\sigma\sigma'} \mathcal{P}_s \left(  \hat{\mathcal{B}}_{i\sigma}^\dagger \hat{\mathcal{B}}_{j\sigma} + \text{H.c.} \right) (1-\mathcal{P}_s) \left(  \hat{\mathcal{B}}_{k\sigma'}^\dagger \hat{\mathcal{B}}_{l\sigma'} + \text{H.c.} \right) \mathcal{P}_s.
\end{aligned}
\end{equation}
In the second-order contribution, the successive action of the projectors $\mathcal{P}_s$, $(1-\mathcal{P}_s)$, and $\mathcal{P}_s$ results in the surviving terms in $H_t$ must map a no-double-occupied configuration to a double-occupied configuration and then back to a no-double-occupied configuration. As such, $H_U^{-1}$ simplifies to $1/U$, and the two links $\langle ij \rangle$ and $\langle kl \rangle$ must share sites. In analogy with the standard derivation of the fermionic $t$–$J$ model, we retain only the leading contributions from pairs of coinciding links $\langle ij \rangle = \langle kl \rangle$, and neglect terms where the links merely overlap at a single site. In this case, the action of $(1 - \mathcal{P}_s)$ can be replaced by an operator $n_i n_j$ acting immediately after the first hopping term, indicating that only configurations with singly-occupied sites $i$ and $j$ can participate in virtual doublon processes. Consequently, the second-order effective Hamiltonian becomes
\begin{equation}
\begin{aligned}
    H_{\text{eff}}^{(2)\prime} = -\frac{t^2}{U}\sum_{\avg{ij}\sigma\sigma'} \mathcal{P}_s \left(  \hat{\mathcal{B}}_{i\sigma}^\dagger \hat{\mathcal{B}}_{j\sigma} + \text{H.c.} \right) \left(  \hat{\mathcal{B}}_{i\sigma'}^\dagger \hat{\mathcal{B}}_{j\sigma'} + \text{H.c.} \right) \hat{n}_i \hat{n}_j \mathcal{P}_s.
\end{aligned}
\end{equation}
The terms in the summation on a single link $\langle ij \rangle$ can be evaluated as
\begin{equation}\label{hardcoreB}
    \mathcal{P}_s\left( \sum_{\sigma} \hat{\mathcal{B}}_{i\sigma}^\dagger \hat{\mathcal{B}}_{j\sigma} + \text{H.c.} \right)^2 \hat{n}_i \hat{n}_j \mathcal{P}_s 
    = \mathcal{P}_s \left(\hat{n}_i + \hat{n}_j + 2 \sum_{\sigma\sigma'} \hat{\mathcal{B}}_{i\sigma}^\dagger \hat{\mathcal{B}}_{i\sigma'} \hat{\mathcal{B}}_{j\sigma'}^\dagger \hat{\mathcal{B}}_{j\sigma} - 4\sum_{\sigma} \hat{n}_{i\sigma}\hat{n}_{j\sigma}\right) \hat{n}_i \hat{n}_j \mathcal{P}_s.
\end{equation}
Compared to the fermionic case, the exchange term $2 \sum_{\sigma\sigma'} \hat{\mathcal{B}}_{i\sigma}^\dagger \hat{\mathcal{B}}_{i\sigma'} \hat{\mathcal{B}}_{j\sigma'}^\dagger \hat{\mathcal{B}}_{j\sigma}$ has a reversed sign, and an additional term $-4\sum_{\sigma} \hat{n}_{i\sigma}\hat{n}_{j\sigma}$ appears. The exchange term can be rewritten using the Schwinger boson representation of spin operators:
\begin{equation}
    \hat{\mathbf{S}}_i=(\hat{S}^x_i, \hat{S}^y_i, \hat{S}^z_i)=\frac{1}{2}\sum_{\sigma\sigma'} \hat{\mathcal{B}}_{i\sigma}^\dagger \boldsymbol{\tau}_{\sigma\sigma'} \hat{\mathcal{B}}_{i\sigma'},
\end{equation}
so that:
\begin{equation}
\begin{aligned}
    2\sum_{\sigma\sigma'} \hat{\mathcal{B}}_{i\sigma}^\dagger \hat{\mathcal{B}}_{i\sigma'} \hat{\mathcal{B}}_{j\sigma'}^\dagger \hat{\mathcal{B}}_{j\sigma} = 4 \hat{\mathbf{S}}_i\cdot \hat{\mathbf{S}}_j + \hat{n}_{i} \hat{n}_{j}.
\end{aligned}
\end{equation}
By using the identity:
\begin{equation}
    2 \delta_{\sigma\rho'} \delta_{\sigma'\rho} = \boldsymbol{\tau}_{\sigma\sigma'} \cdot \boldsymbol{\tau}_{\rho\rho'} + \delta_{\sigma\sigma'}\delta_{\rho\rho'} ,
\end{equation}
where $\boldsymbol{\tau}_{\sigma\sigma'}$ denotes the vector composed of the three Pauli matrices. Hence, the second-order contribution to the effective Hamiltonian is
\begin{equation}\label{eq:Heff_2_spinful}
    H_{\text{eff}}^{(2)\prime} = - J \mathcal{P}_s \left( \hat{\mathbf{S}}_i\cdot \hat{\mathbf{S}}_j + \frac{3}{4} \hat{n}_i \hat{n}_j - \sum_{\sigma} \hat{n}_{i\sigma}\hat{n}_{j\sigma} \right) \mathcal{P}_s,
\end{equation}
where the spin-exchange coupling constant is still $J=4t^2/U$. However, the coupling term $\hat{\mathbf{S}}_i\cdot \hat{\mathbf{S}}_j$ becomes ferromagnetic, which is completely different from the fermionic case. Recombining the terms in Eq.\,\eqref{eq:Heff_2_spinful} using the identity
\begin{equation}
    4 \hat{S}_i^z \hat{S}_j^z + \hat{n}_i \hat{n}_j = \sum_{\sigma\sigma'} (\sigma\sigma'+1) \hat{n}_{i\sigma} \hat{n}_{j\sigma'} = \sum_{\sigma\sigma'} 2\delta_{\sigma\sigma'} \hat{n}_{i\sigma} \hat{n}_{j\sigma'} = 2\sum_{\sigma} \hat{n}_{i\sigma} \hat{n}_{j\sigma},
\end{equation}
will obtain
\begin{equation}
\begin{aligned}
    \hat{\mathbf{S}}_i\cdot \hat{\mathbf{S}}_j + \frac{3}{4} \hat{n}_i \hat{n}_j - \sum_{\sigma} \hat{n}_{i\sigma}\hat{n}_{j\sigma} &= \hat{\mathbf{S}}_i\cdot \hat{\mathbf{S}}_j + \frac{3}{4} \hat{n}_i \hat{n}_j - \frac{1}{2}\left(4 \hat{S}_i^z \hat{S}_j^z + \hat{n}_i \hat{n}_j\right) \\
    &= \hat{S}_i^x \hat{S}_j^x + \hat{S}_i^y \hat{S}_j^y - \left(\hat{S}_i^z \hat{S}_j^z - \frac{1}{4} \hat{n}_i \hat{n}_j\right),
\end{aligned}
\end{equation}
where $\sigma\in\{+1,-1\}$ for $\{\up,\dn\}$ respectively when serving as a coefficient. Therefore, the effective Hamiltonian of the spinful hard-core Bose-Hubbard model at $U\gg t$ up to the second order is 
\begin{equation}\label{eq:btj_minus}
    H_{\text{eff}} = \mathcal{P}_s \left(H_t + H_J^{-} \right) \mathcal{P}_s,
\end{equation}
where
\begin{equation}\label{eq:eff_tj_ferro_xy}
\begin{aligned}
    H_t &= - t\sum_{\avg{ij} \sigma} \left( \hat{\mathcal{B}}^{\dagger}_{i\sigma} \hat{\mathcal{B}}_{j\sigma} + \text{H.c.}\right), \\
    H_J^{-} &= J \sum_{\avg{ij}} \left( - \hat{S}_i^x \hat{S}_j^x - \hat{S}_i^y \hat{S}_j^y + \hat{S}_i^z \hat{S}_j^z - \frac{1}{4} \hat{n}_i \hat{n}_j \right).
\end{aligned}
\end{equation}
This effective Hamiltonian is quite similar to the bosonic $t$-$J$ model except that the in-plane $(\hat{S}^x, \hat{S}^y)$ spin-exchange is ferromagnetic (FM). The root cause is that exchanging two fermions of different spin states yields a negative sign while that for spinful hard-core bosons does not. This is a significant difference compared to the fermionic case where the effective model of the Fermi-Hubbard model at $U\gg t$ is just the fermionic $t$-$J$ model. Note that the spin-exchange coupling along $\hat{S}^z$ in Eq.\,\eqref{eq:eff_tj_ferro_xy} is still antiferromagnetic, in line with the common argument about the Hubbard model, i.e., the nearest-neighboring (NN) spins tend to be antiparallel because NN parallel spins cannot gain kinetic energy from the second-order virtual hopping process due to the Pauli exclusion principle or hard-core property while NN antiparallel spins can.

From the above analysis, it is evident that the effective model of the spinful hard-core Bose-Hubbard model at half-filling deviates from the conventional Heisenberg model. However, on bipartite lattices such as the square lattice, one can restore the Heisenberg limit by performing an on-site unitary (phase) transformation—namely, the Marshall transformation introduced in Eq.~\eqref{MS}, which flips the sign of the in-plane spin-exchange interactions. Under this transformation, the effective Hamiltonian exactly reduces to the bosonic $\sigma t$–$J$ model:
\begin{equation}\label{eq:bosonic_sigma_tj}
    H_{\text{eff}} \rightarrow H_{\sigma t\text{-}J} = \mathcal{P}_s( H_{\sigma t} + H_{J} ) \mathcal{P}_s,
\end{equation}
where
\begin{equation}
\begin{aligned}
    H_{\sigma t} &= - t\sum_{\avg{ij} \sigma} \sigma \left( \hat{\mathcal{B}}^{\dagger}_{i\sigma} \hat{\mathcal{B}}_{j\sigma} + \text{H.c.}\right), \\
    H_J &= J \sum_{\avg{ij}} \left( \hat{\mathbf{S}}_i\cdot \hat{\mathbf{S}}_j - \frac{1}{4} \hat{n}_i \hat{n}_j \right).
\end{aligned}
\end{equation}
Remember that $\sigma\in\{+1,-1\}$ for $\{\up,\dn\}$ respectively when serving as a coefficient. In summary, in the limit $U \gg t$, the spinful hard-core Bose-Hubbard model becomes effectively equivalent to the bosonic $\sigma t$–$J$ model, up to a Marshall transformation that maps the in-plane ferromagnetic spin exchange to its antiferromagnetic counterpart on bipartite lattices.

Importantly, the above derivation demonstrates that the $\sigma t$–$J$ model emerges as the effective theory of the spinful hard-core Bose-Hubbard model in the large-$U$ limit, rather than from the conventional (soft-core) Bose-Hubbard model. This distinction originates from the fundamentally different on-site commutation relations obeyed by hard-core and regular boson operators.

For the conventional spinful Bose-Hubbard model, governed by the Hamiltonian
\begin{equation}
    H_{\text{Bose-Hubbard}}=H_t+H_U=-t \sum_{\langle i j\rangle \sigma}\left(\hat{B}_{i \sigma}^{\dagger} \hat{B}_{j \sigma}+\text { H.c. }\right)+U \sum_i \hat{n}_{i \uparrow} \hat{n}_{i \downarrow},
\end{equation}
the (soft-core) boson operators $\hat{B}_{i \sigma}$ satisfy the canonical commutation relations:
\begin{equation}
    \left[\hat{B}_{i \sigma}, \hat{B}_{j \sigma'}^{\dagger}\right]=\delta_{ij} \delta_{\sigma \sigma'}, \;\;\;\;\left[\hat{B}_{i \sigma}, \hat{B}_{j \sigma'}\right]=0,\;\;\;\;\left[\hat{B}^\dagger_{i \sigma}, \hat{B}^\dagger_{j \sigma'}\right]=0.
\end{equation}
As a result of this relation, the second-order contribution to the effective Hamiltonian on a given bond $\langle ij \rangle$ takes the form:
\begin{equation}\label{normalB}
    \sum_{\langle i j\rangle} \mathcal{P}_s\left(\sum_\sigma \hat{B}_{i \sigma}^{\dagger} \hat{B}_{j \sigma}+\text{H.c.}\right)^2 \hat{n}_i \hat{n}_j \mathcal{P}_s= \sum_{\langle i j\rangle} \mathcal{P}_s\left(2 \sum_{\sigma \sigma^{\prime}} \hat{B}_{i \sigma}^{\dagger} \hat{B}_{i \sigma^{\prime}} \hat{B}_{j \sigma^{\prime}}^{\dagger} \hat{B}_{j \sigma}+\hat{n}_i+\hat{n}_j\right) \hat{n}_i \hat{n}_j \mathcal{P}_s.
\end{equation}
Compared to the hard-core boson case Eq.~(\ref{hardcoreB}), the key difference is the absence of the term $-4 \sum_\sigma \hat{n}_{i \sigma} \hat{n}_{j \sigma}$, which vanishes in Eq.~(\ref{normalB}). This leads to a different second-order effective term:
\begin{equation}
    H_{\mathrm{eff}}^{(2) \prime}=-J \mathcal{P}_s\left(\hat{\mathbf{S}}_i \cdot \hat{\mathbf{S}}_j+\frac{3}{4} \hat{n}_i \hat{n}_j\right) \mathcal{P}_s.
\end{equation}
Combining this with the zeroth and first order terms, $H^{(0)}_{\text{eff}} = \mathcal{P}_s H_U \mathcal{P}s = 0$ and $H^{(1)}{\text{eff}} = \mathcal{P}_s H_t \mathcal{P}_s$, the resulting effective model of the conventional Bose-Hubbard model in the large-$U$ limit becomes:
\begin{equation}
    H_{\text {eff }}=\mathcal{P}_s \left[-t \sum_{\langle i j\rangle \sigma} \left(\hat{B}_{i \sigma}^{\dagger} \hat{B}_{j \sigma}+\text { H.c. }\right)-J \left(\hat{\mathbf{S}}_i \cdot \hat{\mathbf{S}}_j+\frac{3}{4} \hat{n}_i \hat{n}_j\right) \right]\mathcal{P}_s,    
\end{equation}
which resembles a bosonic $t$–$J$ model but features an isotropic ferromagnetic Heisenberg interaction.

\section{\label{SM_Phase_diagram_sigmatj}J. Phase diagram of the bosonic $\sigma t$-$t'$-$J$ model}
\begin{figure}
   \includegraphics[width=0.5\textwidth,angle=0]{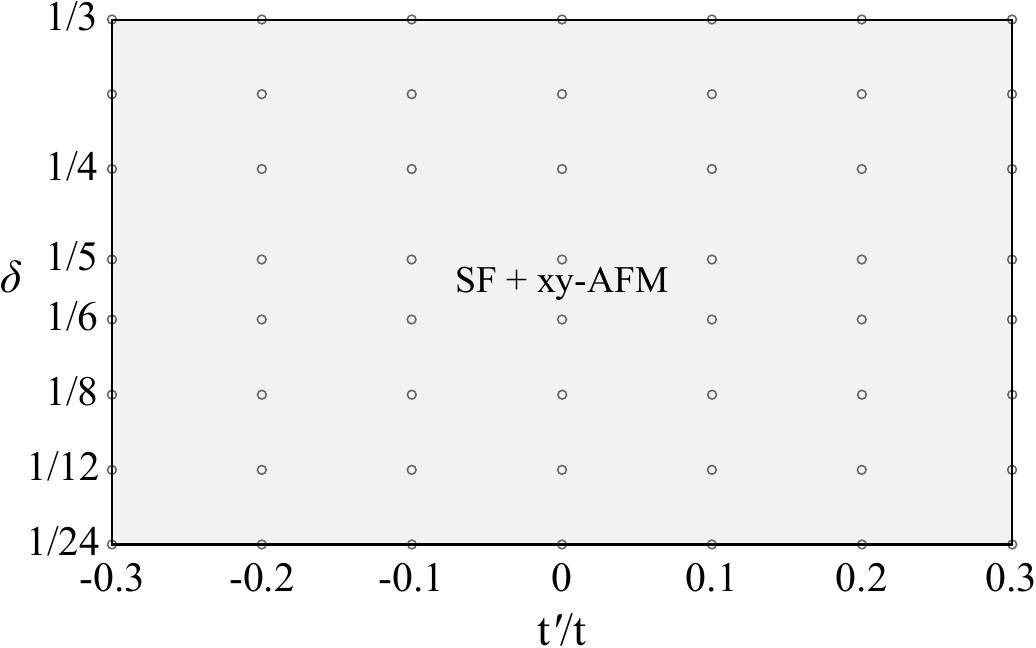}
   \caption{\label{SigmatJ_PD}
   \textbf{Phase diagram of the bosonic $\sigma t$-$t'$-$J$ model on four-leg cylinder.}
   Within $-0.3 \le t' /t \le 0.3$ and $1/24\le \delta \le1/3$, we only identify a SF+$xy$-AFM phase. The symbols denote the calculated parameter points.}
\end{figure}

In the main text, we have highlighted the crucial role of the intrinsic $\mathbb{Z}_{2}$ Berry phase in governing the emergence of all phases in the bosonic $t$-$t'$-$J$ model. To further consolidate this point, we performed the same set of DMRG calculations for the closely related bosonic $\sigma t$-$t'$-$J$ model, in which the $\mathbb{Z}_{2}$ Berry phase is strictly hidden by the opposite hopping amplitudes of up/down spins with holes between NN sites. In the Fig.~\ref{SigmatJ_PD}, we show the phase diagram of the bosonic $\sigma t$-$t'$-$J$ model on four-leg cylinder, in which we find the ground state of the system remains highly stable against doping and hopping, featuring only a single SF phase with in-plane AFM order (SF+$xy$-AFM). This is in sharp contrast to the rich phase diagram of the original bosonic $t$-$t'$-$J$ model in Fig. 1 (a).

\begin{figure}
   \includegraphics[width=0.72\textwidth,angle=0]{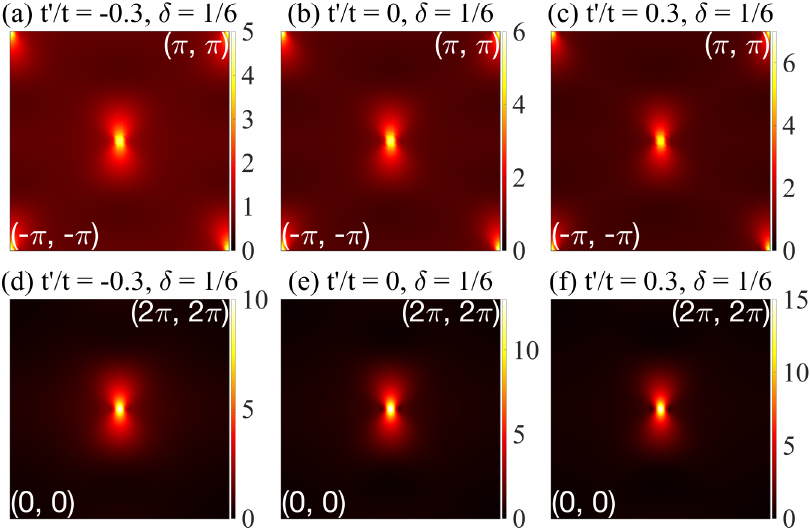}
   \caption{\label{SM_Sigma_Q}
   \textbf{Structure factors in the SF+$xy$-AFM phase.}
   (a-c) are the momentum distributions $n(\bf k)$ at $\delta=1/6$ with varied $t'/t$. (d-f) are the spin structure factors $S(\bf k)$ at $\delta=1/6$ with varied $t'/t$. Here all the structure factors are obtained by taking the Fourier transformation for the all-to-all correlations.
   }
\end{figure}

\begin{figure}
   \includegraphics[width=0.9\textwidth,angle=0]{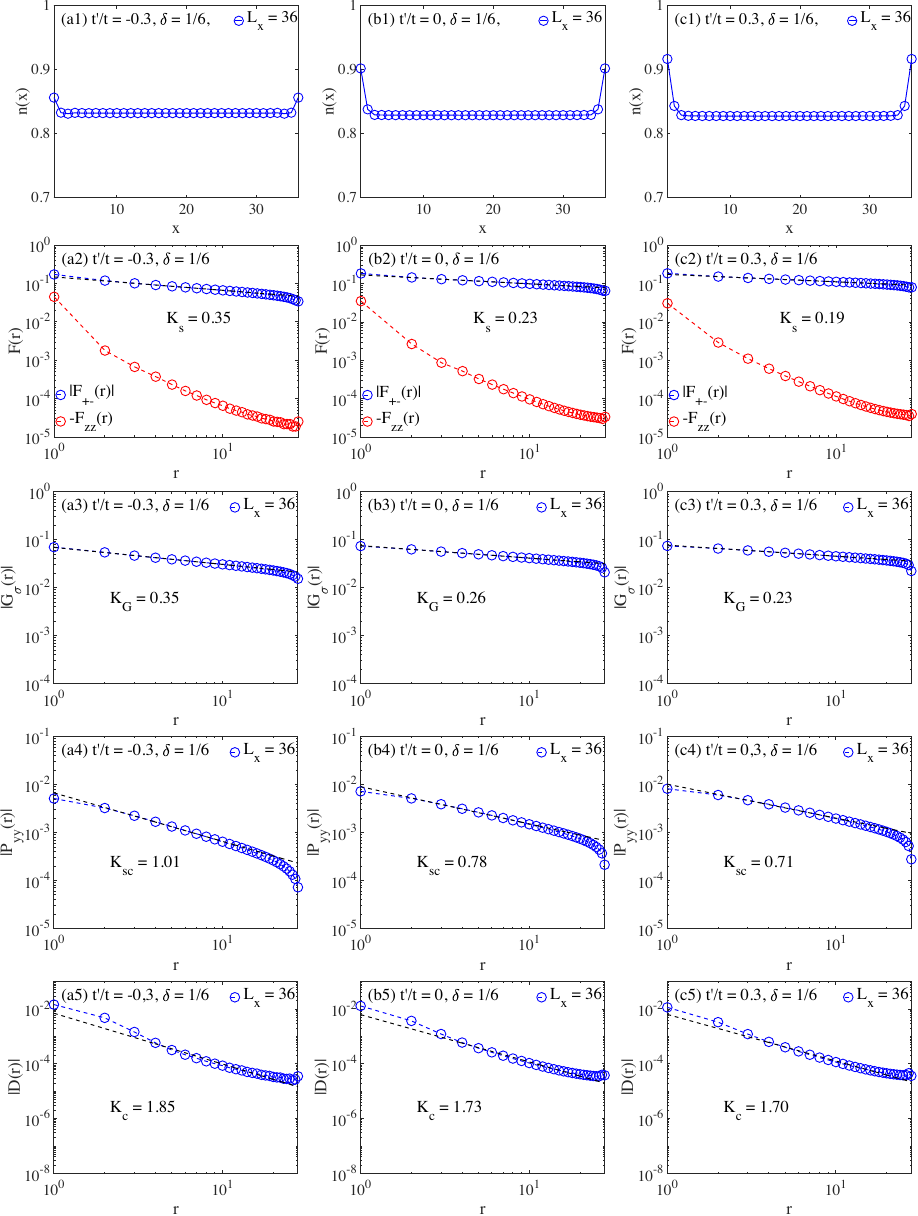}
   \caption{\label{SM_sigma_CDWCorre}
   \textbf{Charge density profiles and correlation functions in the SF+$xy$-AFM phase.} (a1-a5) are respectively the charge density profile $n(x)$, double-logarithmic plot of the spin correlation $F(r)$, double-logarithmic plot of the single-boson correlation $G_{\sigma}(r)$, double-logarithmic plot of the pairing correlation $P_{yy}(r)$, and double-logarithmic plot of the charge density correlation $D(r)$ at $t'/t=-0.3$ with fixed $\delta=1/6$. The power exponents $K_{\mathrm{s}}$, $K_{\mathrm{G}}$, $K_{\mathrm{sc}}$, and $K_{\mathrm{c}}$ are obtained by algebraic fitting with dash line. (b1-b5) and (c1-c5) are the similar plots at $t'/t=0$ and $t'/t=0.3$, respectively.
   }
\end{figure}

In the following, we present additional physical quantities in the SF+$xy$-AFM phase to further support our conclusion. Given the high consistency of the phase diagram across different doping levels, we then only focus on the representative $\delta=1/6$ doping for detailed discussion. In Figs.~\ref{SM_Sigma_Q}(a-c), we show the momentum distribution $n(\bf k)$ for $t'/t$ ranging from $-0.3$ to $0.3$. It is evident that the topology of $n(\bf k)$ remains entirely insensitive to the sign of $t'$. Two divergent peaks are observed at $(0, 0)$ and $(\pi, \pi)$, originating from the condensation of bosons with up and down spins, respectively. In Figs.~\ref{SM_Sigma_Q}(d-f), we also present the corresponding spin structure factor $S(\bf k)$. Similar to the $n(\bf k)$, the behavior of $S(\mathbf{k})$ is also insensitive to the sign of $t'$ and exhibits a sharp peak at $(\pi, \pi)$, which primarily arises from the in-plane AFM spin correlations $F_{+-}(r)$, while the longitudinal spin correlations $F_{zz}(r)$ are very weak [see Figs.~\ref{SM_sigma_CDWCorre}(a2-c2)]. As a further check for other possible competing or intertwined orders in this phase, we also calculate the charge density profiles $n(x)$ and the real-space correlation functions as shown in Figs.~\ref{SM_sigma_CDWCorre}. Particularly, the $n(x)$ [Figs.~\ref{SM_sigma_CDWCorre}(a1-c1)] do not show any charge modulation or nonuniformity, indicating the absence of the CDW. On the other hand, we find the pairing correlations $P_{yy}(r)$ [Figs.~\ref{SM_sigma_CDWCorre}(a4-c4)] and charge density correlations $D(r)$ [Figs.~\ref{SM_sigma_CDWCorre}(a5-c5)] are weaker than the single-boson correlations $G_{\sigma}(r)$ [Figs.~\ref{SM_sigma_CDWCorre}(a3-c3)], which also rule out the other SC and charge-ordered states. To sum up, if we compare Fig.~\ref{SM_sigma_CDWCorre} with Fig.~\ref{SM_FM}, we will naturally find that the SF+$xy$-AFM phase shares many similarities with the SF+$xy$-FM in Fig. 1(a). In particular, both show strong magnetic correlations that are dominated by in-plane spin correlations $F_{+-}(r)$. Although one exhibits $xy$-AFM order and the other $xy$-FM order, their longitudinal spin correlations $F_{zz}(r)$ always show FM correlation in both cases. 

Physically, as discussed in Appendix.~H, the $\sigma t$-hopping does not induce frustration, in contrast to the ordinary $t$-hopping. As a result, the system reduces approximately to a simple band scenario, where bosons tend to condense at the band minimum. When the magnitude of $t'$ is not too large, the band minimum remains at $(0,0)$, leading to a uniform phase diagram across the relevant parameter regime.

\section{\label{SM_Rydberg}K. Rydberg tweezer implementation}

\begin{figure}
   \includegraphics[width=0.9\textwidth,angle=0]{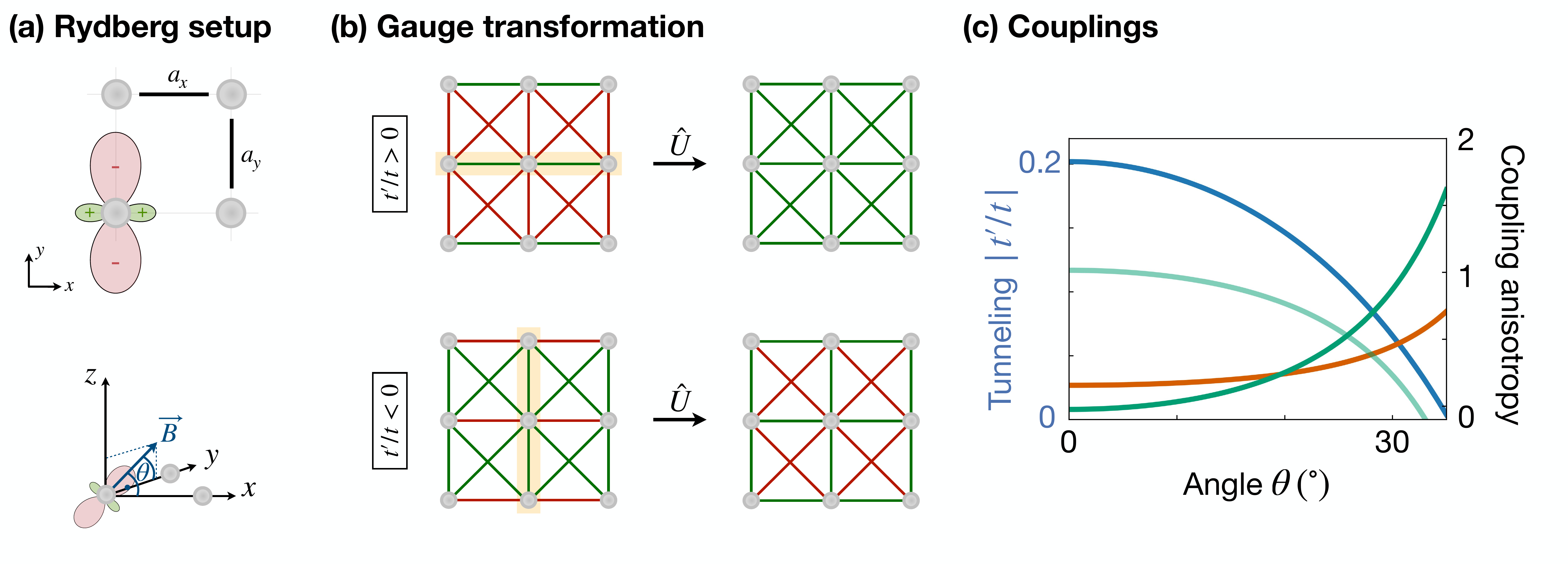}
   \caption{\label{SM_Rydberg}
   \textbf{Implementation of tunable $t'/t$ in Rydberg tweezer arrays.} (a) We propose to use the angular dependency of the dipole-dipole interactions to tune the ratio of $t'/t$. Top: By placing the quantization axis along the $y$-direction of the lattice, the tunnelings have opposite sign along the two lattice direction; to maintain the same overall strength~$|t^x/t^y|=1$ the lattice is distorted and shorted along the $x$-direction. Bottom: When the quantization axis is moved out of plane, the lattice has to be stretched along the $x$-direction. This displacement moves the diagonal atoms closer to the magic angle, where the tunneling vanishes; hence it enables us to smoothly vary the ratio $t'/t$. (b) By changing the overall sign of tunneling, see main text, a different structure of positive tunnelings (green) and negative tunnelings (red) is realized (Top: $\Delta m = \pm 1$, Bottom: $\Delta m = 0$). By applying the gauge transformation~$\hat{U}$, see Eq.~\eqref{eq:gauge_trafo}, on the sites highlighted in yellow, we obtain the desired sign structure of tunnelings with either~$t'/t>0$ or $t'/t<0$. (c) As the angle and lattice spacing is varied, the van-der-Waals interactions also get modified. In our scheme, the tunneling ratio $|t'/t|$ can be tuned to all relevant parameter regimes. We further plot the coupling anisotropy of spin flip-flop~$J_\perp^y/J_\perp^x$ interactions (orange) and density-density interactions~$V^y/V^x$ (light green: $t'/t>0$; dark green: $t'/t<0$).
   }
\end{figure}

The implementation of hard-core bosonic $t$-$t'$-$J$~models in Rydberg arrays utilizes the strong dipole-dipole interactions between highly-excited atomic states. Therefore, by choosing suitable atomic states the global sign of tunneling~$t,t'$ can be changed, as described in the main text. Here, we first provide additional details about the gauge transformation, allowing us to effectively change the sign of $t'/t$. Further, we discuss a scheme to tune the ratio $|t'/t|$ over a broad parameter range.

The resonant dipole-dipole interactions we consider (i.e. the atomic interactions giving rise to tunnelings~$t$) depend on the distance~$r_{ij} = |\vec{r}_{ij}|$ between two atoms~$i$ and $j$, and the angle~$\theta_{ij}$ between the interatomic vector~$\vec{r}$ with the quantization axis set by the magnetic field~$\vec{B}$. The tunneling amplitude is then defined as~$t_{ij} = -\frac{C_3}{r_{ij}^3}(1-3\cos^2\theta_{ij})$; in the following we truncate the long-range tails of the tunneling amplitudes and only consider up to next-nearest neighbor terms.

The conceptually simplest case, realized in Ref.~\cite{Qiao_2025} in two-dimensional arrays, is obtained for~$\theta=90^\circ$, i.e. when the magnetic field is pointing perpendicular to the atomic plane. Then, the Hamiltonian has spatially isotropic interactions with $t,t'>0$ ($\Delta m = \pm 1$) or $t,t'<0$ ($\Delta m = 0$) depending on the difference in total magnetic quantum number $\Delta m$ between the spin and hole atomic states. For $t,t'>0$, the model directly relates to the Hamiltonian studied in this work. For $t,t'<0$, we perform a gauge transformation~$\hat{U}$ on sublattice $A$ and~$B$ with
\begin{align}
\begin{split} \label{eq:gauge_trafo}
    \hat{U}^\dagger \hat{\mathcal{B}}^\dagger_{j,\sigma}\hat{U} &= - \hat{\mathcal{B}}^\dagger_{j,\sigma} ~~~~\mathrm{for}~j\in A \\
    \hat{U}^\dagger \hat{\mathcal{B}}^\dagger_{j,\sigma}\hat{U} &=  \hat{\mathcal{B}}^\dagger_{j,\sigma} ~~~~~~~\mathrm{for}~j\in B.
\end{split}
\end{align}
In the transformed frame, the Hamiltonian~$\hat{U}^\dagger\hat{\mathcal{H}}\hat{U}$ has (anti)ferromagnetic tunnelings between (next-)nearest-neighbor sites, which correspond to the case~$t'/t<0$.

Next, we discuss the relative magnitude~$|t'/t|$. For~$\theta=90^\circ$ the strength of tunneling~$\propto r_{ij}^{-3}$ only depends on the distance between two lattice sites. Therefore, on the square lattice~$|t'/t| \approx 0.35$, and thus this relatively strong next-nearest neighbour tunneling prevents one from accessing e.g. the PDW+AFM phase.
Nevertheless, we can use the angular degree-of-freedom to tune $|t'/t|$ over a broader parameter regime. To achieve this, we propose to set the quantization to~$\vec{B}/|\vec{B}|=(0,\cos\theta,\sin\theta)$, where the 2D array of atoms is in the $x$-$y$~plane, see Fig.~\ref{SM_Rydberg}(a). We set the lattice constant along the $x$-direction to $a_x \equiv 1$ and vary the lattice constant along the $y$-direction as~$a_y = |1-3\cos^2\theta|^{1/3}$, such that nearest neighbour tunnelings have equal magnitude $|t_x|=|t_y|=|t|$. Thus, the geometric constraints set the magnitude of $|t'/t| = 0 ... 0.2$ for $\theta=35^\circ ... 0^\circ$, see Fig.~\ref{SM_Rydberg}(c).

In our scheme, we consider angles~$\theta \lesssim 35^\circ$, where the sign of $t'/t_y > 0$ and $t_x/t_y < 0$; the global sign of the interaction can again be controlled by the choice of the magnetic quantum number. To obtain the correct sign structure of tunnelings, we perform a gauge transformation on every even row (column) if~$t_x>0$ ($t_x<0$), see Fig.~\ref{SM_Rydberg}(b), which allows us to realize both relative signs of $t'/t$ in the transformed frame.

Since the spin-spin interactions~$J$ and density-density interactions~$V$ scale with~$r_{ij}^{-6}$ in the Rydberg setup, the lattice anisotropy leads to spatially anisotropic couplings along the $x$- and $y$-directions; these affect in particular the spin-spin interactions~$J_\perp$ and~$J_z$ as well as the density-density interaction~$V$.
To be explicit, we choose the following set of atomic states in $^{87}$Rb:
\begin{align}
    \ket{60S,J=1/2,m_J=-1/2} \sim \ket{\!\downarrow}  ~~~~~ \ket{61S,J=1/2,m_J=-1/2} \sim \ket{\!\uparrow} ~~~~~ \ket{60P,J=1/2,m_J=\pm 1/2} \sim \ket{0}.
\end{align}
The configuration with $m_J=-1/2$ ($m_J=+1/2$) for the $P$-state implements the model with~$t'/t >0$ ($t'/t <0$). For this set of states, we explicitly compute the Rydberg-Rydberg interactions at a magnetic field of~$B=45\,\mathrm{G}$ using the \textit{pairinteraction} package~\cite{Weber2017}.

In Fig.~\ref{SM_Rydberg}(c), we plot the spatial dependency of the interactions as the angle~$\theta$ is varied. The ratio of tunneling~$|t'/t|$ can be tuned across the entire relevant parameter range to access the phase diagram shown in Fig. 1(a). The coupling anisotropy of spin interactions~$J_\perp^y/J_\perp^x$ and density-density interactions~$V^y/V^x$ along the $x$- and $y$-direction of the lattice exhibits an angular dependence, but remains within reasonable bounds. Future scheme based on microwave dressing of Rydberg states could allow to control the angular dependency of these interactions to achieve fully isotropic interactions.

%\end{CJK*}
\end{document}